\documentclass [preprint,aps,epsfig]{revtex4}
\def\be{\begin{eqnarray} &&}
\def\nonu{\nonumber \\ &&}
\def\ee{\end{eqnarray}}
\def\psla{\rlap \slash}

\input psfig.sty
\usepackage{graphicx}
\usepackage{epsfig}
\usepackage{epstopdf}
\begin{document}
\title{Space-like and time-like pion electromagnetic form factor
and Fock state components within the Light-Front dynamics}
\author{J. P. B. C. de Melo$^a$, T. Frederico$^b$, E. Pace$^c$ and
G. Salm\`e$^d$}
\affiliation{\
$^a$Centro de Ci\^encias Exatas e Tecnol\'ogicas,
 Universidade Cruzeiro do Sul, 08060-070,  and
Instituto de F\'\i sica Te\'orica, Universidade Estadual Paulista
 01405-900, S\~ao Paulo, Brazil \\
$^b$ Dep. de F\'\i sica, Instituto Tecnol\'ogico da Aeron\'autica,
Centro T\'ecnico Aeroespacial, 12.228-900 S\~ao Jos\'e dos
Campos, S\~ao Paulo, Brazil\\
$^c$ Dipartimento di Fisica, Universit\`a di Roma "Tor Vergata" and Istituto
Nazionale di Fisica Nucleare, Sezione Tor Vergata, Via della Ricerca
Scientifica 1, I-00133  Roma, Italy \\
$^d$Istituto  Nazionale di Fisica Nucleare, Sezione Roma I, P.le A. Moro 2,
 I-00185 Roma, Italy  }
\date{\today}
\begin{abstract}
The simultaneous investigation of the pion electromagnetic form factor in the space- and time-like
regions within a light-front model allows one to address the
issue of non-valence components of the pion and
photon wave functions.
Our relativistic approach is based on a microscopic vector meson dominance (VMD) model for the
dressed vertex where a photon decays in a quark-antiquark pair,
and on a simple parametrization for the emission or absorption of a pion
by a quark. The
results show an excellent agreement in the space like region up
to -10 $(GeV/c)^2$, while in time-like region the model produces
reasonable results up to 10 $(GeV/c)^2$.
\end{abstract}
\maketitle
\section{Introduction}
Electroweak properties are widely used as an important source
of information on the structure of hadrons. In particular
within the framework of
light-front  (LF) dynamics \cite{dirac,brodsky,karmanov,kp,LEV}, a large number
of papers has been devoted to the study of nuclei and hadrons
(see e.g.
\cite{chung,we,Jaus90,tob92,carpi,card,salme,sim,nua,Jaus99,choi01,JI01,pion99,ba01,pach02,
Hwang,ba02},
just to give a partial account of previous works with a finite number of constituents).

The LF dynamics allows one to exploit the intuitive language of the Fock space.
Indeed the Fock-space language is particularly meaningful within
LF dynamics, since: "The simplicity of the light-cone Fock
representation as compared to that in equal-time quantization is
directly linked to the fact that the physical vacuum state has a
much simpler structure on the light-cone because the Fock vacuum
is an exact eigenstate of the full Hamiltonian."\cite{brodsky}

Another basic motivation for choosing the LF dynamics is
represented by the striking feature that the Fock decomposition is
stable under LF boosts, since they are of kinematical
nature and therefore  do not change the  number of particles,
i.e., are diagonal in the Fock space.

Therefore the LF dynamics is a suitable framework for the investigation of the Fock
expansion for mesons and baryons, viz
\be
| meson \rangle =  |q\bar{q} \rangle + |q \bar{q} q \bar{q}\rangle +
|q \bar{q} ~g\rangle +
 .....
\nonu
| baryon \rangle = |qqq \rangle +
|qqq~q \bar{q} \rangle +|qqq~g \rangle +
 .....
\ee

 In particular, within the LF dynamics the
electromagnetic form factor of the pion has
been the object of many papers
(see, e.g., Refs.
\cite{chung,tob92,carpi,card,salme,choi01,pion99,Hwang,pach02}).
Indeed the pion  electromagnetic form factor yields a simple tool for
the investigation of pion and photon microscopic
structure in terms of hadronic constituents. In
what follows we will present an approach to investigate in a common
framework the
pion  and  photon vertex functions, with the perspective of an extension of our
approach to the nucleon. The intuitive
language of the Fock space will be widely  exploited
 to analyze the above mentioned vertex functions.

Aim of this work is to give a unified description of the  electromagnetic form
factor of the pion, both in the space-like (SL) and in the time-like (TL) regions,
taking into account the complexities related to the pion and photon
vertex functions, both in the valence and in the nonvalence sectors,
as well as {\em{the fermionic nature of the constituents}}.
A first presentation of our approach was given in Ref. \cite{DFPS}.

The choice of the {\em reference frame} where the form factor analyses
are carried out has a fundamental role, as shown in previous
works in the space-like region \cite{pach98,pion99,ba01,pach02}
and in the time-like one \cite{choi01}.
For a unified description of TL and SL form factors,
a reference frame is needed where the plus component of the momentum transfer, $q^+ = q^0 + q^3$,
is different from zero (otherwise, $q^2=q^+ q^- - q^2_{\perp}$ cannot be positive).
  As a matter of fact, a
reference frame where $q^+\neq 0$ allows one to analyze, in a
common framework \cite{DFPS}, the pair
production process (Z-diagram contribution) \cite{JI01}, i.e. the effect of  multiquark propagation,
as well as the ultrarelativistic effect of the so-called instantaneous contributions and the hadronic
components of the photon wave function \cite{brodsky,ashery}.

In Ref. \cite{LPS} it was shown that, within the Hamiltonian LF
 dynamics (HLFD), a Poincar\'e covariant and conserved current
operator can be obtained
from the matrix elements of the free current, evaluated in the Breit
reference frame,
where the initial and the final total momenta of the system are directed
along the
spin quantization axis, $z$. Following Ref. \cite{LPS}, we calculate
the pion form factor in a
reference frame where ${\bf q}_{\perp}=0$ and $q^+>0$.

Our starting point is the {\em Mandelstam formula} \cite{mandel}. To construct a bridge toward the
Hamiltonian language, the hadron vertex functions will be connected to the LF
wave function of the valence component of the hadron state. Furthermore, the concept of
hadronic valence, i.e. $q\bar q$, component of the photon wave
function will be introduced \cite{brodsky,ashery}.

The main difficulties to be dealt with are: i) how to construct the
photon-hadron coupling when a $q\bar q$ pair is produced by a photon
with $q^+>0$; and ii) how to describe the nonvalence content relevant for the
process under
consideration, both in
the pion and in the photon wave functions.

The first issue is addressed  by using a {\em covariant generalization of
the vector meson dominance approach}
(see, e.g., \cite{Connell}) at the
level of the photon vertex function (see Ref. \cite{DFPS}).

 As a matter of fact, it is necessary to construct the
Green's function of the interacting $q\bar q$ pair in the $1^-$ channel. For the
description of the vector meson vertex functions in the valence sector we  use
the eigenfunctions of the square
mass operator proposed in Refs. \cite{pauli,tobpauli}. The simplified version
of the model that we are going to use \cite{FPZ02} includes
confinement through a harmonic oscillator potential.
The model showed a universal and
satisfactory description of the experimental values of the masses
of both singlet and triplet  $S$-wave mesons and the
corresponding radial excitations \cite{FPZ02}, giving
a natural explanation of the almost linear relationship
between the mass squared of excited states and the radial quantum
number $n$~\cite{iach,ani}. Therefore such a relativistic QCD-inspired
model for pseudoscalar and vector mesons retains the main feature
of the spectra and at the same time allows one to perform simple numerical calculations.

The second issue, i.e. {\em the contribution of the nonvalence
($2q2\bar q$) components} of the pion and photon wave functions,
is addressed using a  model where a quark in the valence
component radiates a pair by a contact interaction \cite{JI01}.
This interaction is described through a pseudoscalar coupling of quark and pion fields,
multiplied by a constant.
In a recent study of meson decay processes
 within LF dynamics \cite{JI01},
this approximation  was shown  to give a good description of the
experimental data. Here, we just follow the above
suggestion to parameterize the radiative pion emission amplitude
from the quark.

Another important point to be treated carefully is the {\em contribution of
the instantaneous terms}, which is strictly related to the fermionic nature of the constituents.
We remind the reader that the Dirac propagator can be
decomposed using the light-front momentum components \cite{brodsky}, as
follows:
\begin{eqnarray}
\frac{\rlap\slash{k}+m}{k^2-m^2+\imath \epsilon} =
\frac{\rlap\slash{k}_{on}+m}{k^+(k^--k^-_{on}+\frac{\imath
\epsilon}{k^+})} +\frac{\gamma^+}{2k^+} \ ,
\label{inst}
\end{eqnarray}
where $\gamma^+ = \gamma^0 + \gamma^3$ and $k^-_{on}=(|{\bf k}_{\perp}|^{2}+m^2)/{k^+}$.
The second term on the right-hand side of Eq. (\ref{inst}) is
an instantaneous term in the light-front time, related to the so-called zero modes.
As already known (see, e.g., \cite{pach02}), the instantaneous contributions play a
 dominant role  in the description
of the pion electromagnetic form factor in the space-like region,
in a reference frame where $q^+>0$. Therefore a
special care is devoted in the present work
to the treatment of the {\it instantaneous}  contributions in the
light-cone representation of the fermion propagators. In particular the contributions of
the zero modes are under control, thanks to the momentum behavior
of the hadron vertex functions. It should be pointed out
that the effects of the instantaneous terms is emphasized  by the small mass of the pion.

Our description contains a small set of parameters: the
oscillator strength, the constituent quark mass,
and the width for the vector mesons. We use  experimental
widths for the vector mesons, when available \cite{pdg}, while for the unknown widths
of the radial excitations we use a single width as a fitting parameter. The
constant involved  in the description of the nonvalence component can be
fixed by the pion charge normalization in the limit of a vanishing pion mass.
The evaluation of the instantaneous vertex functions involves a further
parameter (see Sect. X).

Previously, the elastic time-like form factor was explored in
the light-front quantization in a boson model of $q\overline Q$ mesons
with point-like vertexes \cite{choi01}, which does not exploit the
rich structure of the meson excited states. Here, by studying in a
common framework the pion space- and time-like form factors, we
also access information from the radially excited vector meson
wave functions. Indeed, in our approach, in the time-like region
 the virtual
photon couples directly to the vector meson resonances, which in turn
decay in $\pi^+\pi^-$.  Therefore our microscopical model could represent a useful
tool to address the investigation of the vector meson Green function.

In the present paper in order to simplify the numerical calculations,
{\em we use a massless pion},
i.e. we evaluate the pion form factor at the chiral limit.
In the time-like region the full result for the pion form factor
is always given by the pair-production process ("Z-diagram") alone,
independently of this approximation. In the
space-like region only the  "Z-diagram" contribution
\cite{sawicki,pach98,pion99,ba01,pach02,ba02} survives  for
a massless pion \cite{DFPS}. The importance of the "Z-diagram"
contribution to the electromagnetic current for $q^+>0$ was also
recently investigated in the context of the Bethe-Salpeter
equation within the light-front quantization in Ref. \cite{tiburzi}.

This work is organized as follows. In  Sec. II, we present the
general form of the covariant electromagnetic form factor of the
pion in impulse approximation, which is our starting point, and
our vector-meson-dominance
approach for the dressed photon vertex.
In Sec. III, we first decompose the triangle diagram
in on-shell and instantaneous contributions. Then we
integrate  for $q^+~>~0$ on the light-front energy in the momentum
loop of the triangle diagram, under analytical assumptions for the
vertex functions.

The valence components of the light-front meson and
photon wave functions are defined in Sec. IV.
In Sec. V, we discuss the contribution of the nonvalence component of the
photon to the time-like current, which appears through the vertex
for the radiative emission of pions by a virtual quark inside the
photon. In this section, we also discuss the contribution of the
nonvalence component of the pion wave function to the space-like
current.
In Sec. VI, we introduce the pion and vector
meson wave functions in the expression of the triangle diagram.
The pion time-like and
space-like form factors written in terms of the valence components
of the meson wave functions and of the emission/absorption vertexes are derived
in Sec. VII and Sec. VIII, respectively. In Sec. IX, we briefly revise
the light-front model for the pion and vector mesons and conclude
the derivation of our model for the pion form factor with a discussion
on our treatment of the vertex functions for the instantaneous terms. In Sec. X,
we present the numerical results for the pion form factor in the momentum transfer range
between -10 $(GeV/c)^2$ and +10 $(GeV/c)^2$. In Sec. XI, our conclusions are
presented.

\section{Covariant em form factor of the pion}
Our starting point is the covariant expression for the amplitude of the processes
 $\pi~\gamma^* \rightarrow \pi'$, or
$\gamma^* \rightarrow \pi \pi'$, where the meson $\pi'$ is a pion
in the elastic case or an antipion in the production process,
evaluated in impulse approximation \cite{mandel}
 (see the triangle diagram of Fig. 1). In the time-like region one has (see Fig. 2)
\begin{eqnarray}
j^{\mu} &=& \langle \bar{\pi} \pi | J^{\mu} (q) | 0 \rangle =
~-\imath ~ 2 ~ e ~ \frac{m^2}{f^2_\pi} N_c\int \frac{d^4k}{(2\pi)^4}
~ \Lambda_{\bar{\pi}}(k - P_{\pi},P_{\bar{\pi}}) ~
\overline \Lambda_{\pi}(k,P_{\pi}) ~\times  \nonu
Tr[S(k - P_{\pi}) ~ \gamma^5 ~
S(k-q) ~ \Gamma^\mu(k,q) ~ S(k) ~ \gamma^5 ] \ ,
\label{jmu}
\ee
where
$N_c=3$ is the number of colors; $\displaystyle
S(p)=\frac{1}{\rlap\slash p-m + \imath \epsilon} \,$
is the quark propagator with $m$ the mass of the constituent quark;
$q^{\mu}$ is the virtual photon momentum; $P^{\mu}_{\pi}$ and
 $P^{\mu}_{\bar{\pi}}$ are the pion momenta.
 The factor 2 stems from isospin algebra, since
\begin{eqnarray}
Tr \left [ \frac{\tau _x -  \imath \tau _y} {\sqrt 2} ~ \frac {1 + \tau _z} {2} ~
\frac{\tau _x +  \imath \tau _y} {\sqrt 2} \right ] ~  = ~ 2 \ \ ,
\label{iso}
\ee
where $( 1 + \tau _z ) / 2$ is the isospin factor of the current
and the other isospin factors in Eq. (\ref{iso}) pertain to the pions.

The function $\overline \Lambda_{\pi}(k,P_{\pi})$ is the momentum component of
 the $q\bar{q}$ vertex function for the outgoing
pion, which will be taken as a symmetric function of the $q$,
$\bar{q}$ momenta. In this vertex function, $P_{\pi}$ is the momentum of the
outgoing pion and $k$ is the momentum of the incoming quark (see Fig. 2).
The "bar" notation on the vertex function labels the adjoint Bethe-Salpeter amplitude,
i.e. the solution of a Bethe-Salpeter equation where the
two-body irreducible kernel is placed on the right of the
amplitude, while for the Bethe-Salpeter amplitude it is placed on the left
 \cite{lurie,izuber}. This is a
well known property of time orderings implied by the Mandelstam
formula, for initial and final states \cite{mandel,lurie}.
The vertex function  is defined by the following equation
\begin{eqnarray}
&& \imath \frac{\rlap\slash{k} +m}{k^{ 2}-m^2+\imath \epsilon} ~\gamma_5~
\Lambda_{\pi}(k,P_{\pi}) ~ \frac{\psla k^\prime+m}{k^{\prime 2}- m^2 + \imath \epsilon}
 ~ \delta^4( k^\prime + P_{\pi} - k)
=\frac{1}{(2\pi)^4}\int d^4x~d^4y
\times
\nonu
\exp{i(k^\prime\cdot y - k\cdot x )} ~
\langle 0 | \text{T}\left[ q (x)~\overline
q (y)\right]|P_{\pi}\rangle ~~~,
\label{tpf}
\end{eqnarray}
where $q(x)$ is the quark field.

 To obtain the current matrix element for the
space-like region, $P^\mu_\pi$ should be replaced by $-P^\mu_\pi$
and $\bar{\pi}$ by $\pi'$.
Then the pion vertexes $\overline \Lambda_{\pi}(k,P_{\pi})$ and
$\Lambda_{\bar{\pi}}(k - P_{\pi},P_{\bar{\pi}})$ in Eq. (\ref{jmu})
are to be changed with $\Lambda_{\pi}(-k,P_{\pi})$ and
$\overline \Lambda_{\pi^{\prime}}(k + P_{\pi},P_{\pi^{\prime}})$, respectively (see Fig. 3).
The momentum dependence of the vertex functions
$\Lambda_{\bar{\pi}}(k - P_{\pi},P_{\bar{\pi}})$ and  $
\overline \Lambda_{\pi}(k,P_{\pi})$ is
expected to regularize the integrals of Eq. (\ref{jmu}).

The dressed photon-vertex, $\Gamma^\mu(k,q)$, is related to the
photon Bethe-Salpeter amplitude, which is defined  from the three-point
function in the standard form:
\begin{eqnarray}
&&\imath \frac{\psla k^\prime+m}{k^{\prime 2}-m^2+\imath \epsilon}
\Gamma^\mu (k,q) \frac{\rlap\slash{k} +m}{k^{ 2}-m^2+\imath
\epsilon} \delta^4( k^\prime+q-k)=\frac{1}{(2\pi)^4}\int
d^4x~d^4x^\prime~d^4x^{\prime\prime}
\times
\nonu
\exp{i(k^\prime\cdot x^\prime -k\cdot x + q\cdot
x^{\prime\prime})} R^\mu_3(x,x^\prime,x^{\prime\prime}) \ .
\label{wf1zg}
\end{eqnarray}
The three-point function is given by
\begin{eqnarray}
  R^\mu_3(x,x^\prime,x^{\prime\prime})=
\langle 0 | \text{T}\left[ q (x)~\overline
q(x^{\prime\prime})\gamma^\mu q(x^{\prime \prime})\overline
q(x^\prime)\right]|0\rangle ~~~,
\label{tpf1}
\end{eqnarray}
which is the matrix element between the vacuum states of the time
ordered product of the four quark fields written above
\cite{izuber}.

The central assumption of our paper is the
microscopical description of the dressed photon vertex,
$\Gamma^\mu(k,q)$,
in the processes where a photon with $q^+>0$ decays in a quark-antiquark pair. In these processes
 we use for the  photon vertex,
 dressed by the interaction between the $q\overline q$ pair,
 the following covariant vector meson dominance approximation
(see Fig. 4)
\begin{eqnarray}
\Gamma^{\mu}(k,q) &=& \sqrt{2} \sum_{n}
\left [ -g^{\mu \nu} + {q^{\mu}  q^{\nu} \over M_n^2} \right ]~ \widehat{V}_{n \nu}(k,k-q)
~ \Lambda_{n}(k,q) ~ { f_{Vn} \over \left [ q^2 -
M^2_n + \imath M_n \tilde{\Gamma}_n(q^2)\right ]} \ ,
\label{cur6}
\end{eqnarray}
where
\begin{eqnarray}
 \left [ -g^{\mu \nu} + {q^{\mu}  q^{\nu} \over M_n^2} \right ]
 {1 \over \left [ q^2 -
M^2_n + \imath M_n \tilde{\Gamma}_n(q^2)\right ]} \ ,
\label{prop}
\end{eqnarray}
is the vector meson propagator \cite{Halzen}.
In Eq. (\ref{cur6}) $f_{Vn}$ is the decay constant of the $n{\rm th}$ vector
meson in a virtual photon, $M_n$  the corresponding
 mass, $\Lambda_{n}(k,q)$ gives the momentum dependence
and $\widehat{V}_{n \nu}(k,k-q)$ the Dirac structure of the VM vertex function,
while $\tilde{\Gamma}_n(q^2)$ is the total decay width.
 If we approximate Eq. (\ref{cur6}) considering on-shell quantities
 for the VM in the numerator,
 i.e. if we replace $q^-$ with $P^-_n=(|{\bf q}_{\perp}|^2+M^2_n)/q^+$,
 we have
\begin{eqnarray}
\Gamma^{\mu}(k,q) &=& \sqrt{2} \sum_{n, \lambda}
\left [ \epsilon_{\lambda} (P_n)\cdot \widehat{V}_{n}(k,k-P_n)  \right ]
\Lambda_{n}(k,P_n) ~ { [\epsilon ^{\mu}_{\lambda}(P_n)]^* f_{Vn} \over \left [ q^2 -
M^2_n + \imath M_n \tilde{\Gamma}_n(q^2)\right ]} \ ,
\label{cur7}
\end{eqnarray}
where the quantity
$\left [ \epsilon_{\lambda}(P_n) \cdot \widehat{V}_{n}(k,k-P_n)  \right ]  ~
\Lambda_{n}(k,P_n)$ is the VM vertex function  and $\epsilon_{\lambda}(P_n)$
the VM polarization.  Note that the total momentum for
an on-shell vector meson is
 $P^{\mu}_n \equiv \{P^-_n=(|{\bf q}_{\perp}|^2+M^2_n)/q^+,
{\bf P}_{n \perp}={\bf q}_{\perp}, P^+_{n}=q^+ \}$,
while $q^{\mu}
\equiv \{q^-, {\bf q}_{\perp}, q^+ \}$
and that at the production vertex, see Fig. 4, the light-front
three-momentum is conserved.

In Eq. (\ref{cur6}) the sum runs over
all the possible vector mesons. The
vector meson decay constant, $f_{Vn}$, can be obtained from the definition \cite{Jaus99}
\be
\epsilon^{\mu}_{\lambda} \sqrt{2} f_{V,n} = \langle 0| \bar{q}(0) \gamma^{\mu}
q(0)|\phi _{n,\lambda}\rangle
\label{fVap}
\ee
with $|\phi _{n,\lambda}\rangle$ the vector meson state. A detailed expression for $f_{Vn}$
is given in Appendix A.
The total decay width in the denominator of Eqs. (\ref{prop}) and (\ref{cur7}), $\tilde{\Gamma}_n(q^2)$,
is vanishing in the SL
 region. In the TL region it is assumed to be equal to
\be
\tilde{\Gamma}_n(q^2) = \Gamma_{n} ~ \left[ {p(q^2) \over p(M_{n}^2) } \right]^3
\left[ { M_{n}^2 \over q^2} \right]^{1/2}
\label{Gamma}
\ee
where  $p(q^2) = [q^2 - 4 m_\pi^2]^{1/2}/2$ ~ ~ ~ \cite{Saku,Klingl,Benayoun}.

In Ref. \cite{pach97} the
following expression was used for the Dirac structure, $\widehat{V}_{n}(k,k-P_n)$,
 of the vector meson vertex :
\be
\widehat{V}^{\mu}_{n}(k,k') = \gamma^{\mu}-{M_n
\over 2}{k^{\mu}+k'^{\mu} \over P_n \cdot k +mM_n}
\label{gamV}
\ee
where $k' = k - P_n$.

 Let us consider, instead of Eq. (\ref{gamV}),
 a symmetric form for $\widehat{V}_{n}(k,k-P_n)$:
  \be
\widehat{V}^{\mu}_{n}(k,k-P_n) =
\gamma^{\mu}- M_n ~ {k^{\mu} + k'^{\mu} \over P_n \cdot k - P_n \cdot k' + 2 mM_n} =
 \gamma^{\mu} - {k^{\mu} + k'^{\mu}  \over  M_n + 2 m } \; \; .
\label{gams2}
\ee
 If in Eq. (\ref{gams2}) both the CQ's are taken on their
mass shell (i.e., $k^- = k^-_{on} = (|{\bf k}_{\perp}|^2+m^2)/k^+ $)
and the VM mass, $M_n$, is replaced by the free mass,
$M_0$,
of the quarks in a system of total momentum $q^{\mu}$,
\begin{eqnarray}
M_0 = \left [ (k_{on} + (q-k)_{on}) \cdot (k_{on} +( q-k)_{on}) \right ]^{1/2}
\label{M0}
\ee
one obtains
\begin{eqnarray}
\left [ \widehat{V}^{\mu}_{n}(k,k-q)\right ]_{on}
&& = ~ \gamma^{\mu} -
{k^{\mu}_{on}-(q-k)_{on}^{\mu} \over  M_0 + 2 m } \; \; .
\label{gams1}
\ee
This form coincides with the on-shell expression given in
 Ref. \cite{Jaus90} for the $^3S_1$ vector meson vertex,
 but then $\Gamma^{\mu}(k,q)$ of Eq. (\ref{cur7}) is not anymore a four vector.

  Let us note that to obtain the pion form factor one actually needs
 only one of the components of the current. In the following we will derive the
 pion form factor from the {\em plus} component.

In Ref. \cite{DFPS} to evaluate the pion form factor
we considered the plus component of Eq. (\ref{cur7}),
where  $\Lambda_{n}$  and the VM polarizations were
taken at the vector meson pole, and the
on-shell expression for $\widehat{V}^{\mu}_{n}$, as given by  Eq. (\ref{gams1}), was used,
in order to have the structure of the VM vertex suggested by the Hamiltonian
LF dynamics.

 In Appendix B, starting  from Eq. (\ref{cur6}), we propose a current
 which satisfies current conservation. In the reference frame, where
 $q^+ = M_n > 0$ and ${\bf q }_\perp=0$, the $n{\rm th}$ term of this current has
  exactly the same plus component as the $n{\rm th}$ term of the current defined in Eq. (\ref{cur7}).

 One might wonder that a bare $\gamma ^{\mu}$ coupling term should be added to the current
 defined in
 Eqs. (\ref{cur6}) or (\ref{cur7}). However, as it is shown in Appendix C, in the case of a massless pion
 a bare coupling produces violation of current conservation. Therefore we do not consider this term in
 the present paper.

\section{Approximating the triangle diagram on the light-front }
Our aim is to retain the essential physics contained in  the triangle
diagram (Fig. 1) and at the same time
to construct a bridge toward the Hamiltonian wave function language.
At the same time we wish to go beyond a simple valence description. To
accomplish these goals and to
eliminate the relative light-front time between the
quarks, we perform the $k^-$ integration in Eq. (\ref{jmu}) with
 some  assumptions on the analytical structure of
the $\Lambda$ and $\Gamma$ vertexes  for the pion and the photon.
To be more precise, Eq. (\ref{jmu}) is
evaluated  with the assumptions that: (i) the momentum components,
$\Lambda(k,P)$,
of the vertex functions,
both for the pion and the vector mesons,
vanish in the complex plane $k^-$ for
$|k^-|\rightarrow\infty$; and (ii) the contributions of the possible
singularities of $\Lambda(k,P)$ can be neglected. Furthermore, the Dirac
structures of the vector meson vertex
function, $\widehat{V}^{\mu}_{n}(k,k-q)$,
are assumed to be  regular functions
of the complex variable $k^-$.
The expressions for $\widehat{V}^{\mu}_{n}(k,k')$ given
by Eqs.  (\ref{gams1}) or (\ref{gams2})
obviously fulfill
the requirement that no pole is present in the $k^-$ complex plane.

To make clear the discussion of the $k^-$ integration, it is helpful to first separate
instantaneous and non-instantaneous contributions, using the decomposition of the Dirac
propagator given in Eq. (\ref{inst}). Indeed this decomposition is useful to have a
better control on
possible divergences both in $k^-$ and in $k^+$ integrations. In particular,
as already mentioned, it should be
pointed out the tight relation between the instantaneous terms and the so-called zero modes,
where $k^+ = 0$. We assume that the behavior of the functions $\Lambda(k,P)$ in $k^+$ is able
to regularize the divergences at the $k^+$ end points \cite{SPPF}.

Since three propagators are present in Eq. (\ref{jmu}), one has a total of
eight contributions. The
contribution with three instantaneous terms vanishes because of the property
$\gamma ^+ ~ \gamma ^+ = 0$, since the combination
$\gamma ^+ ~\gamma ^5 ~\gamma ^+$ appears.
Also the three contributions with two instantaneous terms vanish,
as a consequence of our assumptions on $\Lambda(k,P)$.
Indeed, only a single pole from the
propagators is present in these contributions. Then, since we assume that
the functions $\Lambda(k,P)$ go to zero for
$|k^-|\rightarrow\infty$ and disregard their singularities, we can
perform the integration in the $k^-$
complex plane closing the contour in the semiplane where no singularity
in the propagators is present
and we obtain a null result. Moreover, two of these contributions with two instantaneous terms
are also identically vanishing because of the
presence of the combination $\gamma ^+ ~\gamma ^5 ~\gamma ^+$.

Therefore we are left with four contributions : three contributions with  one
instantaneous term only and one contribution with no instantaneous term.

To evaluate the triangle diagram we treat separately the time-like
case and the space-like case.

\subsection{Time-like case}
In the time-like case, one has $q^{\mu} = P^{\mu}_{\pi} +
P^{\mu}_{\bar{\pi}}$, and $q^+ > 0$ .
Equation (\ref{jmu})  written in light-front variables becomes
(the Jacobian for the transformation to the light-front variables is 1/2):
\begin{eqnarray}
&&j^{\mu} = ~-\imath \frac{e} {(2\pi)^4} \frac{m^2}{f^2_\pi} N_c~
\int \frac{dk^- dk^+ d{\bf k}_{\perp}}{(k^+ - P^+_{\pi}) k^+ (k^+
- q^+)} ~ Tr[{\cal O}^{\mu}]  ~ \times \nonu
\frac{\Lambda_{\bar{\pi}}(k - P_{\pi},P_{\bar{\pi}})
\overline\Lambda_{\pi}(k,P_{\pi})}{(k^- - k^-_{on} +  \frac{\imath
\epsilon}{k^+}) (k^- - q^- -(k - q)^-_{on} +
\frac{\imath\epsilon}{k^+ - q^+}) (k^- -P^-_{\pi} - (k -
P_{\pi})^-_{on} + \frac{\imath\epsilon}{k^+ - P^+_{\pi}})} \ .
\label{jmuA}
\ee
The on-mass-shell values of the
minus-components of the momenta in Eq. (\ref{jmuA}) are given by
\begin{eqnarray}
k^-_{on}=\frac{{\bf k}_{\perp}^{2}+m^2}{k^+} \ , \quad
(k-q)^-_{on}=\frac{({\bf k-q})_{\perp}^{2}+m^2}{k^+ - q^+} \ ,
\quad \
 (k - P_{\pi})^-_{on} = \frac{({\bf k - P}_{\pi})_{\perp}^{2}+m^2}{k^+ - P^+_{\pi}} \ ,
 \label{onek}
 \end{eqnarray}
 and  the operator
${\cal O}^{\mu}$ is defined as follows
\begin{eqnarray}
{\cal O}^{\mu} ~ = ~ (\rlap\slash k - \rlap\slash P_{\pi} + m)
~\gamma^5 (\rlap\slash k - \rlap\slash q + m)~ \Gamma^\mu(k,q)
~(\rlap\slash k + m)~ \gamma^5 \ .
\label{O}
\end{eqnarray}

Let us perform the decomposition of the propagators in instantaneous and in on-shell
parts (see Eq. (\ref{inst})), as discussed at the beginning of this section.
Then Eq. (\ref{jmuA}) becomes
\begin{eqnarray}
&&j^{\mu} =  {\cal J}^{\mu}_{on} + {\cal J}^{\mu}_{1} + {\cal J}^{\mu}_{2}
+ {\cal J}^{\mu}_{3}
\label{jmuAA}
\ee
where ${\cal J}^{\mu}_{on}$ represents the on-shell contribution
 and  ${\cal J}^{\mu}_{i} (i=1,2,3)$ represent the contributions with one instantaneous term.
 Then we have
\begin{eqnarray}
&&{\cal J}^{\mu}_{on} = -\imath \frac{e} {(2\pi)^4} \frac{m^2}{f^2_\pi} N_c~
\int \frac{dk^- dk^+ d{\bf k}_{\perp}}{(k^+ - P^+_{\pi}) k^+ (k^+
- q^+)} ~  \Lambda_{\bar{\pi}}(k - P_{\pi},P_{\bar{\pi}}) ~
\overline\Lambda_{\pi}(k,P_{\pi}) ~ {\cal T}^{\mu}_{on}
\label{jmuAon}
\ee
and
\begin{eqnarray}
&&{\cal J}^{\mu}_{i} = -\imath \frac{e} {(2\pi)^4} \frac{m^2}{f^2_\pi} N_c~
\int \frac{dk^- dk^+ d{\bf k}_{\perp}}{(k^+ - P^+_{\pi}) k^+ (k^+
- q^+)} ~  \Lambda_{\bar{\pi}}(k - P_{\pi},P_{\bar{\pi}}) ~
\overline\Lambda_{\pi}(k,P_{\pi}) ~ {\cal T}^{\mu}_i
\label{jmuAi}
\ee
where
\begin{eqnarray}
{\cal T}^{\mu}_{on} =
\frac{Tr[[(\rlap\slash k - \rlap\slash P_{\pi})_{on} + m]
~\gamma^5 [(\rlap\slash k - \rlap\slash q)_{on} + m]~ \Gamma^\mu(k,q)
~(\rlap\slash k_{on} + m)~ \gamma^5 ]}
{(k^- - k^-_{on} +  \frac{\imath \epsilon}{k^+})
 (k^- - q^- -(k - q)^-_{on} + \frac{\imath\epsilon}{k^+ - q^+})
 (k^- -P^-_{\pi} - (k -
P_{\pi})^-_{on} + \frac{\imath\epsilon}{k^+ - P^+_{\pi}})}
\label{TmuAon}
\ee
\begin{eqnarray}
&&{\cal T}^{\mu}_{1} =
\frac{Tr[ \gamma^+
~\gamma^5 ~ [(\rlap\slash k - \rlap\slash q)_{on} + m]~ \Gamma^\mu(k,q)
~(\rlap\slash k_{on} + m)~ \gamma^5 ]}{2 ~ (k^- - k^-_{on} +  \frac{\imath
\epsilon}{k^+}) (k^- - q^- -(k - q)^-_{on} +
\frac{\imath\epsilon}{k^+ - q^+}) }
\label{jmuA1}
\ee
\begin{eqnarray}
&&{\cal T}^{\mu}_{2} =
\frac{Tr[[(\rlap\slash k - \rlap\slash P_{\pi})_{on} + m]
~\gamma^5 ~ \gamma^+ ~ \Gamma^\mu(k,q)
~(\rlap\slash k_{on} + m)~ \gamma^5 ]}{2 ~ (k^- - k^-_{on} +  \frac{\imath
\epsilon}{k^+})  (k^- -P^-_{\pi} - (k -
P_{\pi})^-_{on} + \frac{\imath\epsilon}{k^+ - P^+_{\pi}})}
\label{jmuA2}
\ee
\begin{eqnarray}
&&{\cal T}^{\mu}_{3} =
\frac{Tr[[(\rlap\slash k - \rlap\slash P_{\pi})_{on} + m]
~\gamma^5 ~ [(\rlap\slash k - \rlap\slash q)_{on} + m]~ \Gamma^\mu(k,q)
~ \gamma^+ ~ \gamma^5 ]}{2 ~ (k^- - q^- -(k - q)^-_{on} +
\frac{\imath\epsilon}{k^+ - q^+}) (k^- -P^-_{\pi} - (k -
P_{\pi})^-_{on} + \frac{\imath\epsilon}{k^+ - P^+_{\pi}})} \ .
\label{jmuA3}
\ee

In Eqs. (\ref{jmuAon}, \ref{jmuAi}) the three propagators of the triangle diagram
generate three poles:
\begin{eqnarray}
k^-_{(1)} &=& k^-_{on}~ - ~ \frac{\imath \epsilon}{k^+} \ , \nonumber \\
k^-_{(2)} &=& q^- ~+ ~(k - q)^-_{on}~ - ~\frac{\imath\epsilon}{k^+
- q^+} \ ,
\nonumber \\
k^-_{(3)} &=& P^-_{\pi} ~+~ (k - P_{\pi})^-_{on} ~ - ~
\frac{\imath\epsilon}{k^+ - P^+_{\pi}} \ .
\label{po}
\end{eqnarray}

Within our assumptions on the vertex functions,
 $\Lambda(k,P)$, if $k^+ < 0$ there are no
poles in the lower complex semi-plane of $k^-$ (cf. Eq. (\ref{po})).
Therefore, if the $k^-$ integration is performed
by closing the contour of integration in the lower
semi-plane, a vanishing result is obtained. Furthermore, if $k^+ >
q^+$, there are no poles in the upper complex semi-plane and a
vanishing result is obtained by closing the contour in the upper
semi-plane. Then, the integrals (\ref{jmuAon}, \ref{jmuAi})
 have contributions only for $k^+$ in
the range $0 < k^+ < q^+ $. The integration range can be
decomposed in two intervals, $0 < k^+ < P^+_{\pi}$ and $P^+_{\pi}
< k^+ < q^+$. In the first one, if the $k^-$ integration contour is
closed in the lower semi-plane, only the pole $k^-_{(1)}$ falls
within the integration contour, while in the second one,  if the
integration contour is closed in the upper semi-plane, only the
pole $k^-_{(2)}$ falls within the integration contour. Then
in the range $0 < k^+ < P^+_{\pi}$ one has contributions from
${\cal J}^{\mu}_{on}$, ${\cal J}^{\mu}_{1}$ and
${\cal J}^{\mu}_{2}$, while in the range $P^+_{\pi} < k^+ < q^+$ one has contributions from
${\cal J}^{\mu}_{on}$, ${\cal J}^{\mu}_{1}$ and
${\cal J}^{\mu}_{3}$.

Let us introduce the free mass, $M_{0}(k^+, {\bf
k}_{\perp}; P^+, {\bf P}_{\perp} )$, of a $q\bar{q}$
 system of mass $M$, total momentum $P$, kinematical
momenta $(k^+, {\bf k}_{\perp})$ and $(P^+-k^+, {\bf P}_\perp
-{\bf k}_{\perp})$ for the pair of quarks:
\begin{eqnarray}
M_{0}^2(k^+, {\bf k}_{\perp}; P^+, {\bf P}_{\perp} )
= \frac{{\bf k}^2_\perp+m^2}{x} + \frac{
({\bf P-k})^2_\perp + m^2}{1-x}- {\bf P}^2_\perp \
\label{M0q}
\end{eqnarray}
where $x = k^+/P^+$, with $0 \ \le \ x \ \le \ 1$. Using this
definition of the free mass, the following equations hold:
\begin{eqnarray}
 & &{1 \over (P^{-} -(P-k)^-_{on}- k^-_{on})}={P^+ \over (M^2 + {\bf P}^2_{\perp} -
 \frac{({\bf P-k})_{\perp}^{2}+m^2}{(1-x)}-  \frac{{\bf k}_{\perp}^{2}+m^2}{x})}=
 \nonumber
 \\ & & ={P^+ \over (M^2 - M^2_0(k^+, {\bf k}_{\perp}; P^+, {\bf P}_{\perp}))}
 \label{M01}
 \end{eqnarray}
 \begin{eqnarray}
 & &{ 1 \over \left [ P^{\prime -} - (P^{\prime} - (k - P))^-_{on}
 - (k - P)^-_{on} \right ] } =
{ P^{\prime +}
  \over (M^2 + {\bf P^{\prime}}^{2}_{\perp} -
 \frac{({\bf P - k})_{\perp}^{2} + m^2}{x^{\prime}} -
\frac{({\bf P^{\prime} - ({\bf k - P})})_{\perp}^{2} +
m^2}{(1-x^{\prime})})   } = \nonumber
\\ & & ={P^{\prime +} \over \left [ M^2 -
M^2_0((k - P)^+, ({\bf k - P})_{\perp}; P^{\prime +}, {\bf
P^{\prime}}_{\perp}) \right ]}
\label{M02}
\end{eqnarray}
where
$x^{\prime} = (k^+ - P^+)/P^{\prime +}$.

Then, performing the $k^-$ integration and using Eqs. (\ref{M01}, \ref{M02}),
 from Eq. (\ref{jmuAA}) we obtain
\begin{eqnarray}
j^{\mu} &=&
 \frac{ e } {(2\pi)^3} \frac{m^2}{f^2_\pi} N_c~\int_0^{q^+} \frac{ dk^+
d{\bf k}_{\perp}~ }{(k^+ - P^+_{\pi}) k^+ (q^+ -
k^+)}~
\left \{ \Theta (P^+_{\pi} -k^+) ~ I^{\mu}_1
+  \Theta (k^+ - P^+ _{\pi} ) ~ I^{\mu}_2  \right \} \ .
\label{jmuB}
\ee
The quantities $I^{\mu}_1$ and $I^{\mu}_2$ in Eq. (\ref{jmuB}) are defined as follows
\begin{eqnarray}
I^{\mu}_1 &=&
\left [ \overline\Lambda_{\pi}(k,P_{\pi})
\Lambda_{\bar{\pi}}(k - P_{\pi},P_{\bar{\pi}})  \right ] _{k^-=k^-_{on}}
\left [ T^{\mu}_{on, (1)} ~ + ~  T^{\mu}_{1, (1)}  ~ +  T^{\mu}_{2, (1)} \right ]
\label{jmu1}
\ee
\begin{eqnarray}
I^{\mu}_2 &=&
\left [ \overline\Lambda_{\pi}(k,P_{\pi})
 \Lambda_{\bar{\pi}}(k - P_{\pi},P_{\bar{\pi}}) \right ] _{k^-=q^- + (k - q)^-_{on}}
\left [ T^{\mu}_{on, (2)} ~ + ~  T^{\mu}_{1, (2)}  ~ +  T^{\mu}_{3, (2)} \right ]
\label{jmu2}
\ee
where
\begin{eqnarray}
T^{\mu}_{on, (1)} = ~ q^+ ~
P^+_{\pi} ~ \frac{ Tr[[(\rlap\slash k - \rlap\slash P_{\pi})_{on} + m]
~\gamma^5 [(\rlap\slash k - \rlap\slash q)_{on} + m]~
\Gamma^{\mu}(1)
~(\rlap\slash k_{on} + m)~ \gamma^5 ] }
{ \left [ M^2_0(k^+, {\bf k}_{\perp}; q^+, {\bf q}_{\perp}) - q^2
-\imath \epsilon \right]~
\left [ ~ M^2_0(k^+, {\bf k}_{\perp}; P^+_{\pi}, {\bf P}_{\pi \perp}) ~ - m^2_\pi ~ \right ] }
\label{Ton1}
\ee
\begin{eqnarray}
T^{\mu}_{on, (2)} = \frac{  q^+ ~ P^+_{\bar{\pi}} ~ ~
 Tr[[(\rlap\slash k - \rlap\slash P_{\pi})_{on} + m]
~\gamma^5 [(\rlap\slash k - \rlap\slash q)_{on} + m]~
\Gamma^{\mu}(2)
~(\rlap\slash k_{on} + m)~ \gamma^5 ] }
{\left [  M^2_0(k^+, {\bf k}_{\perp}; q^+, {\bf q}_{\perp}) - q^2
- \imath \epsilon \right]
\left [ ~ m^2_\pi - M^2_0((k^+ - P^+ _{\pi} ), ({\bf k - P_{\pi}})_{\perp};
P^+_{\bar{\pi}}, {\bf P}_{{\bar{\pi}} \perp}) ~ \right ] }
\label{Ton2}
\ee
\begin{eqnarray}
T^{\mu}_{1, (i)} = q^+ ~
 \frac{ Tr[ \gamma^+
~\gamma^5 [(\rlap\slash k - \rlap\slash q)_{on} + m]~
\Gamma^{\mu}(i)
~(\rlap\slash k_{on} + m)~ \gamma^5 ] }{ 2 ~
\left [ M^2_0(k^+, {\bf k}_{\perp}; q^+, {\bf q}_{\perp}) - q^2
 - \imath \epsilon \right]} \quad \quad ( i = 1, 2 )
\label{T11}
\ee
\begin{eqnarray}
T^{\mu}_{2, (1)} &=&
P^+_{\pi} ~ \frac{ Tr[[(\rlap\slash k - \rlap\slash P_{\pi})_{on} + m]
~\gamma^5 ~ \gamma^+ ~
\Gamma^{\mu}(1)
~(\rlap\slash k_{on} + m)~ \gamma^5 ] }
{2 ~ \left [ ~ M^2_0(k^+,{\bf k}_{\perp}; P^+_{\pi}, {\bf P}_{\pi \perp}) ~ - m^2_\pi ~ \right ]}
\label{T21}
\ee
\begin{eqnarray}
T^{\mu}_{3, (2)} &=&
P^+_{\bar{\pi}} ~ \frac{ Tr[[(\rlap\slash k - \rlap\slash P_{\pi})_{on} + m]
~\gamma^5 [(\rlap\slash k - \rlap\slash q)_{on} + m]~
\Gamma^{\mu}(2)
~ \gamma^+ ~ \gamma^5 ] }{ 2 ~
\left [ ~ M^2_0((k^+ - P^+ _{\pi} ), ({\bf k - P_{\pi}})_{\perp};
P^+_{\bar{\pi}}, {\bf P}_{{\bar{\pi}} \perp}) ~ - m^2_\pi ~  \right ]}
\label{T32}
\ee
with
\begin{eqnarray}
\Gamma^{\mu}(i) = \Gamma^\mu(k^+,{\bf k}_{\perp}, k^-=k^-_{(i)},q)
 \quad \quad ( i = 1, 2 )
\label{Gi}
\ee

The first term of Eq. (\ref{jmuB}), with $k^+ - P^+_{\pi} \le 0$,
 and the second term, with $k^+ - P^+_{\pi} \ge 0$,  are represented in Fig. 2
by the diagrams (a), with the arrow of $k^+ - P^+_{\pi}$ from the left to the right,
 and (b), with the arrow of $k^+ - P^+_{\pi}$ from the right to the left, respectively.
 In the first term only the vertex function
 $\overline\Lambda_{\pi}(k,P_{\pi})$ has the momentum fraction $k^+ / P^+ _{\pi}$
in the {\em valence-sector} range $[0, 1]$ and in the second term only the vertex function
 $\Lambda_{\bar{\pi}}(k - P_{\pi},P_{\bar{\pi}})$ has the momentum fraction
 $(k^+ - P^+ _{\pi}) / P^+_{\bar{\pi}}$
 in the {\em valence-sector} range $[0, 1]$.

\subsection{Space-like case}
In the space-like case, one has $P^{\mu}_{\pi^{\prime}} =
P^{\mu}_{\pi} + q^{\mu}$. Then the expression for the triangle
diagram can be obtained from Eq. (\ref{jmu}) replacing
$-P^{\mu}_{\pi}$ with $P^{\mu}_{\pi}$,
$\bar{\pi}$ with $\pi'$ and
 the pion vertexes $\overline \Lambda_{\pi}(k,P_{\pi})$ and
$\Lambda_{\bar{\pi}}(k - P_{\pi},P_{\bar{\pi}})$
 with $\Lambda_{\pi}(-k,P_{\pi})$ and
$\overline \Lambda_{\pi^{\prime}}(k + P_{\pi},P_{\pi^{\prime}})$, respectively : \
\begin{eqnarray}
j^{\mu} &=&-\imath 2 e \frac{m^2}{f^2_\pi} N_c \int
\frac{d^4k}{(2\pi)^4} Tr[S(k+P_{\pi}) \gamma^5
S(k-q)~\Gamma^\mu(k,q)~S(k) \gamma^5 ]
\overline\Lambda_{\pi^{\prime}}(k + P_{\pi},P_{\pi^{\prime}})
\Lambda_{\pi}(-k, P_{\pi})   \
\nonumber \\
&=& -\imath \frac{ e} {(2\pi)^4} \frac{m^2}{f^2_\pi} N_c~ \int
\frac{dk^- dk^+ d{\bf k}_{\perp}}{(k^+ + P^+_{\pi}) k^+ (k^+ - q^+)} ~
Tr[{\cal O ^{\prime}}^{\mu}]  ~
\times \nonu
\frac{\overline\Lambda_{\pi^{\prime}}(k +
P_{\pi},P_{\pi^{\prime}}) \Lambda_{\pi}(-k,P_{\pi}) }{(k^- -
k^-_{on} + \frac{\imath \epsilon}{k^+}) (k^- - q^- -(k - q)^-_{on}
+ \frac{\imath\epsilon}{k^+ - q^+}) (k^- + P^-_{\pi} - (k +
P_{\pi})^-_{on} + \frac{\imath\epsilon}{k^+ + P^+_{\pi}})} \ ,
\label{jmuE}
\ee
where
\begin{eqnarray}
{\cal O ^{\prime}}^{\mu} ~ = ~ (\rlap\slash k + \rlap\slash P_{\pi} + m)
~\gamma^5 (\rlap\slash k - \rlap\slash q + m)~ \Gamma^\mu(k,q)
~(\rlap\slash k + m)~ \gamma^5 \ .
\label{O1}
\end{eqnarray}

As in the time-like case, let us decompose the propagators in on-shell and instantaneous parts.
Then Eq. (\ref{jmuE}) becomes
\begin{eqnarray}
&&j^{\mu} =  {\cal J'}^{\mu}_{on} + {\cal J'}^{\mu}_{1} + {\cal J'}^{\mu}_{2}
+ {\cal J'}^{\mu}_{3}
\label{jmuAS}
\ee
where ${\cal J'}^{\mu}_{on}$ represents the on-shell contribution
and  ${\cal J'}^{\mu}_{i} (i=1,2,3)$ represent the contributions with one instantaneous term.
 Then we have
\begin{eqnarray}
{\cal J'}^{\mu}_{on} = -\imath \frac{e} {(2\pi)^4} \frac{m^2}{f^2_\pi} N_c~
\int \frac{dk^- dk^+ d{\bf k}_{\perp}}{(k^+ + P^+_{\pi}) k^+ (k^+
- q^+)} ~  \overline\Lambda_{\pi^{\prime}}(k +
P_{\pi},P_{\pi^{\prime}}) \Lambda_{\pi}(-k,P_{\pi}) ~ {\cal T'}^{\mu}_{on}
\label{jmuSon}
\ee
and
\begin{eqnarray}
{\cal J'}^{\mu}_{i} = -\imath \frac{e} {(2\pi)^4} \frac{m^2}{f^2_\pi} N_c~
\int \frac{dk^- dk^+ d{\bf k}_{\perp}}{(k^+ + P^+_{\pi}) k^+ (k^+
- q^+)} ~  \overline\Lambda_{\pi^{\prime}}(k +
P_{\pi},P_{\pi^{\prime}}) \Lambda_{\pi}(-k,P_{\pi}) ~ {\cal T'}^{\mu}_i
\label{jmuSi}
\ee
where
\begin{eqnarray}
 {\cal T'}^{\mu}_{on} =
\frac{Tr[[(\rlap\slash k + \rlap\slash P_{\pi})_{on} + m]
~\gamma^5 [(\rlap\slash k - \rlap\slash q)_{on} + m]~ \Gamma^\mu(k,q)
~(\rlap\slash k_{on} + m)~ \gamma^5 ]}
{(k^- - k^-_{on} +  \frac{\imath
\epsilon}{k^+}) (k^- - q^- -(k - q)^-_{on} +
\frac{\imath\epsilon}{k^+ - q^+}) (k^- + P^-_{\pi} - (k +
P_{\pi})^-_{on} + \frac{\imath\epsilon}{k^+ + P^+_{\pi}})}
\label{tmuSon}
\ee
\begin{eqnarray}
&&{\cal T'}^{\mu}_{1} = {\cal T}^{\mu}_{1} =
\frac{Tr \left [ \gamma^+
~\gamma^5 ~ [(\rlap\slash k - \rlap\slash q)_{on} + m]~ \Gamma^\mu(k,q)
~(\rlap\slash k_{on} + m)~ \gamma^5 \right ]}{2 ~ (k^- - k^-_{on} +  \frac{\imath
\epsilon}{k^+}) (k^- - q^- -(k - q)^-_{on} +
\frac{\imath\epsilon}{k^+ - q^+}) }
\label{jmuS1}
\ee
\begin{eqnarray}
&&{\cal T'}^{\mu}_{2} =
\frac{Tr  \left [[(\rlap\slash k + \rlap\slash P_{\pi})_{on} + m]
~\gamma^5 ~ \gamma^+ ~ \Gamma^\mu(k,q)
~(\rlap\slash k_{on} + m)~ \gamma^5 \right ]}{2 ~ (k^- - k^-_{on} +  \frac{\imath
\epsilon}{k^+})  (k^- + P^-_{\pi} - (k +
P_{\pi})^-_{on} + \frac{\imath\epsilon}{k^+ + P^+_{\pi}})}
\label{jmuS2}
\ee
\begin{eqnarray}
&&{\cal T'}^{\mu}_{3} =
\frac{Tr \left [[(\rlap\slash k + \rlap\slash P_{\pi})_{on} + m]
~\gamma^5 ~ [(\rlap\slash k - \rlap\slash q)_{on} + m]~ \Gamma^\mu(k,q)
~ \gamma^+ ~ \gamma^5 \right ]}{2 ~ (k^- - q^- -(k - q)^-_{on} +
\frac{\imath\epsilon}{k^+ - q^+}) (k^- + P^-_{\pi} - (k +
P_{\pi})^-_{on} + \frac{\imath\epsilon}{k^+ + P^+_{\pi}})} \quad \ .
\label{jmuS3}
\ee

In Eq. (\ref{jmuE}) the quark propagators generate three poles :
\begin{eqnarray}
k^-_{(1)} &=& k^-_{on}~ - ~ \frac{\imath \epsilon}{k^+} \ , \nonumber \\
k^-_{(2)} &=& q^- ~+ ~(k - q)^-_{on}~ - ~\frac{\imath\epsilon}{k^+
- q^+} \ ,
\nonumber \\
k^-_{(4)} &=& -P^-_{\pi} ~+~ (k + P_{\pi})^-_{on} ~ - ~
\frac{\imath\epsilon}{k^+ + P^+_{\pi}} \quad \ ,
\label{pos}
\end{eqnarray}
where
\begin{eqnarray}
(k + P_{\pi})^-_{on} = \frac{({\bf k + P}_{\pi})_{\perp}^{2}+m^2}{k^+ + P^+_{\pi}}
\label{posi}
\end{eqnarray}
Let us assume $q^+\ge 0$. Therefore, if $k^+ > q^+$ and within the
hypotheses stated at the beginning of this Section, there are no
poles in the upper complex semi-plane and a vanishing result is
obtained by closing the $k^-$ integration contour in this semi-plane.
Furthermore, if $k^+ < -P^+_{\pi}$, there are no poles in the lower
complex semi-plane of $k^-$. Therefore, one obtains a vanishing
result by closing the contour of integration in the lower
semi-plane. Then, the integral has contributions only for $k^+$ in
the range $-P^+_{\pi} < k^+ < q^+ $. The integration range can be
decomposed in two intervals, $-P^+_{\pi} \le k^+ \le 0 $ and
$0 < k^+ \le q^+$. In the first one, if the integration contour is closed
in the lower semi-plane, only the pole $k^-_{(4)}$ falls within
the integration contour, while in the second one, if the
integration contour is closed in the upper semi-plane, only the
pole $k^-_{(2)}$ falls within the integration contour.
Then
in the range $-P^+_{\pi} \le k^+ \le 0 $  one has contributions from
${\cal J'}^{\mu}_{on}$, ${\cal J'}^{\mu}_{2}$ and
${\cal J'}^{\mu}_{3}$, while in the range
$0 < k^+ \le q^+$ one has contributions from
${\cal J'}^{\mu}_{on}$, ${\cal J'}^{\mu}_{1}$ and
${\cal J'}^{\mu}_{3}$.

As a consequence, $j^{\mu}$ can be decomposed as follows
\begin{eqnarray}
j^{\mu} = j^{(I) \mu} + j^{(II) \mu} \ ,
\label{jIeII}
\end{eqnarray}
where $j^{(I) \mu}$ has the integration on $k^+$  constrained by
$-P^+_{\pi} \le k^+ \le 0$, while $j^{(II) \mu}$ has the
integration on $k^+$ in the interval $0 < k^+ < q^+$. As illustrated below,
the valence
component of the pion contributes to  $j^{(I) \mu}$ only, while
  $j^{(II) \mu}$ is the contribution
of the pair production mechanism from an incoming virtual photon
with $q^+ \ > \ 0$ \cite{sawicki,pach98,pion99,ba01,DFPS,pach02,ba02}.
Performing the $k^-$ integration, the two contributions to $j^{\mu}$
are given by the following expressions
\begin{eqnarray}
j^{(I) \mu} &=&
 ~ \frac{ e } {(2\pi)^3} \frac{m^2}{f^2_\pi} N_c~\int_{-P^+_{\pi}}^{0}
\frac{ dk^+ d{\bf k}_{\perp}}{(k^+ + P^+_{\pi}) ~ k^+ ~ (q^+ - k^+)} ~
\left [ T^{\prime \mu}_{on, (4)} ~ + ~ T^{\prime \mu}_{2, (4)} ~ + ~ T^{\prime \mu}_{3, (4)} \right ] ~
 \times \nonu \nonu ~
\left [ \overline\Lambda_{\pi \prime}(k + P_{\pi}, P_{\pi \prime})\Lambda_{\pi}(-k, P_{\pi})
 \right ] _{k^- = -P^-_{\pi} + (k + P_{\pi})^-_{on}}
\label{jmuF}
\ee
\begin{eqnarray}
j^{(II) \mu} &=&
 - \frac{ e } {(2\pi)^3} \frac{m^2}{f^2_\pi} N_c~\int_{0}^{q+}
\frac{ dk^+ d{\bf k}_{\perp}}{(k^+ + P^+_{\pi}) ~ k^+ ~ (q^+ - k^+)}~
\left [ T^{\prime \mu}_{on, (2)} ~ + ~ T^{\prime \mu}_{1, (2)} ~ + ~ T^{\prime \mu}_{3, (2)} \right ]
\times
\nonu \nonu
\left [  \overline\Lambda_{\pi \prime}(k + P_{\pi}, P_{\pi \prime})
 \Lambda_{\pi}(-k, P_{\pi})
 \right ] _{k^- = q^- + (k - q)^-_{on}} \quad \ ,
\label{jmuG}
\ee
where
\begin{eqnarray}
T^{\prime \mu}_{on, (4)} =
 \frac{ Tr \left [[(\rlap\slash k + \rlap\slash P_{\pi})_{on} + m]
~\gamma^5 [(\rlap\slash k - \rlap\slash q)_{on} + m]~
\Gamma^{\mu}(4)
~(\rlap\slash k_{on} + m)~ \gamma^5 \right ] }
{ \left [ P^-_{\pi} ~-~ (k + P_{\pi})^-_{on} + k^-_{on} \right] ~
\left [ P^-_{\pi \prime} ~-~ (k + P_{\pi})^-_{on} +
(k - q)^-_{on} \right] }
\label{TonI}
\ee
\begin{eqnarray}
T^{\prime \mu}_{on, (2)} =
\frac{ Tr \left [[(\rlap\slash k + \rlap\slash P_{\pi})_{on} + m]
~\gamma^5 [(\rlap\slash k - \rlap\slash q)_{on} + m]~
\Gamma^{\mu}(2)
~(\rlap\slash k_{on} + m)~ \gamma^5 \right ] }
{\left [ q^- + (k - q)^-_{on} - k^-_{on} + \imath \epsilon \right ] ~
\left [ P^-_{\pi \prime} ~-~ (k + P_{\pi})^-_{on} +
(k - q)^-_{on} \right ]}
\label{TonII}
\ee
\begin{eqnarray}
T^{\prime \mu}_{1, (2)} =
 \frac{ Tr \left [ \gamma^+
~\gamma^5 [(\rlap\slash k - \rlap\slash q)_{on} + m]~
\Gamma^{\mu}(2)
~(\rlap\slash k_{on} + m)~ \gamma^5 \right ] }
{ 2 ~ \left [ q^- + (k - q)^-_{on} - k^-_{on} + \imath \epsilon \right]}
\label{T1II}
\ee
\begin{eqnarray}
T^{\prime \mu}_{2, (4)} &=&
 - ~ \frac{ Tr \left [[(\rlap\slash k + \rlap\slash P_{\pi})_{on} + m]
~\gamma^5 ~ \gamma^+ ~
\Gamma^{\mu}(4)
~(\rlap\slash k_{on} + m)~ \gamma^5 \right ] }
{ 2 ~ \left [ ~  P^-_{\pi} ~-~ (k + P_{\pi})^-_{on} + k^-_{on} ~ \right ]}
\label{T2I}
\ee
\begin{eqnarray}
T^{\prime \mu}_{3, (4)} &=&
 - ~ \frac{ Tr \left [[(\rlap\slash k + \rlap\slash P_{\pi})_{on} + m]
~\gamma^5 [(\rlap\slash k - \rlap\slash q)_{on} + m]~
\Gamma^{\mu}(4)
~ \gamma^+ ~ \gamma^5 \right ] }
{ 2 ~ \left [ ~ P^-_{\pi \prime} ~-~ (k + P_{\pi})^-_{on} +
(k - q)^-_{on} ~  \right ]}
\label{T3I}
\ee
\begin{eqnarray}
T^{\prime \mu}_{3, (2)} &=&
 \frac{ Tr \left [[(\rlap\slash k + \rlap\slash P_{\pi})_{on} + m]
~ \gamma^5 [ (\rlap\slash k - \rlap\slash q)_{on} + m ] ~
\Gamma^{\mu}(2)
~ \gamma^+ ~ \gamma^5 \right ] }
{ 2 ~ \left [ ~  P^-_{\pi \prime} ~-~ (k + P_{\pi})^-_{on} +
(k - q)^-_{on} ~  \right ]}
\label{T3II}
\ee
with
\begin{eqnarray}
\Gamma^{\mu}(4) = \Gamma^{\mu}(k^+,{\bf k}_{\perp}, k^-=k^-_{(4)},q)  \quad \ .
\label{G4}
\ee

The contributions $j^{(I) \mu}$ and $j^{(II) \mu}$ are represented
by diagrams (a) and (b) of Fig. 3, respectively.

\subsubsection{Valence region contribution}
Let us change integration variables in Eq. (\ref{jmuF}) for the
valence contribution, defining $k^{\prime +} = k^+ + P^+_{\pi}$
and ${\bf k}^{\prime}_{\perp} ={\bf k}_{\perp} + {\bf P}_{\pi
\perp}$, with $(k^{\prime +},{\bf k}^{\prime}_{\perp})$ the light-front momentum
of a quark in the valence range. Then $j^{(I) \mu}$ acquaints the following more familiar expression
\begin{eqnarray}
j^{(I) \mu} &=&
 ~ \frac{ e } {(2\pi)^3}   \frac{m^2}{f^2_\pi} N_c~\int_{0}^{P^+_{\pi}}
\frac{ dk^{\prime +} d{\bf k}^{\prime}_{\perp}} {(k^{\prime+}-
P^+_{\pi}) k^{\prime+}(P_{\pi \prime}^+ - k^{\prime+})}~
\left [ T^{\prime \mu}_{on, (4)} ~ + ~ T^{\prime \mu}_{2, (4)}  ~ + ~ T^{\prime \mu}_{3, (4)} \right ]
\times \nonu \nonu ~
\left [  \overline \Lambda_{\pi
\prime}(k^{\prime}, P_{\pi \prime})
\Lambda_{\pi}(P_{\pi}-k^{\prime}, P_{\pi})
 \right ] _{k^{\prime -} = k^{\prime -}_{on}} ~ ,
\label{jmuI}
\ee
where
\begin{eqnarray}
k ^{\prime -}_{on} = \frac{({\bf k ^{\prime}_{\perp}}^{2} + m^2)}{k^{\prime +} }
\label{k'}
\end{eqnarray}
and we have defined $k^- + P^-_{\pi} ~=~ k^{\prime -}$.
The quantities $T^{\prime \mu}_{on, (4)}$, $T^{\prime \mu}_{2, (4)}$,
$T^{\prime \mu}_{3, (4)}$ can now be expressed as follows :
\begin{eqnarray}
T^{\prime \mu}_{on, (4)} =
[~P^+_{\pi \prime} ~ P^+_{\pi}~] \frac{ Tr \left [(\rlap\slash {k^{\prime}}_{on} + m)
~\gamma^5 [(\rlap\slash {k^{\prime}} - \rlap\slash {P_{\pi \prime}})_{on} + m]~
\Gamma^\mu(4)
~ [(\rlap\slash {k^{\prime}} -
\rlap\slash {P_{\pi}})_{on} + m]~ \gamma^5 \right ] }
{ \left [ m^2_{\pi \prime} - M^2_0(k^{\prime +},
{\bf k}^{\prime}_{\perp}; P^+_{\pi \prime}, {\bf P}_{\pi \prime \perp}) \right ] ~
 \left [ m^2_\pi - M^2_0(k^{\prime +}, {\bf k}^{\prime}_{\perp};
P^+_{\pi}, {\bf P}_{\pi \perp}) \right ] }
\label{TonI'}
\ee
\begin{eqnarray}
T^{\prime \mu}_{2, (4)} &=&
 - P^+_{\pi} ~ \frac{ Tr \left [(\rlap\slash {k^{\prime}}_{on} + m)
~\gamma^5 ~ \gamma^+ ~
\Gamma^\mu(4)
~[(\rlap\slash {k^{\prime}} -
\rlap\slash {P_{\pi}})_{on} + m]~ \gamma^5 \right ] }
{ 2 ~ \left [ m^2_\pi - M^2_0(k^{\prime +}, {\bf k}^{\prime}_{\perp};
P^+_{\pi}, {\bf P}_{\pi \perp}) \right ]}
\label{T2I'}
\ee
\begin{eqnarray}
T^{\prime \mu}_{3, (4)} &=&
 - P^+_{\pi \prime} ~ \frac{ Tr \left [(\rlap\slash {k^{\prime}}_{on} + m)
~\gamma^5 [(\rlap\slash {k^{\prime}} - \rlap\slash {P_{\pi \prime}})_{on} + m]~
\Gamma^\mu(4)
~ \gamma^+ ~ \gamma^5 \right ] }{ 2 ~
\left [ m^2_{\pi \prime} - M^2_0(k^{\prime +},
{\bf k}^{\prime}_{\perp}; P^+_{\pi \prime}, {\bf P}_{\pi \prime \perp}) \right ]} \quad \ .
\label{T3I'}
\ee
In Eq. (\ref{jmuI}) both the vertex functions have the quark momentum fractions
$k^{\prime +} / P^+_{\pi  \prime}$ and $(P_{\pi} - k^{\prime })^+/P_{\pi}^+$
in the valence sector $[0,1]$. Note that the on-shell momenta in Eq. (\ref{TonI'}) allow one
to retrieve the relativistic spin coupling factors with the spin 1/2 Melosh
rotations automatically included \cite{Jaus90,tob92,araujo99}.

\subsubsection{Pair-production contribution}
By making use of Eq. (\ref{M01}) the quantities
$T^{\prime \mu}_{on, (2)}$, $T^{\prime \mu}_{1, (2)}$, $T^{\prime \mu}_{3, (2)}$
in the pair-production contribution (Eq. (\ref{jmuG})) become
\begin{eqnarray}
T^{\prime \mu}_{on, (2)} =
 P^+_{\pi \prime} ~ \frac{ Tr \left [[(\rlap\slash k + \rlap\slash P_{\pi})_{on} + m]
\gamma^5 [(\rlap\slash k - \rlap\slash q)_{on} + m] ~
\Gamma^\mu(2)
~(\rlap\slash k_{on} + m)~ \gamma^5 \right ] }
{\left [ q^- - q^-_0 + \imath \epsilon \right ] ~
\left [ m^2_{\pi \prime} - M^2_0(k^{\prime +},
{\bf k}^{\prime}_{\perp}; P^+_{\pi \prime}, {\bf P}_{\pi \prime \perp})  \right] }
\label{TonII'}
\ee
\begin{eqnarray}
T^{\prime \mu}_{1, (2)} =
 \frac{ Tr \left [ \gamma^+
~\gamma^5 [(\rlap\slash k - \rlap\slash q)_{on} + m]~
\Gamma^\mu(2)
~(\rlap\slash k_{on} + m)~ \gamma^5 \right ] }
{ 2 ~ \left [ q^- - q^-_0 + \imath \epsilon \right]}
\label{T1II'}
\ee
\begin{eqnarray}
T^{\prime \mu}_{3, (2)} =
 P^+_{\pi \prime} ~ \frac{ Tr \left [[(\rlap\slash k + \rlap\slash P_{\pi})_{on} + m]
~\gamma^5 [(\rlap\slash k - \rlap\slash q)_{on} + m]~
\Gamma^\mu(2)
~ \gamma^+ ~ \gamma^5 \right ] }{ 2 ~
\left [  m^2_{\pi \prime} - M^2_0(k^{\prime +},
{\bf k}^{\prime}_{\perp}; P^+_{\pi \prime}, {\bf P}_{\pi \prime \perp})  \right ]}
\label{T3II'}
\ee
where
$q^-_0 = k^-_{on} + (q-k)^-_{on}$ and
$k^{\prime +} = k^+ + P^+_{\pi}$,
 ${\bf k}^{\prime}_{\perp} ={\bf k}_{\perp} + {\bf P}_{\pi \perp}$.
 In Eq. (\ref{jmuG}) only the vertex function $\overline\Lambda_{\pi \prime}$ has
 the quark momentum fraction $(P_{\pi} + k )^+/P_{\pi \prime}^+$ in the range $[0,1]$.

\section{Valence component of the light-front wave function}
\subsection{Meson wave functions}
\subsubsection{Pion}
To fully interpret the terms that appear in Eq. (\ref{jmuB})
and in Eqs. (\ref{jmuF}), (\ref{jmuG}), we
have to discuss valence and nonvalence components of the
light-front wave function. Let us first begin with the valence
component, which can be derived from the Bethe-Salpeter
amplitude \cite{sales1}. In the present model the pion
Bethe-Salpeter amplitude is given by
\begin{eqnarray}
\Psi_\pi (k,P_\pi) =
\frac{m}{f_\pi}\frac{\rlap\slash{k}+m}{k^2-m^2+\imath \epsilon}
\gamma^5 ~ \Lambda_\pi (k,P_\pi)
\frac{\rlap\slash{k}-\rlap\slash{P_\pi} + m}{(k - P_\pi)^2 - m^2 + \imath
\epsilon}  \ \ ,
\label{bsa}
\end{eqnarray}
where the pion vertex is $\frac{m}{f_\pi} ~ \gamma^5 \Lambda_\pi (k,P_\pi)$.

The valence component of the light-front wave function
can be obtained from the Bethe-Salpeter amplitude (\ref{bsa})
in the valence sector, $0 \leq k^+ \leq P_{\pi}^+$,
disregarding the instantaneous terms in Eq. (\ref{bsa}),
multiplying $\Psi_\pi$ by the factor $[k^+ ~ ( k^+ - P^+_{\pi} )]/(2\pi \imath)$
and integrating over $k^-$ :
\begin{eqnarray}
&&\phi_\pi(k^+,\vec k_\perp; P^+_{\pi},\vec P_{\pi \perp})=\nonu
=-\imath
\frac{m}{f_\pi}k^+(k^+-P^+_{\pi} ) \int \frac{dk^-}{2\pi}
\frac{\rlap\slash{k}_{on}+m}{k^2-m^2+\imath \epsilon} \gamma^5
\Lambda_\pi (k,P_\pi)
\frac{(\rlap\slash{k}-\rlap\slash{P_\pi})_{on}+m}{(k-P_\pi)^2-m^2+\imath
\epsilon} \  \, .
\label{wf1}
\end{eqnarray}

 Two poles, $k_{(1)}$ and $k_{(3)}$, appear in Eq. (\ref{wf1}), respectively
in the lower and in the upper $k^-$ semiplanes.
We perform the $k^-$ integration in the lower complex
semi-plane disregarding the contributions that arise from
the singularities of the vertex $\Lambda_\pi (k,P_\pi)$
 (cf. the assumptions (i) and (ii) at the beginning of Sect. III).
Then the pion wave function becomes
\begin{eqnarray}
\phi_\pi(k^+, {\bf k}_{\perp}; P^+_{\pi}, {\bf
P}_{\pi \perp}) = ~ \frac{m}{f_\pi} ~(\psla k_{on} + m) ~ \gamma^5
\frac{P^+_{\pi} ~  [ \Lambda_{\pi}(k,P_{\pi}) ]_{[k^- = k^-_{on}]}}
{[m^2_\pi - M^2_0(k^+, {\bf k}_{\perp}; P^+_{\pi},
{\bf P}_{\pi \perp})]} ~
\left [(\psla k - \psla P_{\pi})_{on} + m \right ]  ~~ ~~ .
\label{wf2}
\end{eqnarray}
 If the  $k^-$
integration is done in the upper
semi-plane within the same assumptions, one has:
\be
\phi_{\pi}(k^+, {\bf k}_{\perp}; P^+_{\pi}, {\bf P}_{\pi \perp})
=\nonu
= ~ \frac{m}{f_\pi} ~ (\psla k_{on} + m) ~ \gamma^5
\frac{P^+_{\pi} ~ [ \Lambda_{\pi}(k,P_{\pi}) ] _{[k^- = P^-_\pi - (P_\pi-k)^-_{on}]}}
{[m^2_\pi - M^2_0(P^+_\pi-k^+, {\bf
P}_{\pi\perp}-{\bf k}_{\perp}; P^+_{\pi}, {\bf P}_{\pi
\perp})]} ~\left [(\psla k - \psla P_{\pi})_{on} + m \right ]  ~.
\label{wf22}
\ee

In principle, the elimination of the relative light-front time
between the quark and the antiquark in the pion Bethe-Salpeter
amplitude by the $k^-$ integration in Eq. (\ref{wf1}), should  give
a unique answer, which defines the valence component of the wave
function in the range $0 \leq k^+ \leq P^+_\pi$,
 with both the quarks on their mass shell.
 Therefore in order to
require consistency within our model, we will assume
 $[ \Lambda_{\pi}(k,P_{\pi}) ]_{[k^- = k^-_{on}]} $ and
$[ \Lambda_{\pi}(k,P_{\pi}) ]_{[k^- =P^-_\pi - (P_\pi-k)^-_{on}]}$
to be equal in that kinematical range
(note that $M^2_0(k^+, {\bf k}_{\perp}; P^+_{\pi},
{\bf P}_{\pi \perp})$ is  equal to
$M^2_0(P^+_\pi-k^+, {\bf P}_{\pi\perp}-{\bf k}_{\perp}; P^+_{\pi}, {\bf P}_{\pi\perp})$).
This assumption produces a
momentum component of the
valence light-front wave function symmetrical for the exchange of the quark momenta,
since the vertex function $\Lambda_{\pi}(k,P_{\pi})$ is assumed to be symmetrical.

 Within a Bethe-Salpeter approach, the function $\phi_{\pi}$ fulfills a two-body Schroedinger-like
equation, with the proper Melosh structure represented by the matrix
$(\psla k_{on} + m) ~ \gamma^5
\left [(\psla k - \psla P_{\pi})_{on} + m \right ] $ \cite{Jaus90}.
Therefore, when the plus component
of the quark momentum is in
the interval  $0\leq k^+ \leq P^+_\pi$, $\phi_{\pi}$ will be identified
in our approach with
 the HLFD pion wave function, with momentum component
 $\psi_{\pi}(k^+, {\bf k}_{\perp};
P^+_{\pi}, {\bf P}_{\pi \perp})$ (see Ref. \cite{Jaus90}):
 \begin{eqnarray}
 ~\phi _{\pi}(k^+, {\bf k}_{\perp}; P^+_{\pi}, {\bf P}_{\pi
\perp}) = (\psla k_{on} + m) ~ \gamma^5
\left [(\psla k - \psla P_{\pi})_{on} + m \right ]
\psi_{\pi}(k^+, {\bf k}_{\perp};
P^+_{\pi}, {\bf P}_{\pi \perp}) ~~  ~~.
\label{wfp}
\end{eqnarray}

\subsubsection{Vector meson}
In analogy with the pion case,
one can define the light-front VM wave function,
which describes the valence component of the meson state  $|n \lambda \rangle$.
Indeed, starting from the Bethe-Salpeter amplitude for a vector meson
\begin{eqnarray}
\Psi_{n \lambda} (k,P_n) =
\frac{\rlap\slash{k} + m}{k^2 - m^2 + \imath \epsilon}
\left [ \epsilon_{\lambda}(P_n) \cdot \widehat{V}_{n}(k,k-P_n)  \right ]  ~
\Lambda_n(k,P_n) ~
\frac{\rlap\slash{k}-\rlap\slash{P_n} + m}{(k - P_n)^2 - m^2 + \imath
\epsilon} \ ,
\label{bsan}
\end{eqnarray}
 the valence component of the light-front wave function can be defined
from $\Psi_{n \lambda} (k,P_n)$ integrating over $k^-$,
disregarding the instantaneous terms and multiplying by the factor
$[k^+ ~ ( k^+ - P^+_{\pi} )]/(2\pi \imath)$, as we already did for the pion.
Furthermore, in this case one has to take on their mass shell
both the quark momenta in the Dirac structure
of the VM vertex function, $\widehat{V}_{n}(k,k-P_n)$ :
\begin{eqnarray}
&&\phi_{n \lambda} (k^+,\vec k_\perp; P^+_n,\vec P_{n \perp})=
 -\imath k^+ (k^+-P^+_{n} ) \times
 \nonu
 \nonu
\int \frac{dk^-}{2\pi}
\frac{\rlap\slash{k}_{on}+m}{k^2 - m^2 + \imath \epsilon} ~
\Lambda_n (k,P_n)
~ \left [ \epsilon_{\lambda} (P_n) \cdot [ \widehat{V}_{n}(k,k-P_n) ]_{on} \right ]  ~
\frac{(\rlap\slash{k}-\rlap\slash{P_n})_{on}+m}{(k-P_n)^2-m^2 + \imath
\epsilon} \  \,
\label{wf1n}
\end{eqnarray}
where $[ \widehat{V}_{n}(k,k-P_n) ]_{on}$ is defined by Eq. (\ref{gams1})
in order to retrieve the  $^3S_1$ vector meson vertex of Ref. \cite{Jaus90}.
Assuming that $\Lambda_n(k,P_n)$
does not diverge in the complex plane $k^-$ for
$|k^-|\rightarrow\infty$, and
 neglecting the contributions of its singularities
 in the $k^-$ integration, the valence VM wave function is
\begin{eqnarray}
&&\phi _{n \lambda}(k^+, {\bf k}_{\perp}; P^+_{n}, {\bf P}_{n \perp}) =
 \nonu
= ~ P^+_{n} ~ (\psla k_{on} + m) ~
\frac{\left [ \epsilon_{\lambda}(P_n) \cdot [ \widehat{V}_{n}(k,k-P_n) ]_{on} \right ] }
{[M^2_n - M^2_0(k^+, {\bf k}_{\perp}; P^+_{n}, {\bf P}_{n \perp})]} ~
 [ \Lambda_{n}(k,P_{n}) ]_{[k^- = k^-_{on}]} ~
 \left [(\psla k - \psla P_{n})_{on} + m \right ]~~ ~~ .
\label{wfn}
\end{eqnarray}
 In analogy with the pion case, we assume
$[\Lambda_{n}(k,P_n)]_{[k^- = k^-_{on}]} =
[\Lambda_{n}(k,P_n)]_{[k^- = P^-_n - (P_n - k)^-_{on}]}$
 in the valence sector, $0 \leq k^+ \leq P^+_n$.

As for the pion, the function $\phi _{n \lambda}$, with the plus component
of the quark momentum in
the interval  $0\leq k^+ \leq P^+_\pi$, will be identified with
  the HLFD vector meson wave function, with
 momentum component $\psi_{n}(k^+, {\bf k}_{\perp}; P^+_{n}, {\bf P}_{n \perp})$,
   \cite{Jaus90}:
 \begin{eqnarray}
 &&\phi _{n \lambda}(k^+, {\bf k}_{\perp}; P^+_{n}, {\bf P}_{n \perp}) =
 \nonu
= ~ (\psla k_{on} + m) ~
 \left [ \epsilon_{\lambda}(P_n) \cdot [ \widehat{V}_{n}(k,k - P_n) ]_{on} \right ]
 ~\left [(\psla k - \psla P_{n})_{on} + m \right ]
\psi_{n}(k^+, {\bf k}_{\perp}; P^+_{n}, {\bf P}_{n \perp})   ~~.
\label{wfpn}
\end{eqnarray}

In conclusion, Eqs. (\ref{wf2}, \ref{wfp}) and (\ref{wfn}, \ref{wfpn}) establish a link
between the momentum part of the meson HLFD wave functions and
the momentum part of
  the meson vertex functions.

 The valence component of the VM wave function are
 normalized to the probability of the valence component of the meson state  $|n \lambda \rangle$
 (see Appendix D).
 This probability is estimated in a schematic model in Appendix E.

 The corresponding normalization for the pion wave function is included in an overall
 normalization constant for the pion form factor.

\subsection{Photon wave function}
One can define as well the valence component of the hadronic contribution
to the photon wave function, starting from
the Bethe-Salpeter amplitude of the photon, which can be written as:
\begin{eqnarray}
\Psi^\mu_\gamma (k,q) =
\frac{\rlap\slash{k+m}}{k^2 - m^2 + \imath \epsilon}
\Gamma^\mu (k,q) \frac{\rlap\slash{k} - \rlap\slash{q} + m}
{(k-q)^2 - m^2 + \imath \epsilon} \ ,
\label{bsag}
\end{eqnarray}
where $\Gamma^\mu (k,q)$ is the photon vertex amplitude (see Eq. (\ref{wf1zg})).

In analogy with Eq. (\ref{wf1n}), the valence component of the virtual photon light-front wave
function can be obtained from the Bethe-Salpeter amplitude
(\ref{bsag}) in the valence sector, $0\leq k^+\leq q^+$,
separating out the instantaneous terms of Eq. (\ref{bsag}), integrating in $k^-$
 and multiplying by the factor $[k^+ ~ ( k^+ - P^+_{\pi} )]/(2\pi \imath)$.

 Then,
 using our explicit expression for $\Gamma^{\mu}(k,q)$ given by Eq. (\ref{cur6}), the
  light-front wave function of the photon can be defined by
\be
\phi^\mu_\gamma(k^+,{\bf k}_\perp; q^2,q^+,{\bf q}_\perp)=
\nonu
= -\imath k^+(k^+-q^+) \int \frac{dk^-}{2\pi}
\frac{\rlap\slash{k}_{on} + m}{k^2 - m^2 + \imath \epsilon}
\left [ \Gamma^\mu (k, q) \right ]_{on}
\frac{(\psla k - \psla q)_{on} + m}{(k-q)^2 - m^2 + \imath \epsilon}
\ \ ,
\label{wf1g}
\ee
where the label "$on$" in  $\left [ \Gamma^\mu (k, q) \right ]_{on}$ means that,
as in the VM case, the Dirac structures of the photon vertex
amplitude, $\Gamma^\mu (k,q)$, have to be taken with both the quark
momenta on their mass shell. Therefore, in analogy with Eq. (\ref{wf1n}),
a possible choice for the quantity
$\widehat{V}_{n}(k,k-q)$ of Eq. (\ref{cur6}) is given by the quantity
$\left [ \widehat{V}_{n}(k,k-q) \right ]_{on}$ as defined by Eq. (\ref{gams1}).

Then, performing the $k^-$ integration with the assumptions given at the beginning of Section 3,
in the range $0\leq k^+\leq q^+$
Eq. (\ref{wf1g}) becomes

\begin{eqnarray}
 ~\phi^\mu _{\gamma}(k^+, {\bf k}_{\perp}; q^2,q^+, {\bf q}_{\perp}) =
 (\psla k _{on} + m ) ~
\psi^\mu_{\gamma}(k^+, {\bf k}_{\perp};q^2, q^+, {\bf q}_{\perp}) ~
\left[(\psla k-\psla q)_{on} + m\right] ~~  ~~,
\label{wf3g}
\end{eqnarray}
where the function $\psi^\mu_{\gamma}(k^+, {\bf k}_{\perp};q^2, q^+, {\bf q}_{\perp})$,
which includes the
Dirac structures of the photon vertex, is defined by
\begin{eqnarray}
\psi^\mu_{\gamma}(k^+, {\bf k}_{\perp};q^2, q^+, {\bf q}_{\perp}) =
[ \Gamma^\mu(k,q) ]_{on} ~
\frac {q^+}{[ q^2 - M^2_0(k^+, {\bf k}_{\perp}; q^+, {\bf q}_{\perp}) + \imath \epsilon]} ~~ ~~ .
\label{wf2g}
\end{eqnarray}
As in the previous meson case, to have consistency in our
virtual photon wave function model, we will not distinguish between
$[ \Lambda_{n}(k,q) ]_{[k^- = k^-_{on}]} $
and $[ \Lambda_{n}(k,q) ]_{[k^- =q^- - (q-k)^-_{on}]}$
in the valence sector, $0 \leq k^+ \leq q^+$.
Therefore we obtain
for $\psi^\mu_{\gamma}(k^+, {\bf k}_{\perp};q^2, q^+, {\bf q}_{\perp})$
the same result
when the $k^-$ integration is performed both in the lower or in the upper $k^-$ complex semiplane.

The valence  wave function,
 $ \phi^\mu_\gamma(k^+,{\bf k}_\perp; q^2,q^+,{\bf q}_\perp)$,
and the function
$\psi^\mu_{\gamma}(k^+, {\bf k}_{\perp};q^2, q^+, {\bf q}_{\perp})$
depend  on the value of $q^2$ carried by the virtual photon.
Note that in the time-like case
a singularity appears in the photon valence wave function
(see Eq. (\ref{wf2g})).

If, as in Eq. (\ref{cur7}), the photon vertex $[ \Gamma^\mu(k,q) ]_{on}$
is taken with on-shell quantities for
the vector mesons in the numerator, i.e. if we take
\begin{eqnarray}
\left [\Gamma^{\mu}(k,q) \right ]_{on} = \sqrt{2} \sum_{n, \lambda}
 \epsilon_{\lambda} (P_n)\cdot \left [ \widehat{V}_{n}(k,k - P_n) \right ]_{on}
\left [ \Lambda_{n}(k,P_n) \right ]_{[k^- = k^-_{on}]}
 { [\epsilon ^{\mu}_{\lambda}(P_n)]^* f_{Vn} \over \left [ q^2 -
M^2_n + \imath M_n \tilde{\Gamma}_n(q^2)\right ]} \
\label{cur7b}
\end{eqnarray}
 and identify Eqs. (\ref{wfn}) and (\ref{wfpn}), then
  the function
$\psi^\mu_{\gamma}(k^+, {\bf k}_{\perp};q^2, q^+, {\bf q}_{\perp})$
can be written as follows:
\begin{eqnarray}
&& \psi^\mu_{\gamma}(k^+, {\bf k}_{\perp};q^2, q^+, {\bf q}_{\perp})
= \sqrt{2} ~ \sum_{n, \lambda}
\left [ \epsilon_{\lambda} (P_n)\cdot \left [ \widehat{V}_{n}(k,k-P_n) \right ]_{on}  \right ]
\times \nonu
\frac {\left [ M^2_n - M^2_0(k^+, {\bf k}_{\perp}; P^+_{n}, {\bf P}_{n \perp})\right ]}
{\left [ q^2 - M^2_0(k^+, {\bf k}_{\perp}; q^+, {\bf q}_{\perp}) + \imath \epsilon \right ]} ~
\psi_{n}(k^+, {\bf k}_{\perp}; P^+_{n}, {\bf P}_{n \perp})
~ { [\epsilon ^{\mu}_{\lambda}(P_n)]^* f_{Vn} \over \left [ q^2 -
M^2_n + \imath M_n \tilde{\Gamma}_n(q^2)\right ]}
\label{wfni1}
\end{eqnarray}
in terms of the momentum part of the HLFD vector meson  wave functions,
$\psi_{n}(k^+, {\bf k}_{\perp}; P^+_{n}, {\bf P}_{n \perp})$.

\section{Contribution of nonvalence components to the current }

\subsection{Time-like case: the photon nonvalence component}

The process of pion-antipion production is shown in Fig. 2, where
the dashed lines (both in (a) and in (b)) represent two different
light-front times. At the first time (the one on the right) the hadronic valence component
of the virtual photon is represented, while at the second one the
$2q2\overline q$ photon nonvalence component is depicted (see  also Fig. 5).
The two parts of Fig. 2, i.e. (a) and  (b), differ by
the emission vertex of an antipion or of a pion (see also Fig. 5 (b)), respectively.
The corresponding quark amplitudes for the
radiation of an antipion or a pion are given in
Eq. (\ref{jmuB}) by
the antipion vertex
$\Lambda_{\overline\pi}(k-P_\pi; P_{\overline\pi})$, evaluated at $k^-=k^-_{on}$
for $(k - P_\pi)^+ < 0$, and  by
the pion  vertex $\overline\Lambda _{\pi}(k; P_{\pi})$,
evaluated at $k^-=q^- + (k - q)^-_{on}$ for $k^+ > P^+_{\pi}$, respectively.

Once the interaction that couples the
valence to the $2q2\overline q$ component is known (see Fig. 2), the amplitude for the photon
decay in a $\pi{\overline\pi}$ pair can be constructed. To this end,
let us introduce a kernel operator ${\cal K}$ which realizes this coupling.
Then, we can write the following
equation to relate the valence component of the pion wave function,
$\psi_{\pi}(k^{\prime +},
{\bf k}^\prime_{\perp}; P^+_{\pi}, {\bf P}_{\pi \perp})$,
to the vertex function $ \overline\Lambda _{\pi}(k; P_{\pi})$ at ${k^-=q^- + (k -
q)^-_{on}}$, which is the amplitude for the pion emission (see Fig. 2 (b) and Fig. 6):
\be
 \overline{{\cal{D}}}_{\pi} := \frac{m}{f_\pi} \overline\Lambda
_{\pi}(k; P_{\pi})_{[k^-=q^- + (k - q)^-_{on}]}=
\nonu=\frac14
\sum_{\alpha^\prime\beta^\prime\alpha\beta}\int_0^{P^+_\pi}
\frac{dk^{\prime +}d{\bf k}^\prime_\perp}{k^{\prime
+}(P^+_\pi-k^{\prime+})} ~
 \psi^* _{\pi}(k^{\prime +},
{\bf k}^\prime_{\perp}; P^+_{\pi}, {\bf P}_{\pi \perp})\times
\nonu~
 (\gamma^5)^{\beta\alpha}(\gamma^5)_{\beta^\prime\alpha^\prime} ~
{\cal K}^{\alpha^\prime\beta^\prime}_{\alpha\beta}\left(k^{\prime
+},{\bf k}^\prime_\perp~;~k^{ +},{\bf k}_\perp~;~q^-,q^+,{\bf
q}_\perp\right) \ .
\label{piemi}
\ee
  For simplicity, the example of a
$\gamma^5$ structure was used in Eq. (\ref{piemi}), just to be
consistent with our assumption of a pseudoscalar pion model.

 One can write an analogous expression for the emission of $\overline
\pi$
:
\be
{\cal{D}}_{\overline{\pi}} :=\frac{m}{f_\pi} [\Lambda _{\overline\pi}(k-P_\pi;
P_{\overline\pi})]_{(k^-=k^-_{on})} =
\nonu=\frac14
\sum_{\alpha^\prime\beta^\prime\alpha\beta}\int_0^{P^+_{\overline\pi}}
\frac{dk^{\prime +}d{\bf k}^\prime_\perp}{k^{\prime
+}(P^+_\pi-k^{\prime+})} ~
 \psi^* _{\overline\pi}(k^{\prime +},
{\bf k}^\prime_{\perp}; P^+_{\overline\pi}, {\bf P}_{\overline\pi \perp})\times
\nonu~
(\gamma^5)^{\beta\alpha}(\gamma^5)_{\beta^\prime\alpha^\prime} ~
{\cal K}^{\alpha^\prime\beta^\prime}_{\alpha\beta}\left(k^{\prime
+},{\bf k}^\prime_\perp~;~k^{ +}-P^+_\pi,{\bf k}_\perp-{\bf
P}_{\pi\perp}~;~q^-,q^+,{\bf q}_\perp\right) \ .
\label{pibemi}
\ee
In our model calculation, both pion emission vertexes will be
substituted by a constant, following Ref. \cite{JI01}.

\subsection{Space-like case: the pion nonvalence component}

A part the direct photon coupling to the quark line,
in the space-like region the nonvalence component of the final pion wave
function  appears for $q^+>0$ in both the
 contributions of the current obtained after the $k^-$ integration, and
given by Eqs. (\ref{jmuF}) and (\ref{jmuG}) (see Fig. 7).
 On one hand the valence component of the final pion is coupled
 to the nonvalence $2q2\overline q$ component (see Fig. 7 (b)), through
 an interaction
 kernel ${\cal H}$, which contributes
to  the quark-photon absorption vertex of Eq. (\ref{jmuF}),
given by $\Gamma^\mu(4) = \Gamma^\mu(k^{\prime} - P_{\pi},q)$ with
$k^{\prime-}=k^{\prime-}_{on}$. On the other hand
the  vertex $\Lambda _{\pi}(-k; P_\pi)$ in Eq. (\ref{jmuG}),
evaluated at $k^-=q^- + (k - q)^-_{on}$ for $- k^+ < 0$,
describes the quark-pion absorption through another
 interaction kernel ${\cal K'}$ and generates
the nonvalence $2q2\overline q$ component of the final pion (see Fig. 7 (c)).
We identify the kernel ${\cal K'}$ with the kernel ${\cal {K}}$, already used in the previous
subsection A for the description of the pion emission (Fig. 6).

Equation (\ref{jmuF}) gives a contribution to the SL form factor
where the initial and the final pion valence components appear (diagram (a) of
 Fig. 3).
The plus component of the quark-photon absorption vertex, given by
$\Gamma^+(k^{\prime} - P_{\pi},q)$ with $k^{\prime-}=k^{\prime-}_{on}$ which enters in
 Eq. (\ref{jmuF}), is represented by an empty circle in
 diagram (a) of Fig. 3 and is approximated by the sum of i) the bare photon vertex
 multiplied by a renormalization constant, $a$, (diagram (a) of Fig. 7) and ii)
  the
 contribution due to the $2q2\overline q$ component of the final pion wave
function, which is represented by diagram  (b) of Fig. 7.

Therefore, we can make
the following identification:
\be
\left[ [\Gamma^+(k^{\prime} - P_{\pi},q)]_{(k^{-} = k^{-}_{on})} \right]_{\alpha\beta} =
a ~ (\gamma^+)_{\alpha\beta} + \sum_{\alpha^\prime\beta^\prime}\int_0^{q^+}
\frac{dk^{\prime \prime +}d{\bf k}^{\prime \prime}_\perp}{k^{\prime \prime +}
(q^+-k^{\prime \prime +})} \times
\nonu
{\cal H}^{\alpha^\prime\beta^\prime}_{\alpha\beta}
\left( k^{\prime +} - P^+_\pi,{\bf k^\prime}_\perp-{\bf P}_{\pi\perp}~;
~k^{\prime \prime +},{\bf k}^{\prime \prime}_\perp~;
~P^-_{\pi\prime},P^+_{\pi\prime},{\bf P}_{\pi\prime\perp}\right)
\left[\psi^+_{\gamma}(k^{\prime \prime +}, {\bf k}^{\prime \prime}_{\perp};
q^-,q^+, {\bf q}_{\perp}) \right ]_{\alpha^\prime\beta^\prime} .
\label{phabs}
\ee

As already discussed in Sec. II, we do not consider the bare term
photon vertex in the present paper, since it violates current conservation
for a massless pion (see Appendix C).
Therefore, disregarding the  bare photon vertex in
the right-hand side of Eq. (\ref{phabs}), we can formally write:
\be
[\Gamma^+(k^{\prime} - P_{\pi},q)]_{(k^{-}=k^{-}_{on})} \simeq {\cal H}
\psi^+_{\gamma} \ .
\label{phabs1}
\ee
One could try to interpret
Eq. (\ref{phabs1}) in terms of constituent quark form factors. However,
we have to point out that the absorption vertex of
Eq. (\ref{phabs}) does not depend only on $q^2$, as one could
naively think, but it depends on the virtuality of the quark, and
therefore depends on the hadron where this process occurs.

Let us note that, within our assumption of a vanishing pion mass, the
 contribution of Eq. (\ref{jmuF}) is also vanishing (see Sect. VIII)
 and therefore there is no contribution from
 $[\Gamma^+(k^{\prime} - P_{\pi},q)]_{(k^{-} = k^{-}_{on})}$.

  Equation (\ref{jmuG}) represents the pair-production term
  (Z-diagram) and is depicted in Fig. 7 (c). The quark-pion absorption vertex,
given by $\Lambda _{\pi}(-k; P_\pi)$ evaluated at $k^-=q^- + (k - q)^-_{on}$,
which appears in Eq. (\ref{jmuG}) can be written as
\be
{\cal{D}}_{\pi} := \frac{m}{f_\pi}[\Lambda _{\pi}(-k ;
P_{\pi})]_{(k^-=q^- + (k - q)^-_{on})} = \frac14
\sum_{\alpha^\prime\beta^\prime\alpha\beta} \int_0^{P^+_\pi}
\frac{dk^{\prime +} d{\bf k}^\prime_\perp}{k^{\prime+} (P^+_\pi-k^{\prime+})}
  \times \nonu~
(\gamma^5)_{\beta\alpha}(\gamma^5)^{\beta^\prime\alpha^\prime}
{\cal K}_{\alpha^\prime\beta^\prime}^{\alpha\beta}
\left(k^{+},{\bf k}_\perp;~k^{\prime +},{\bf k}^\prime_\perp~;
~P^-_{\pi\prime},P^+_{\pi\prime},{\bf P}_{\pi\prime\perp}\right)
\psi_{\pi}(k^{\prime +}, {\bf k}^\prime_{\perp}; P^+_{\pi}, {\bf P}_{\pi \perp}) \
\label{pibabs}\ .
\ee
 For our purpose this quark-pion absorption vertex will be taken
  constant, as we do in the TL case for the quark-pion emission vertex,
  as was  proposed in Ref. \cite{JI01}.

\section{Triangle diagram and pion LF wave function}
\subsection{Time-like case}
Let us insert into Eq. (\ref{jmuB}) the photon
vertex of Eq. (\ref{cur7}). Furthermore, whenever the full expression for
the light-front pion wave function
$\phi_{\pi}(k^+, {\bf k}_{\perp}; P^+_{\pi}, {\bf P}_{\pi \perp})$,
given by Eq. (\ref{wf2}), appears in Eq. (\ref{jmuB})
and the momentum fraction is in the valence-sector range $[0,1]$, let us replace it
with the expression of Eq. (\ref{wfp}), i.e. let us write the pion vertex in terms
of the  momentum component of the HLFD pion wave function.
This means that we introduce in Eq. (\ref{jmuB}) the wave functions
$\psi_{\pi}(k^+, {\bf k}_{\perp}; P^+_{\pi}, {\bf P}_{\pi \perp})$ and
$\psi_{\bar{\pi}}((k^+ - P^+ _{\pi} ), ({\bf k -
P_{\pi}})_{\perp}; P^+_{\bar{\pi}}, {\bf P}_{{\bar{\pi}} \perp})$
when these wave functions have the correct support.
Then the triangle diagram can be expressed as follows:
\begin{eqnarray}
j^{\mu} &=&
 \frac{ e } {(2\pi)^3} \frac{m}{f_\pi} N_c~\int_0^{q^+}
 \frac{ dk^+ d{\bf k}_{\perp}}{(k^+ - P^+_{\pi}) k^+ (q^+ - k^+)}~
\sum_{n, \lambda}
~ {\sqrt{2} ~ [\epsilon ^{\mu}_{\lambda}(P_n)]^* f_{Vn} \over
\left [ q^2 - M^2_n + \imath M_n \tilde{\Gamma}_n(q^2) \right ]}
\times \nonu
\left \{ \Theta (P^+_{\pi} -k^+) ~ I_{1, n, \lambda}  ~
+ ~ \Theta (k^+ - P^+ _{\pi} ) ~ I_{2, n, \lambda}  \right \} \ .
\label{jmuD}
\ee
The quantities $I_{1, n, \lambda}$ and $I_{2, n, \lambda}$ in Eq. (\ref{jmuD})
are defined as follows
\begin{eqnarray}
I_{1, n, \lambda} &=&
\left [
\Lambda_{\bar{\pi}}(k - P_{\pi},P_{\bar{\pi}}) ~ \Lambda_{n}(k,P_n) \right ]_{k^- = k^-_{on}}
~ ~ \times \nonumber  \\
&& \left \{ \frac{ q^+ }
{ \left [ q^2 - M^2_0(k^+, {\bf k}_{\perp}; q^+, {\bf q}_{\perp}) + \imath \epsilon \right ]} ~
\left [ T_{on, (1, n, \lambda)} ~ + ~  T_{1, (1, n, \lambda)} \right ] ~
+ ~ T_{2, (1, n, \lambda)} \right \}
\label{jmu1n}
\ee
\begin{eqnarray}
I_{2, n, \lambda} &=&
\left [ \overline\Lambda_{\pi}(k,P_{\pi}) ~ \Lambda_{n}(k,P_n) \right ]_{k^- =  q^- + (k - q)^-_{on}}
~ ~ \times \nonumber  \\
&& \left \{  \frac{ q^+ }
{\left [ q^2 - M^2_0(k^+, {\bf k}_{\perp}; q^+, {\bf q}_{\perp}) + \imath \epsilon \right ] }
~ \left [ T_{on, (2, n, \lambda)} ~ + ~  T_{1, (2, n, \lambda)} \right ] ~
+ ~ T_{3, (2, n, \lambda)} \right \}
\label{jmu2n}
\ee
where
\begin{eqnarray}
&& T_{on, (1, n, \lambda)} = ~
 \psi^* _{\pi}(k^+, {\bf k}_{\perp}; P^+_{\pi}, {\bf P}_{\pi \perp})
~ \times \quad \quad \quad \quad \quad \quad
\label{Ton1n} \\
\nonumber  \\
&&  Tr  \left [[(\rlap\slash k - \rlap\slash P_{\pi})_{on} + m]
~\gamma^5 [(\rlap\slash k - \rlap\slash q)_{on} + m]~
\left [ \epsilon_{\lambda} (P_n) \cdot \widehat{V}_{n}(k,k-P_n) ~ \right ] _{k^- = k^-_{on}}
~(\rlap\slash k_{on} + m)~ \gamma^5 \right ] \quad \ ,
\nonumber
\ee

\begin{eqnarray}
&& T_{on, (2, n, \lambda)} = -  ~ \psi^* _{\bar{\pi}}((k^+ - P^+ _{\pi} ), ({\bf k -
P_{\pi}})_{\perp}; P^+_{\bar{\pi}}, {\bf P}_{{\bar{\pi}} \perp})
~  ~
\times \quad \quad \quad \quad
\label{Ton2n} \\ \nonumber \\
&&  Tr \left [ [(\rlap\slash k - \rlap\slash P_{\pi})_{on} + m]
~ \gamma^5 [(\rlap\slash k - \rlap\slash q)_{on} + m]~
\left [ \epsilon_{\lambda} (P_n) \cdot \widehat{V}_{n}(k,k-P_n) ~ \right ]
_{k^- = q^- + (k - q)^-_{on}}
(\rlap\slash k_{on} + m)~ \gamma^5 \right ] ~~  ,
\nonumber
\ee

\begin{eqnarray}
&& T_{1, (1, n, \lambda)} = - ~\frac{1} { 2 } ~ \frac{m}{f_\pi} ~
 \left[\overline\Lambda _{\pi}(k; P_{\pi}) \right ]_{k^- = k^-_{on}}  ~
\times \quad \quad
\nonumber \\
\nonumber \\
&&  Tr \left[ \gamma^+
~\gamma^5 [(\rlap\slash k - \rlap\slash q)_{on} + m]~
\left [ \epsilon_{\lambda} (P_n) \cdot \widehat{V}_{n}(k,k-P_n) ~ \right ] _{k^- = k^-_{on}}
~(\rlap\slash k_{on} + m)~ \gamma^5 \right ]  \quad  ,
\label{T11n}
\ee

\begin{eqnarray}
&& T_{1, (2, n, \lambda)} = -  ~\frac{1}{ 2 } ~ \frac{m}{f_\pi} ~
 \left [ \Lambda_{\bar{\pi}}(k - P_{\pi},P_{\bar{\pi}})
 \right ]_{k^- = q^- + (k - q)^-_{on}}
~ ~ \times \quad \quad  \quad \quad
\nonumber \\
\nonumber \\
&&  Tr \left [ \gamma^+
~\gamma^5 [(\rlap\slash k - \rlap\slash q)_{on} + m]~
\left [ \epsilon_{\lambda} (P_n) \cdot \widehat{V}_{n}(k,k-P_n) ~ \right ]
_{k^- = q^- + (k - q)^-_{on}}
~(\rlap\slash k_{on} + m)~ \gamma^5 \right ]
  ~~ ,
\label{T12n}
\ee

\begin{eqnarray}
&& T_{2, (1, n, \lambda)} =
- ~ \frac{1}{2} ~ \psi^* _{\pi}(k^+, {\bf k}_{\perp}; P^+_{\pi}, {\bf P}_{\pi \perp}) ~
~ \times  \quad \quad
\nonumber \\
\nonumber \\
 &&  Tr \left [[(\rlap\slash k - \rlap\slash P_{\pi})_{on} + m]
~\gamma^5 ~ \gamma^+ ~
\left [ \epsilon_{\lambda} (P_n) \cdot \widehat{V}_{n}(k,k-P_n) ~ \right ] _{k^- = k^-_{on}}
~(\rlap\slash k_{on} + m)~ \gamma^5 \right ] ~~ ,
\label{T21n}
\ee

\begin{eqnarray}
&& T_{3, (2, n, \lambda)} = - ~  \frac{1}{2} ~ \psi^* _{\bar{\pi}}((k^+ - P^+ _{\pi} ), ({\bf k -
P_{\pi}})_{\perp}; P^+_{\bar{\pi}}, {\bf P}_{{\bar{\pi}} \perp})
~
~ \times \quad \quad
\label{T32n} \\
\nonumber \\
&& Tr \left [[(\rlap\slash k - \rlap\slash P_{\pi})_{on} + m]
~\gamma^5 [(\rlap\slash k - \rlap\slash q)_{on} + m]~
\left [ \epsilon_{\lambda} (P_n) \cdot \widehat{V}_{n}(k,k-P_n) ~ \right ]
_{k^- = q^- + (k - q)^-_{on}}
~ \gamma^+ ~ \gamma^5 \right ] \quad .
\nonumber
\ee
Let us notice that the momentum component of the LF pion wave function does not appear in
the instantaneous terms $T_{1, (1, n, \lambda)}$ and $T_{1, (2, n, \lambda)}$,
 because  in these terms the propagator
$[(\rlap\slash k - \rlap\slash P_{\pi})_{on} + m]/
(k^- -P^-_{\pi} - (k -
P_{\pi})^-_{on} + \frac{\imath\epsilon}{k^+ - P^+_{\pi}})$
is replaced by $\gamma^+/2$. Indeed the amplitude
\begin{eqnarray}
 ~\frac{1} { 2 } ~ \frac{m}{f_\pi} ~(\rlap\slash k_{on} + m)~ \gamma^5 ~
 \left[\overline\Lambda _{\pi}(k; P_{\pi}) \right ]_{k^- = k^-_{on}} ~
 \gamma^+
 \label{ampist}
 \ee
  does not obey the same two-body Schroedinger-like  equation as the
  light-front pion wave function
  $\phi_\pi(k^+,\vec k_\perp; P^+_{\pi},\vec P_{\pi \perp})$ does.

As already noted at the end of Sect. III A,
the first and the second term of Eq. (\ref{jmuD}) are represented in Fig. 2 by
the diagrams (a) and (b), respectively.  Note that, due to the $\Theta$ functions,
the final pion or antipion wave functions enter into the
first or the second term of Eq. (\ref{jmuD}), respectively.
In Eq. (\ref{jmuD})
 the pion vertexes $[\Lambda _{\overline\pi}(k-P_\pi; P_{\overline\pi})]$, evaluated at
$k^-=k^-_{on}$, and
 $[\overline\Lambda _{\pi}(k; P_{\pi})]$,
 evaluated at $k^-=q^- + (k - q)^-_{on}$, have the momentum fraction
outside the valence-sector range $[0,1]$ and can be related to the quark amplitudes for radiative
 antipion or pion emission, respectively (see Figs. 2 (a), 2 (b)).
 The presence of these
vertexes gives rise to the contribution of the nonvalence
component of the virtual-photon wave function, relevant for the process under
consideration. In
the space-like region the analogous processes can be interpreted for
$q^+>0$ as the contribution of the nonvalence component of the
pion wave function in the final state \cite{JI01}. These points have already been illustrated in
 Section V.

If we choose to take the Dirac structures in the photon vertex with both
 quarks on their
 mass shell, i.e. $\Gamma^{\mu}(k,q)  = \left [\Gamma^{\mu}(k,q) \right ]_{on}$
 (see Eq. (\ref{cur7b})), then
 whenever the full expression for the light-front vector meson wave function
 $\phi_{n \lambda} (k^+,\vec k_\perp; P^+_n,\vec P_{n \perp})$,
given by Eq. (\ref{wfn}), appears in Eq. (\ref{jmuD}),
we can take advantage of our identification of Eqs. (\ref{wfn}) and (\ref{wfpn})
to express the quantities $I_{1, n, \lambda}$ and $I_{2, n, \lambda}$
through the momentum component of the HLFD VM wave function.
However,
in the instantaneous contributions to $I_{1, n, \lambda}$ and $I_{2, n, \lambda}$ which are
proportional to the
quantities $T_{2, (1, n, \lambda)}$ and $T_{3, (2, n, \lambda)}$
we do not express the VM vertex functions
$\left [ \Lambda_{n}(k,P_n) \right ]_{k^- = k^-_{on}}$ and
$\left [ \Lambda_{n}(k,P_n) \right ]_{k^- =  q^- + (k - q)^-_{on}}$
through the momentum component of the HLFD VM wave function,
because the full expression for this function given by Eq. (\ref{wfpn})
does not appear in these instantaneous terms.

Then we obtain
\begin{eqnarray}
 I_{1, n, \lambda} &=&
\left [ \Lambda_{\bar{\pi}}(k - P_{\pi},P_{\bar{\pi}})^{~}_{~}  \right ]_{k^- = k^-_{on}}
~ ~ \times \nonumber  \\
\nonumber  \\
&& \left \{ \frac{ \psi_{n}(k^+, {\bf k}_{\perp}; P^+_{n}, {\bf P}_{n \perp}) ~
[M^2_n - M^2_0(k^+, {\bf k}_{\perp}; P^+_{n}, {\bf P}_{n \perp})] }
{ \left [ q^2 - M^2_0(k^+, {\bf k}_{\perp}; q^+, {\bf q}_{\perp}) + \imath \epsilon \right ]} ~
\left [ T_{on, (1, n, \lambda)} ~ + ~  T_{1, (1, n, \lambda)} \right ] ~ + \right .
\nonumber  \\
\nonumber  \\
&&\left . \left [ \Lambda_{n}(k,P_n)^{~}_{~} \right ]_{k^- = k^-_{on}}
~ T_{2, (1, n, \lambda)} \right \}
\label{j1n}
\ee

\begin{eqnarray}
 I_{2, n, \lambda} &=&
\left [ \overline\Lambda_{\pi}(k,P_{\pi}) \right ]_{k^- =  q^- + (k - q)^-_{on}}
~ ~ \times \nonumber  \\
\nonumber  \\
&& \left \{  \frac{ \psi_{n}(k^+, {\bf k}_{\perp}; P^+_{n}, {\bf P}_{n \perp}) ~
[M^2_n - M^2_0(k^+, {\bf k}_{\perp}; P^+_{n}, {\bf P}_{n \perp})] }
{\left [ q^2 - M^2_0(k^+, {\bf k}_{\perp}; q^+, {\bf q}_{\perp}) + \imath \epsilon \right ] }
~ \left [ T_{on, (2, n, \lambda)} ~ + ~  T_{1, (2, n, \lambda)} \right ] ~ + \right .
\nonumber  \\
\nonumber  \\
&&\left . \left [ \Lambda_{n}(k,P_n) \right ]_{k^- =  q^- + (k - q)^-_{on}} ~
T_{3, (2, n, \lambda)} \right \}
\label{j2n}
\ee

  The quantities
$T_{on, (1, n, \lambda)} , ~  T_{1, (1, n, \lambda)}, ~ T_{2, (1, n, \lambda)}$
and the quantities $ T_{on, (2, n, \lambda)} , ~ T_{1, (2, n, \lambda)}, ~
T_{3, (2, n, \lambda)} $  in Eq. (\ref{j1n}) and in Eq. (\ref{j2n})
have the same expressions as in Eqs. (\ref{Ton1n}), (\ref{T11n}), (\ref{T21n})
and in Eqs. (\ref{Ton2n}),  (\ref{T12n}), and (\ref{T32n}), respectively, with
$\left [ \epsilon_{\lambda} (P_n) \cdot \widehat{V}_{n}(k,k-P_n) ~ \right ] _{k^- = k^-_{on}}$
and $\left [ \epsilon_{\lambda} (P_n) \cdot \widehat{V}_{n}(k,k-P_n) ~ \right ]
_{k^- = q^- + (k - q)^-_{on}}$
both replaced by
$\epsilon_{\lambda} (P_n) \cdot \left [ \widehat{V}_{n}(k,k-P_n) ~ \right ]_{on}$
 (see Eq. (\ref{gams1}) for the definition of
 $\left [ \widehat{V}_{n}(k,k-P_n) ~ \right ]_{on}$).

Note that
 the region of integration over $k^+$ in Eq. (\ref{jmuD})
(a consequence of the non vanishing
integration in the $k^-$ complex plane) agrees with the
support of the wave functions $\psi_{\pi}$ and $\psi_{\bar{\pi}}$ of Eqs.
(\ref{Ton1n}) and (\ref{Ton2n}), respectively. Furthermore,
in agreement with the above assumptions,  the vertex associated with the virtual photon
and consequently the wave function $\psi_{n}(k^+, {\bf k}_{\perp}; P^+_{n}, {\bf P}_{n \perp}) $
in Eqs. (\ref{j1n}) and (\ref{j2n}) have  the intrinsic fraction of the plus-component
of the quark momentum, $k^+/q^+ = k^+/P_n^+$,  in the interval [0, 1].

To be able to evaluate the TL pion form factor
we have still to assign a value,  in the instantaneous terms, to the VM vertex functions
 $\left [ \Lambda_{n}(k,P_n)^{~}_{~} \right ]_{k^- = k^-_{on}}$ and
 $\left [ \Lambda_{n}(k,P_n) \right ]_{k^- =  q^- + (k - q)^-_{on}}$, as well as to the
pion vertex functions $\left[\overline\Lambda _{\pi}(k; P_{\pi}) \right ]_{k^- = k^-_{on}}$
and $ \left [ \Lambda_{\bar{\pi}}(k - P_{\pi},P_{\bar{\pi}})
 \right ]_{k^- = q^- + (k - q)^-_{on}}$.

\subsection{Space-like case}
As in the time-like case, let us replace in Eqs. (\ref{jmuI}) and (\ref{jmuG})
the pion vertex function with its expression in terms of the momentum component of
the HLFD pion wave function, whenever the full expression for
the LF pion wave function,
(Eq. (\ref{wf2})) appears,
 taking advantage of our identification of Eqs. (\ref{wf2})
 and (\ref{wfp}).

\subsubsection{Valence region contribution}
 Substituting in Eq. (\ref{jmuI}) the pion initial and final wave
functions
and noting that for the final pion
the "bar" vertex gives the complex conjugate wave function, while
in the initial state the vertex gives the initial pion wave
function, in the valence region one gets (see Fig. 3):
\begin{eqnarray}
j^{(I) \mu}  &=&
 ~ \frac{ e } {(2\pi)^3}  N_c~\int_{0}^{P^+_{\pi}}
\frac{ dk^{\prime +} d{\bf k}^{\prime}_{\perp}} {(k^{\prime+}-
P^+_{\pi}) k^{\prime+}(P_{\pi \prime}^+ - k^{\prime+})}~  \times
\nonu
\nonu
\left [ {\overline T}^{\mu}_{on, (4)} ~ ~ \psi^*_{\pi\prime}(k^{\prime +}, {\bf k}^\prime_{\perp};
P^+_{\pi\prime}, {\bf P}_{\pi \prime \perp})
~ \psi_{\pi}(P^+_\pi-k^{\prime +}, {\bf P}_{\pi\perp}-{\bf
k}^\prime_{\perp}; P^+_{\pi}, {\bf P}_{\pi \perp}) ~ ~ + \right .
\nonu
\nonu
\left . + ~ ~ {\overline T}^{\mu}_{2, (4)} ~ ~
\psi_{\pi}(P^+_\pi-k^{\prime +}, {\bf P}_{\pi\perp}-{\bf k}^\prime_{\perp};
P^+_{\pi}, {\bf P}_{\pi \perp}) ~
\left [  \overline \Lambda_{\pi \prime}(k^{\prime}, P_{\pi \prime})
 \right ] _{k^{\prime -} = k^{\prime -}_{on} } ~ ~ + \right .
\nonu
\nonu
\left . + ~ ~ {\overline T}^{\mu}_{3, (4)} ~ ~
\psi^*_{\pi\prime}(k^{\prime +}, {\bf k}^\prime_{\perp};
P^+_{\pi\prime}, {\bf P}_{\pi \prime \perp}) ~
\left [ \Lambda_{\pi}(P_{\pi}-k^{\prime}, P_{\pi})
 \right ] _{k^{\prime -} = k^{\prime -}_{on}} \right ] ~~ ,
\label{jmuIa}
\ee
where
\begin{eqnarray}
{\overline T}^{\mu}_{on, (4)} =
  Tr \left [(\rlap\slash {k^{\prime}}_{on} + m)
~\gamma^5 [(\rlap\slash {k^{\prime}} - \rlap\slash {P_{\pi \prime}})_{on} + m]~
\Gamma^\mu(4)
~ [(\rlap\slash {k^{\prime}} -
\rlap\slash {P_{\pi}})_{on} + m]~ \gamma^5 \right ]
\label{TonIb}
\ee
\begin{eqnarray}
{\overline T}^{\mu}_{2, (4)} &=&
 - ~ \frac{1}{2} ~ Tr \left [(\rlap\slash {k^{\prime}}_{on} + m)
~\gamma^5 ~ \gamma^+ ~
\Gamma^\mu(4)
~[(\rlap\slash {k^{\prime}} -
\rlap\slash {P_{\pi}})_{on} + m]~ \gamma^5 \right ]
\label{T2Ib}
\ee
\begin{eqnarray}
{\overline T}^{\mu}_{3, (4)} &=&
 - ~ \frac{1}{2} ~ Tr \left [(\rlap\slash {k^{\prime}}_{on} + m)
~\gamma^5 [(\rlap\slash {k^{\prime}} - \rlap\slash {P_{\pi \prime}})_{on} + m]~
\Gamma^\mu(4)
~ \gamma^+ ~ \gamma^5 \right ] ~~ .
\label{T3Ib}
\ee
In the instantaneous terms proportional to ${\overline T}^{\mu}_{2, (4)} $ and
to ${\overline T}^{\mu}_{3, (4)}$  we do not express the pion vertex functions
$\left [  \overline \Lambda_{\pi \prime}(k^{\prime}, P_{\pi \prime})
 \right ] _{k^{\prime -} = k^{\prime -}_{on} }$ and
 $\left [ \Lambda_{\pi}(P_{\pi}-k^{\prime}, P_{\pi})
 \right ] _{k^{\prime -} = k^{\prime -}_{on}}$
  in terms of the  momentum component of the HLFD pion wave function, because of the presence
  of $\gamma^+$ instead of
  $[(\rlap\slash {k^{\prime}} - \rlap\slash {P_{\pi \prime}})_{on} + m]$ or
  $[(\rlap\slash {k^{\prime}} -
\rlap\slash {P_{\pi}})_{on} + m]$, respectively.

In Eq. (\ref{jmuIa}) the photon  vertex, $\Gamma^\mu(4) = \Gamma^\mu(k^{\prime} -
P_{\pi},q)$, evaluated at $k^{\prime-}=k^{\prime-}_{on}$, is the
amplitude for the photon absorption by a quark.
As discussed in Sect. V B, the photon absorption operator can be decomposed in a bare vertex,
i.e. $\gamma^\mu$ (Fig. 7 (a)), plus other terms. From the expansion in the
light-front Fock-space, the next term relevant for the process we are analyzing
is due to a
contribution of the nonvalence $2q2\overline q$ component of the
final pion wave function, see diagram (b) of Fig. 7. This
contribution
can be thought of as an expectation value
of an operator between the valence component of the wave functions
for the initial and final pions. The operator can be constructed
by applying  to the
virtual photon wave-function  the kernel, $\cal H$, which produces the
nonvalence pion component from the valence one (see Eq. (\ref{phabs})).

\subsubsection{Pair-production contribution}
Also the pair-production contribution to the current, can be rewritten
in terms of the momentum component of the HLFD pion
wave function (see Eqs. (\ref{wf2}),
(\ref{wfp}))  when the light-front pion wave function appears. Then Eq. (\ref{jmuG}) becomes
(see Fig. 3 (b)):
\begin{eqnarray}
j^{(II) \mu} =
 - \frac{ e  N_c} {(2\pi)^3} \frac{m}{f_\pi} \int_0^{q^+}
 \frac{ dk^+ d{\bf k}_{\perp} ~ \left [\Lambda _{\pi}(-k; P_\pi)  \right] _{k^-=q^- + (k - q)^-_{on}} }
 {(k^+ + P^+_{\pi}) ~  k^+ ~ (q^+ - k^+)} ~
 \sum_{n, \lambda}
~ {\sqrt{2}  ~ [\epsilon ^{\mu}_{\lambda}(P_n)]^* f_{Vn}
 \over
\left [ q^2 - M^2_n + \imath M_n \tilde{\Gamma}_n(q^2) \right ]} ~
\times &&
\nonumber \\
\left [ \Lambda_{n}(k,P_n) \right] _{k^- = q^- + (k - q)^-_{on}}
\left \{ \frac{ q^+ }
{ \left [ q^2 - M^2_0(k^+, {\bf k}_{\perp}; q^+, {\bf q}_{\perp}) + \imath \epsilon \right ]}
\left [ T^{\prime}_{on, (2,n)} ~ + ~ T^{\prime}_{1, (2,n)} \right ] ~ + ~ T^{\prime}_{3, (2,n)}  \right \}
\quad ~
\label{jmuO1}
\ee
where
\begin{eqnarray}
&&T^{\prime}_{on, (2,n)} = \psi^* _{\pi\prime}((k^+ + P^+ _{\pi} ), ({\bf k} +
{\bf P}_{\pi})_{\perp}; P^+_{ \pi\prime}, {\bf P}_{\pi\prime \perp})  ~
\times \quad \quad \quad
\label{TonIIn}  \\
&& Tr \left [[(\rlap\slash k + \rlap\slash P_{\pi})_{on} + m]
\gamma^5 ~ [(\rlap\slash k - \rlap\slash q)_{on} + m]
\left [ \epsilon_{\lambda} (P_n) \cdot \widehat{V}_{n}(k,k-P_n)  \right ]_{k^-=q^- + (k - q)^-_{on}}
~(\rlap\slash k_{on} + m)~ \gamma^5 \right ]
\nonumber
\ee

\begin{eqnarray}
&& T^{\prime}_{1, (2,n)} =
 \frac{1} { 2 } ~ \frac{m}{f_\pi} ~
  \left [ \overline\Lambda_{\pi \prime}(k + P_{\pi}, P_{\pi \prime})
 \right ] _{k^- = q^- + (k - q)^-_{on}}
 \times \quad \quad \quad
  \label{T1IIn} \\
&& Tr \left [ \gamma^+ ~ \gamma^5 ~ [(\rlap\slash k - \rlap\slash q)_{on} + m]~
\left [ \epsilon_{\lambda} (P_n) \cdot \widehat{V}_{n}(k,k-P_n)  \right ]_{k^-=q^- + (k - q)^-_{on}}
~(\rlap\slash k_{on} + m)~ \gamma^5 \right ]
\nonumber
\ee

\begin{eqnarray}
&& T^{\prime}_{3, (2,n)} = \frac{1} { 2 } ~ \psi^* _{\pi\prime}((k^+ + P^+ _{\pi} ), ({\bf k} +
{\bf P}_{\pi})_{\perp}; P^+_{ \pi\prime}, {\bf P}_{\pi\prime \perp}) ~
\times \quad \quad \quad
\label{T3IIn} \\
&& Tr \left [[(\rlap\slash k + \rlap\slash P_{\pi})_{on} + m]
~\gamma^5 ~ [(\rlap\slash k - \rlap\slash q)_{on} + m]~
\left [ \epsilon_{\lambda} (P_n) \cdot \widehat{V}_{n}(k,k-P_n)  \right ]_{k^-=q^- + (k - q)^-_{on}}
~ \gamma^+ ~ \gamma^5 \right ]
\nonumber
\ee

As for the time-like case, we have used the expression of Eq. (\ref{cur7}) for the
photon vertex with the virtual photon going into a $q\bar{q}$ pair.
The "bar" vertex $ \overline\Lambda_{\pi \prime}$ implies that the final pion wave
function in the above expressions has to be complex conjugated.
If we take the Dirac structures in the photon vertex with both the quarks on their
 mass shell, as in the time-like case
 (see Eq. (\ref{cur7b})), then using Eqs. (\ref{wfn}) and (\ref{wfpn})
we can express $j^{(II) \mu}$ through the momentum component of the
HLFD vector meson wave functions,
when the LF VM wave function is present, i.e. in the terms given by Eqs.
 (\ref{TonIIn}) and  (\ref{T1IIn}):

\begin{eqnarray}
&&j^{(II) \mu} =
 - \frac{ e  N_c } {(2\pi)^3} \frac{m}{f_\pi} \int_0^{q^+}
 \frac{ dk^+ d{\bf k}_{\perp} ~\left[\Lambda _{\pi}(-k; P_\pi) \right] _{k^-=q^- + (k - q)^-_{on}} }
 {(k^+ + P^+_{\pi}) ~ k^+ ~ (q^+ - k^+)} ~
\sum_{n, \lambda}
~ {\sqrt{2} ~ [\epsilon ^{\mu}_{\lambda}(P_n)]^* ~ f_{Vn} \over
\left [ q^2 - M^2_n + \imath M_n \tilde{\Gamma}_n(q^2) \right ]} ~
\times \nonumber \\
&&\left \{ \frac{ \psi_{n}(k^+, {\bf k}_{\perp}; P^+_{n}, {\bf P}_{n \perp}) ~
[M^2_n - M^2_0(k^+, {\bf k}_{\perp}; P^+_{n}, {\bf P}_{n \perp})] }
{ \left [ q^2 - M^2_0(k^+, {\bf k}_{\perp}; q^+, {\bf q}_{\perp}) + \imath \epsilon \right ]} ~
\left [ T^{\prime}_{on, (2,n)} ~ + ~ T^{\prime}_{1, (2,n)} \right ] ~ \right .
+
\nonumber \\
\nonumber \\
&& \left . \left [ \Lambda_{n}(k,P_n) \right ] _{k^-=q^- + (k - q)^-_{on}} ~
T^{\prime}_{3, (2,n)} \right \}
\label{jmuO1b}
\ee
with
\begin{eqnarray}
 T^{\prime}_{on,(2,n)} &=& \psi^* _{\pi\prime}((k^+ + P^+ _{\pi} ), ({\bf k} +
{\bf P}_{\pi})_{\perp}; P^+_{ \pi\prime}, {\bf P}_{\pi\prime \perp}) ~ \times
\label{TonIInb}  \\
&& Tr \left [[(\rlap\slash k + \rlap\slash P_{\pi})_{on} + m]
\gamma^5 ~ [(\rlap\slash k - \rlap\slash q)_{on} + m] ~
 \epsilon_{\lambda} (P_n) \cdot \left [\widehat{V}_{n}(k,k-P_n)  \right ]_{on}
(\rlap\slash k_{on} + m)~ \gamma^5 \right ]
 \nonumber
\ee

\begin{eqnarray}
 T^{\prime}_{1, (2,n)} &=&
 \frac{1} { 2 } ~ \frac{m}{f_\pi} ~
   \left [ \overline\Lambda_{\pi \prime}(k + P_{\pi}, P_{\pi \prime})
 \right ] _{k^- = q^- + (k - q)^-_{on}} ~
   ~
 \times \quad \quad
 \nonumber \\
&& Tr \left [ \gamma^+ ~ \gamma^5 ~ [(\rlap\slash k - \rlap\slash q)_{on} + m] ~
\epsilon_{\lambda} (P_n) \cdot \left [ \widehat{V}_{n}(k,k-P_n)  \right ]_{on}
~(\rlap\slash k_{on} + m) ~ \gamma^5 \right ]
  \label{T1IInb}
\ee

\begin{eqnarray}
T^{\prime}_{3, (2,n)} &=& \frac{1} { 2 } ~
\psi^* _{\pi\prime}((k^+ + P^+ _{\pi} ), ({\bf k} +
{\bf P}_{\pi})_{\perp}; P^+_{ \pi\prime}, {\bf P}_{\pi\prime \perp}) ~
\times \quad \quad \quad
\label{T3IInb} \\
&& Tr \left [[(\rlap\slash k + \rlap\slash P_{\pi})_{on} + m]
~\gamma^5 ~ [(\rlap\slash k - \rlap\slash q)_{on} + m]~
 \epsilon_{\lambda} (P_n) \cdot \left [ \widehat{V}_{n}(k,k-P_n)  \right ]_{on}
~ \gamma^+ ~ \gamma^5 \right ]
\nonumber
\ee

The vertex $\Lambda _{\pi}(-k; P_\pi)$ evaluated at
$k^-=q^- + (k - q)^-_{on}$ represents the pion absorption amplitude by an on-shell quark. The
presence of this process can be also interpreted as a $2q2\overline q$
component in the final pion wave function (see Figs. 3 (b) and 7 (c)), as
 illustrated in Sect. V.

\section{ Time-like em form factor of the pion}

We have pointed out in the Introduction that,
for a unified description of TL and SL form factors, it is necessary to choose a reference frame
where the plus component of the momentum transfer, $q^+$, is different from zero (otherwise,
$q^2 = q^+ q^- - {\bf q}_{\perp}^2$ cannot be positive). Therefore,
as in Ref. \cite{LPS}, in order to calculate  the pion form factor we adopt a
reference frame where ${\bf q}_{\perp}=0$ and $q^+>0$.

The decay of a time-like virtual photon is written in terms of the
time-like form factor of the pion as follows
\be
j^{\mu} =\langle \pi \bar{\pi}| \bar{q}(0) \gamma^{\mu}q(0)
|0\rangle = e ~ \left (P^{\mu}_{\pi} -P^{\mu}_{\bar{\pi}} \right )~F_{\pi}(q^2)
\label{dec1}
\ee
where $q^{\mu} =P^{\mu}_{\pi}+P^{\mu}_{\bar{\pi}}~$ is the four momentum of the
virtual photon.
In Fig. 2, the diagrammatic analysis of the virtual-photon decay
 in a $\pi \bar{\pi}$ pair is shown.

The virtual-photon decay amplitude can be obtained from Eq.
(\ref{dec1}) by evaluating the plus-component of the matrix
element, $j^{\mu}$.  To be able to evaluate
the matrix element $j^{+}$ from Eq. (\ref{jmuD}), we
substitute in Eq. (\ref{jmuD}) constant values for the  vertexes,
$ \bar{\cal D}_{\pi}$ and ${\cal D}_{\bar\pi}$, namely for pion or antipion
radiation by a quark, Eqs. (\ref{piemi}) and
(\ref{pibemi}), respectively (see also Eqs. (\ref{jmu1n}, \ref{jmu2n})). Then,
it remains to
specify the values of the instantaneous
vertex functions
$\left[\overline\Lambda _{\pi}(k; P_{\pi}) ~ \right ] _{k^- = k^-_{on}}$ in Eq. (\ref{T11n}),
$\left [ \Lambda_{\bar{\pi}}(k - P_{\pi},P_{\bar{\pi}})
\right ] _{k^- = q^- + (k - q)^-_{on}}$ in Eq. (\ref{T12n}),
 and  $\left [ \Lambda_{n}(k,P_n) \right ]$
in Eqs. (\ref{j1n}, \ref{j2n}). This will be thoroughly discussed in Sect. IX.

  By using Eqs. (\ref{dec1})
and (\ref{jmuD}) one can obtain the pion form factor
$F_{\pi}(q^2)$ from the plus-component of the current :
\be
 F_{\pi}(q^2) = \sum_n ~ {f_{Vn}   \over
 \left [ q^2 - M^2_n + \imath M_n \tilde{\Gamma}_n(q^2) \right ]} ~~ g^+_{Vn}(q^2)
\label{tlff}
\ee
where $g^+_{Vn}(q^2)$, for $q^2 > 0$, is the form factor for the VM decay in a pair of pions,
as expected from the VMD approximation. The characteristic feature of our approach
is that we aim at a microscopic description of both $f_{Vn} $ and $g^+_{Vn}(q^2)$.

Let us now evaluate $\sum_{ \lambda}$ of Eq. (\ref{jmuD}).
The momentum of the vector meson is $P^{\mu}_n \equiv \{
P^-_n=(|{\bf q}_{\perp}|^2+M^2_n)/q^+,{\bf q}_{\perp},q^+ \}$ and
the momentum of the virtual photon is $q^{\mu} \equiv \{
q^-,{\bf q}_{\perp},q^+ \}$
(as already noted, see Fig. 4, at the production vertex the LF
three-momentum is conserved).
 In a frame where ${\bf q}_{\perp}=0$,
one has $q^- = q^2/q^+$ for the photon, while for the vector meson $P^{-}_n
= M^2_n/q^+$. Using Cartesian components for the four vectors,
i.e. $a^{\mu} \equiv \left [ a^0,{\bf a} \right ]$, in this frame
the three polarization four-vectors are given by
\be
\epsilon^{\mu}_x \equiv [0,1,0,0 ], \quad \epsilon^{\mu}_y
\equiv [0,0,1,0 ], \quad \epsilon^{\mu}_z \equiv [P_{n z}/M_n,0,0,\sqrt{1+\eta} ]
\ee
where  $\eta=P^2_{n z}/M^2_n$.
.
Let us recall that, in the frame we are adopting,
 $P_{n z}=(q^+ - P^-_n)/2=({q^+}^2 - M^2_n )/ 2q^+$.
Therefore, in the reference frame defined by ${\bf q}_{\perp}=0$ and
$q^+>0$ the polarization four-vector $\epsilon^{\mu}_z$
does not have a defined sign for the zero component.
The plus-component of $\epsilon^{\mu}_z$ is
given by
\be
\epsilon^{+}_z =
{P_{n z} \over M_n} + \sqrt{1 + {P_{n z}^2 \over M^2_n}} =
{{q^+}^2 - M^2_n \over 2q^+ M_n} +
\sqrt{1+ \left ({{q^+}^2 - M^2_n \over 2q^+ M_n} \right )^2}
\label{eps1}
\ee
 The plus-component of the other polarization
four-vectors are vanishing (i.e.,
$\epsilon^{+}_x=\epsilon^{+}_y=0$) and therefore we have
 $\sum_{\lambda} \left [ \epsilon
^{+}_{\lambda}(P_n) \right ]^* \epsilon _{\lambda}(P_n) \cdot
\widehat{V}_{n} = \left [ \epsilon ^{+}_{z}(P_n) \right ]^*
\epsilon _{z}(P_n) \cdot \widehat{V}_{n} $.

Each term of $\sum_{n}$ in Eq. (\ref{tlff}) is invariant under LF boosts,
that are  kinematical, and therefore
  to simplify the calculations it can be evaluated
  in the rest frame of the corresponding resonance. In the rest frame
  of the $n{\rm th}-$resonance
  one has $q^+=M_n$ and $q^- = q^2/M_n$ for the photon, while
 $P^{+}_n= P^{-}_n=M_n$ for the vector meson.
This means that we choose a different frame for
each resonance, but all these frames are related by kinematical LF boosts
along the $z$ axis to each other, and to the Breit frame where $q^+ = - q^- = \sqrt{-q^2}$,
adopted in previous analyses of the SL region
(we have always ${\bf q}_{\perp}=0$) \cite{pach02,LPS}.

Then, in the evaluation of the sum in Eq. (\ref{tlff})
  one has   $\epsilon^+_z=1$ and
\begin{eqnarray}
 \sum_{\lambda} \left [ \epsilon
^{+}_{\lambda}(P_n) \right ]^* \epsilon _{\lambda}(P_n) \cdot
\widehat{V}_{n} =  - \widehat{V}_{n z} \quad
\label{polz}
\ee
for the contribution of any resonance!
In conclusion we have

\begin{eqnarray}
g^+_{Vn}(q^2) &=&  ~ {1 \over P^+_{\bar{\pi}}-P^+_{{\pi}} } ~
 \frac{ N_c} {(2\pi)^3} ~ ~ \times
\nonumber \\
&&\int_0^{q^+}  \frac{ \sqrt{2} ~ dk^+ d{\bf k}_{\perp}}{(k^+ - P^+_{\pi}) k^+ (q^+ - k^+)}
\left \{ \Theta (P^+_{\pi} -k^+) ~ \overline{I}_{1, n}
+  \Theta (k^+ - P^+ _{\pi} ) ~ \overline{I}_{2, n}  \right \} \ \ , \quad
\label{jmuT}
\ee
where the quantities $\overline{I}_{1, n}$ and $\overline{I}_{2, n}$
can be obtained  from Eqs. (\ref{j1n}), (\ref{j2n}) replacing
$m/f_{\pi} ~
\left [ \Lambda_{\bar{\pi}}(k - P_{\pi},P_{\bar{\pi}})^{~}_{~}  \right ]_{k^- = k^-_{on}}$ and
$m/f_{\pi} ~ \left [ \overline\Lambda_{\pi}(k,P_{\pi}) \right ]_{k^- =  q^- + (k - q)^-_{on}}$
with ${\cal{D}}_{\overline{\pi}}$ and $\overline{{\cal{D}}}_{\pi}$,
respectively (see Eqs. (\ref{pibemi}) and (\ref{piemi})):

\begin{eqnarray}
 \overline{I}_{1, n} &=&
{\cal{D}}_{\overline\pi} ~
\left \{ \frac{ \psi_{n}(k^+, {\bf k}_{\perp}; P^+_{n}, {\bf 0}) ~
[M^2_n - M^2_0(k^+, {\bf k}_{\perp}; P^+_{n},  {\bf 0}] }
{ \left [ q^2 - M^2_0(k^+, {\bf k}_{\perp}; q^+, {\bf 0}) + \imath \epsilon \right ]} ~
\left [ {\cal{T}}_{on, (1, n)} ~ + ~  {\cal{T}}_{1, (1, n)} \right ] ~
+  \right .
\nonumber \\
&& \left . \left [ \Lambda_{n}(k,P_n) \right ]_{k^- = k^-_{on}} ~ {\cal{T}}_{2, (1, n)} \right \}
\label{jmu1nT}
\ee

\begin{eqnarray}
\overline{I}_{2, n} &=&
\overline{{\cal{D}}}_{\pi} ~
\left \{ \frac{
 \psi_{n}(k^+, {\bf k}_{\perp}; P^+_{n},  {\bf 0}) ~
[M^2_n - M^2_0(k^+, {\bf k}_{\perp}; P^+_{n},  {\bf 0})]}
{\left [ q^2 - M^2_0(k^+, {\bf k}_{\perp}; q^+,  {\bf 0}) + \imath \epsilon \right ]~
}  ~ \left [ {\cal{T}}_{on, (2, n)} ~ +  {\cal{T}}_{1, (2, n)} \right ] ~
+  \right .
\nonumber \\
 && \left .  \left [ \Lambda_{n}(k,P_n) \right ]_{k^- =  q^- + (k - q)^-_{on}} ~
 {\cal{T}}_{3, (2, n)} \right \}
\label{jmu2nT}
\ee

with
\begin{eqnarray}
{\cal{T}}_{on, (1, n)} &=& - ~
 \psi^* _{\pi}(k^+, {\bf k}_{\perp}; P^+_{\pi}, {\bf P}_{\pi \perp})
~
~ \times \quad \quad
\label{Ton1VT} \\
&&  Tr  \left [[(\rlap\slash k - \rlap\slash P_{\pi})_{on} + m]
~\gamma^5 [(\rlap\slash k - \rlap\slash q)_{on} + m]~
\left [ \widehat{V}_{nz}(k,k-P_n) ~ \right ] _{on}
~(\rlap\slash k_{on} + m)~ \gamma^5 \right ]
\nonumber
\ee

\begin{eqnarray}
{\cal{T}}_{on, (2, n)} &=&  ~\psi^* _{\bar{\pi}}((k^+ - P^+ _{\pi} ), ({\bf k -
P_{\pi}})_{\perp}; P^+_{\bar{\pi}}, {\bf P}_{{\bar{\pi}} \perp}) ~ ~
\times
\label{Ton2VT}    \\
&&  Tr \left [ [(\rlap\slash k - \rlap\slash P_{\pi})_{on} + m]
~ \gamma^5 [(\rlap\slash k - \rlap\slash q)_{on} + m]~
\left [ \widehat{V}_{nz}(k,k-P_n) ~ \right ] _{on}
~(\rlap\slash k_{on} + m)~ \gamma^5 \right ]
\nonumber
\ee

\begin{eqnarray}
{\cal{T}}_{1, (1, n)} &=& \frac{1} { 2 }  ~ \frac{m}{f_\pi} ~
 \left[\overline\Lambda _{\pi}(k; P_{\pi}) ~
\right ] _{k^- = k^-_{on}} ~
\times
\nonumber \\
&&  Tr \left[ \gamma^+
~\gamma^5 [(\rlap\slash k - \rlap\slash q)_{on} + m]~
\left [ \widehat{V}_{nz}(k,k-P_n) ~ \right ] _{on}
~(\rlap\slash k_{on} + m)~ \gamma^5 \right ]
\label{T11VT}
\ee

\begin{eqnarray}
{\cal{T}}_{1, (2, n)} &=& \frac{1} { 2 } ~ \frac{m}{f_\pi} ~
 \left [ \Lambda_{\bar{\pi}}(k - P_{\pi},P_{\bar{\pi}})
\right ]
_{k^- = q^- + (k - q)^-_{on}} ~
~ ~ \times
\nonumber \\
&&  Tr \left [ \gamma^+
~\gamma^5 [(\rlap\slash k - \rlap\slash q)_{on} + m]~
\left [ \widehat{V}_{nz}(k,k-P_n) ~ \right ] _{on}
~(\rlap\slash k_{on} + m)~ \gamma^5 \right ]
\label{T12VT}
\ee

\begin{eqnarray}
{\cal{T}}_{2, (1, n)} &=& ~ \frac{1}{2} ~
\psi^* _{\pi}(k^+, {\bf k}_{\perp}; P^+_{\pi}, {\bf P}_{\pi \perp}) ~
~ \times  \quad \quad
\nonumber \\
&&  Tr \left [[(\rlap\slash k - \rlap\slash P_{\pi})_{on} + m]
~\gamma^5 ~ \gamma^+ ~
\left [ \widehat{V}_{nz}(k,k-P_n) ~ \right ] _{on}
~(\rlap\slash k_{on} + m)~ \gamma^5 \right ]
\label{T21nT}
\ee

\begin{eqnarray}
{\cal{T}}_{3, (2, n)} &=& ~  \frac{1}{2} ~ \psi^* _{\bar{\pi}}((k^+ - P^+ _{\pi} ), ({\bf k -
P_{\pi}})_{\perp}; P^+_{\bar{\pi}}, {\bf P}_{{\bar{\pi}} \perp})
~
~ \times \quad \quad
\nonumber \\
&& Tr \left [[(\rlap\slash k - \rlap\slash P_{\pi})_{on} + m]
~\gamma^5 [(\rlap\slash k - \rlap\slash q)_{on} + m]~
\left [ \widehat{V}_{nz}(k,k-P_n) ~ \right ] _{on}
~ \gamma^+ ~ \gamma^5 \right ]
\label{T32nT}
\ee

The $^3S_1$ vector meson vertex $[\widehat{V}_{n}(k,k-P_n)]_{on}$
given by Eq. (\ref{gams1}), as it was used in previous
calculations \cite{Jaus90}, is completely determined by the
kinematical momenta of the individual quark and antiquark.
In the $^3S_1$ vector meson intrinsic frame, where $q^+=M_n$ and
$\bf q_{\perp}=0$, one has
\be
\left [ \widehat{V}_{nz}(k,k - P_n) ~ \right ] _{on} =
 \left (\gamma^3-
{ k^{3}_{on} - (P_n^{3} - k^{3})_{on} \over M_{0n}+ 2m} \right )
\label{tetap}
\ee
where $M_{0n}^2 = (|{\bf k}_{\perp}|^2+m^2)/x (1-x)$ ($x=k^+/q^+=k^+/M_n$).

The kinematics for the final two-pion state in the particular
frame where the photon has momentum ${\bf q}_{\perp}=0$, $q^+=M_n$
and $q^-=q^2/M_n$ can be derived from the energy-momentum
conservation, which yields
\be
q^+ = M_n = P^+_{\bar{\pi}} + P^+_{{\pi}},
\quad \quad {\bf P}_{\bar{\pi} \perp} = -{\bf
P}_{{\pi} \perp}
\label{kine}
\ee
and thus
\be q^- = P^-_{\bar{\pi}} + P^-_{{\pi}} =
{|{\bf P}_{{\pi} \perp}|^2 + m^2_{\pi} \over
P^+_{\bar{\pi}}} +{|{\bf P}_{{\pi} \perp}|^2 + m^2_{\pi} \over
P^+_{{\pi}}} = {1 \over q^+} {|{\bf P}_{{\pi} \perp}|^2 + m^2_{\pi}
\over x_{{\pi}}~(1-x_{{\pi}})}
\label{kine1}
\ee
where
$x_{\pi}=P^+_{{\pi}}/q^+$ and $x_{\bar{\pi}} = P^+_{\bar{\pi}}/q^+ =1-x_{\pi}$.  Eqs.
(\ref{kine}) and (\ref{kine1}) put in evidence the relation
between the kinematical variables of the virtual photon and the ones of both
pion and antipion. In the time-like region the value
of  $q^2$ does not fully determine the values for the four-momenta
of the pion and the antipion in the final state of the $\pi \bar{\pi}$  pair.
In order to reduce the
freedom we make the purely longitudinal choice, i.e.
${\bf P}_{\bar{\pi} \perp} = -{\bf P}_{{\pi} \perp}={\bf 0} $.  Then,
from Eqs. (\ref{kine}) and (\ref{kine1}), one obtains
\be
x_\pi=\frac12\pm\sqrt{\frac14-{m^2_\pi\over q^2}}.
\label{xpi}
\ee
Let us note that the minimum allowed value for $q^2$ is $4 m^2_\pi$.
At this threshold value one has $x_\pi = 1/2$. Therefore, since
\be
P^+_{\bar{\pi}} - P^+_{{\pi}} =
q^+ - 2 P^+_{{\pi}} = q^+ (1 - 2x_\pi) = M_n (1 - 2x_\pi) ~,
\label{piu}
\ee
one cannot evaluate Eq. (\ref{jmuT}) exactly at threshold, unless an exact cancellation
occurs between vanishing numerator and denominator of Eq. (\ref{jmuT}).
For finite
values of $m_\pi$, the values $x_\pi=1$ or $x_\pi=0$ are possible
only for an infinite value of the momentum transfer and imply an
infinite value of $P_{{\pi z}}$ or $P_{\bar{\pi} z}$,
respectively.

In the limit of {\em a vanishing pion mass} ($m_\pi=0$), Eq. (\ref{xpi}) gives $x_\pi=1$ or $ 0$,
which implies that one of the terms of Eq. (\ref{jmuT}) vanishes due to
the $\Theta$ function. To simplify our calculations, in the following we make the approximation
$m_\pi=0$ and adopt the choice  $x_\pi=0$,
which implies $P^+_\pi = 0$, $P^-_\pi = q^- = q^2/M_n$,
$P^+_{\bar{\pi}} = q^+ = M_n$, and $P^-_{\bar{\pi}} = 0$. Then only the second term
of Eq. (\ref{jmuT}), containing the quantity $\overline{I}_{2, n}$, gives a
contribution to the TL pion form factor.

 Furthermore, for  $m_\pi=0$ one has ${\cal {T}}_{on, (2, n)} = 0 ~$. Indeed, in this limit
 $~(\rlap\slash k - \rlap\slash P_{\pi})_{on} = \rlap\slash k_{on}$ and
\be
(\rlap\slash k_{on} + m) ~ \gamma^5 ~ (\rlap\slash k_{on} + m) =
 (\rlap\slash k_{on} + m) ~ (- ~\rlap\slash k_{on} + m) ~ \gamma^5 =
 (- ~ \rlap\slash k_{on} \rlap\slash k_{on} + m^2) ~ \gamma^5 = 0 ~ \quad
\label{dec3}
\ee
Therefore only the instantaneous contributions
${\cal {T}}_{1, (2, n)}$ and ${\cal {T}}_{3, (2, n)}$ survive in the limit of a vanishing pion mass
and can be written as follows :

\begin{eqnarray}
{\cal {T}}_{1, (2, n)} &=& - ~ \frac{1}{2} ~ \frac{m}{f_\pi} ~
 \left [ \Lambda_{\bar{\pi}}(k - P_{\pi},P_{\bar{\pi}})
\right ] _{k^- = q^- + (k - q)^-_{on}} ~
~ ~ \times
\nonumber \\
&&  Tr \left [ \gamma^+
~ [(\rlap\slash k - \rlap\slash q)_{on} + m]~
\left [ \widehat{V}_{nz}(k,k-P_n) ~ \right ] _{on}
~(\rlap\slash k_{on} + m) \right ]
\label{T12VTL}
\ee

\begin{eqnarray}
{\cal {T}}_{3, (2, n)} &=& ~ \frac{1}{2} ~
\psi^* _{\bar{\pi}}(k^+, {\bf k }_{\perp}; M_n, {\bf 0}_{ \perp})
 ~
~ \times \quad \quad
\nonumber \\
&& Tr \left [[ - \rlap\slash k_{on} + m]
~ [(\rlap\slash k - \rlap\slash q)_{on} + m]~
\left [ \widehat{V}_{nz}(k,k-P_n) ~ \right ] _{on}
~ \gamma^+ \right ] \quad  \ .
\label{T32nTL}
\ee

To evaluate the time-like pion form factor we  have still to
specify the values of the instantaneous
vertex functions
$\left [ \Lambda_{\bar{\pi}}(k - P_{\pi},P_{\bar{\pi}}) \right ] _{k^- = q^- + (k - q)^-_{on}}$
in Eq. (\ref{T12VTL})
 and  $\left [ \Lambda_{n}(k,P_n) \right ]_{k^- = q^- + (k - q)^-_{on}}$
in Eq. (\ref{jmu2nT}) that, as already explained, cannot be directly related to
$\psi_{\bar{\pi}}$ and $\psi_n$.

\section{  Space-like em form factor of the pion}

The space-like form factor of the pion can be obtained from the plus
component of the proper current matrix element
\be
j^{\mu} =\langle \pi | \bar{q}(0) \gamma^{\mu}q(0)
| \pi \prime |\rangle = e ~ \left (P^{\mu}_{\pi} + P^{\mu}_{\pi \prime} \right ) ~ F_{\pi}(q^2)
\label{dec2}
\ee
where $q^{\mu} = P^{\mu}_{\pi \prime} - P^{\mu}_{\pi}~$.

In our reference frame, where ${\bf q}_{\perp}=0$ and $q^+>0$, the minus-component of the
four-momentum transfer is given by $q^- = q^2/q^+$, which is negative in the space-like
region. Let us note that
\be
q^- = {|{\bf P}_{{\pi} \prime
\perp}|^2 + m^2_{\pi} \over P^+_{\pi \prime}} - {|{\bf P}_{{\pi}
\perp}|^2 + m^2_{\pi} \over P^+_{{\pi}} } ~ ~.
\label{qmneg}
\ee
Hence, the constraint $q^- < 0$ is obviously fulfilled for any
value of ${\bf P}_{{\pi} \perp}$, since $|{\bf P}_{{\pi} \prime
\perp}| = |{\bf P}_{{\pi} \perp}|$ and $P^+_{\pi \prime} = q^+ +
P_{\pi}^+ > P_{\pi}^+$. From  Eq. (\ref{qmneg}) one has
\be
q^2 = - (q^+)^2 ~
{|{\bf P}_{{\pi} \perp}|^2 + m^2_{\pi} \over P_{\pi}^+ ~(q^+ +
P_{\pi}^+)} =
 - {|{\bf P}_{{\pi} \perp}|^2 + m^2_{\pi}
\over x_{{\pi}}~(1 + x_{{\pi}})} ~ ~,
\label{qSL}
\ee
where $x_{\pi}=P^+_{{\pi}}/q^+$.
Therefore, once a value for $|{\bf P}_{{\pi} \perp}|$ is chosen,
$P_{\pi}^+$ and $P^+_{\pi \prime}$ are fixed. For a purely longitudinal motion
of the pions, i.e. ${\bf P}_{\pi \perp}= {\bf P}_{\pi \prime \perp} = {\bf 0} $,
 it is easy to obtain from Eq. (\ref{qSL}) that
\begin{eqnarray}
{P}^+_{\pi} = q^+ \left ( -\frac12 +
\sqrt{\frac14-{m^2_\pi\over q^2}}\right) \ \ \text{and} \ \
{P}^+_{\pi \prime} = q^+ \left(\frac12 +
\sqrt{\frac14-{m^2_\pi \over q^2}}\right) \ .
\label{kinsl}
\end{eqnarray}
In the {\em{limit of}} $m_\pi=0$, the longitudinal momenta of the pions
according to Eq. (\ref{kinsl}) are
\begin{eqnarray}
{P}^+_\pi=0  \ \ \text{and} \ \ {P}^+_{\pi \prime} = q^+
\label{kinsl0}
\end{eqnarray}
for any value of the momentum transfer.

 In a frame where $q^+ \neq 0$, the electromagnetic current $j^+$ in
 the space-like region, Eq. (\ref{jIeII}), receives contributions from the valence component
of the wave function, $j^{(I)+}$ given by Eq. (\ref{jmuIa}), as well as
from the nonvalence components, $j^{(II)+}$ of
Eq. (\ref{jmuO1b}), i.e. from the Z-diagram  contribution (see Fig. 7).

The contribution of the pion valence wave
function to the current can be calculated from Eq. (\ref{jmuIa}) introducing
the plus component of the operator
$\Gamma^\mu(k - P_{\pi},q)$ for $k^{ -} = k^{ -}_{on}$, as
given in Eq. (\ref{phabs}) and discussed in Sect. V B,
once  the values of the pion vertex functions
$\left [  \overline \Lambda_{\pi \prime}(k^{\prime}, P_{\pi \prime})
 \right ] _{k^{\prime -} = k^{\prime -}_{on} }$ and
 $\left [ \Lambda_{\pi}(P_{\pi} - k^{\prime}, P_{\pi})
 \right ] _{k^{\prime -} = k^{\prime -}_{on}}$  in the instantaneous terms have been specified.

In the limit of zero pion mass,
according to Eq. (\ref{jmuIa})  the valence contribution to
the space-like pion form factor vanishes,  since ${P}^+_\pi=0$. Then,
 only the Z-diagram contribution survives in this limit,
 as in the time-like region.

The contribution of the  Z-diagram to the elastic pion form factor
can be obtained from Eq. (\ref{dec2}) by substituting
in Eq. (\ref{jmuO1b}) the pion absorption vertex of Eq.
(\ref{pibabs}). The result can be written as follows:
\be
F^{II}(q^2) = \sum_n~{f_{Vn} \over q^2 -M^2_n}~ f^{II}_n(q^2) \ .
\label{ffpi1}  \ee
Since $f^{II}_n(q^2)$ is invariant under
kinematical LF boosts, we choose to evaluate the
contribution of each vector meson, $f^{II}_n(q^2)$, in the same
reference frame that we used in the time-like region, i.e., we adopt
the rest frame for each resonance ($q^+ = M_n$, ${\bf q}_{\perp} =
0$ ; $P^+_n = q^+ = M_n$, $P^-_n = M^2_n/q^+ = M_n$).

Then for a finite value of the pion mass we have:
\begin{eqnarray}
f^{II}_n(q^2) &=& \sqrt{2} ~ {N_c \over 8 \pi^3}
{\epsilon^{+}_{z} \over P^+_{\pi\prime} + P^+_{{\pi}} } \int_0^{q^+} {dk^+
\over k^+ ~ (q^+-k^+) ~ (P^+_{\pi} +k^+)} \int d{\bf k}_{\perp} ~ ~ {\cal{D}}_\pi ~ \times
\nonu \nonu
\left \{ { \psi_{n}(k^+, {\bf k}_{\perp}; P^+_{n}, {\bf 0}_{ \perp}) ~ ~
[M^2_n - M^2_0(k^+, {\bf k}_{\perp}; P^+_{n}, {\bf 0}_{\perp})] \over
\left [ q^2-M^2_0(k^+, {\bf k}_{\perp}; q^+, {\bf
0}_{\perp}) + i\epsilon \right ]
} ~
\left [ {\cal T}^{\prime}_{on, (2,n)} ~ + ~ {\cal T}^{\prime}_{1, (2,n)} \right ] ~ + \right .
\nonu
~ \left . \left [ \Lambda_{n}(k,P_n) \right ] _{k^-=q^- + (k - q)^-_{on}} ~ {\cal T}^{\prime}_{3, (2,n)}  \right \}
\quad  \ ,
\label{ffpi2}
\ee
with
\begin{eqnarray}
 {\cal T}^{\prime}_{on, (2,n)} &=&
 \psi^* _{\pi\prime}((k^+ + P^+ _{\pi} ), ({\bf k} +
{\bf P}_{\pi})_{\perp}; P^+_{ \pi\prime}, {\bf P}_{\pi\prime \perp}) ~
 ~
\times
\label{TonIInbb}   \\
&& Tr \left [[(\rlap\slash k + \rlap\slash P_{\pi})_{on} + m] ~
\gamma^5 ~ [(\rlap\slash k - \rlap\slash q)_{on} + m]
\left [ \widehat{V}_{nz}(k,k - P_n) ~ \right ] _{on}
(\rlap\slash k_{on} + m)~ \gamma^5 \right ]
\nonumber
\ee

\begin{eqnarray}
{\cal T}^{\prime}_{1, (2,n)} &=& ~
 \frac{1} { 2 } ~ \frac{m}{f_\pi} ~
 \left [ \overline\Lambda_{\pi \prime}(k + P_{\pi}, P_{\pi \prime})
 \right ] _{k^- = q^- + (k - q)^-_{on}}
 ~ ~
 \times
\nonumber \\
 && Tr \left [ \gamma^+ ~ \gamma^5 ~ [(\rlap\slash k - \rlap\slash q)_{on} + m] ~
\left [ \widehat{V}_{nz}(k,k - P_n) ~ \right ] _{on}
~(\rlap\slash k_{on} + m) ~ \gamma^5 \right ]
\label{T1IInbb}
\ee

\begin{eqnarray}
{\cal T}^{\prime}_{3, (2,n)} &=& \frac{1} { 2 } ~
\psi^* _{\pi\prime}((k^+ + P^+ _{\pi} ), ({\bf k} +
{\bf P}_{\pi})_{\perp}; P^+_{ \pi\prime}, {\bf P}_{\pi\prime \perp}) ~ ~
\times
 \nonumber \\
&& Tr \left [[(\rlap\slash k + \rlap\slash P_{\pi})_{on} + m]
~\gamma^5 ~ [(\rlap\slash k - \rlap\slash q)_{on} + m]~
\left [ \widehat{V}_{nz}(k,k - P_n) ~ \right ] _{on}
~ \gamma^+ ~ \gamma^5 \right ]
\label{T3IInbb}
\ee
The Dirac structure, $\left [ \widehat{V}_{nz}(k,k - P_n) ~ \right ] _{on} $,   for
the $^3S_1$ meson state is given by Eq. (\ref{tetap}).   As already noted, in our
reference frame  one has $\epsilon^+_z=1$.
Equations (\ref{T1IInbb}) and (\ref{T3IInbb}) represent the instantaneous contributions.
Analogously to the time-like case, for a vanishing pion mass one has
${\cal T}^{\prime}_{on, (2,n)} = 0$ (see Eq. (\ref{dec3})).

Let us now evaluate
$f^{II}_n(q^2)$  at $q^2 \rightarrow 0^-$ for a finite value of the mass of the pion.
To begin with, we consider: i) a constant value for ${\cal{D}}_\pi$, ii)
a simple form for the LF
pion wave function  \cite{tobpauli}
\begin{eqnarray}
\psi_{\pi \prime}[(k^+ + P_{\pi}^+), ({\bf k}_{\perp} + {\bf P}_{\pi \perp});
 P_{\pi \prime}^+, {\bf P}_{\pi \prime \perp}]
 = {m \over f_{\pi}}
{ P^+_{\pi \prime} \over
m^2_{\pi} -
M^2_{0 \pi \prime}(k^+ + P_{\pi}^+, {\bf k}_{\perp} + {\bf P}_{\pi \perp};
 P_{\pi \prime}^+, {\bf P}_{\pi \prime \perp}) } \ ,
\label{model3}
\end{eqnarray}
and iii) in the instantaneous term (\ref{T1IInbb})) take
$\left [ \overline\Lambda_{\pi \prime}(k + P_{\pi}, P_{\pi \prime})
 \right ] _{k^- = q^- + (k - q)^-_{on}}$
 proportional to
$\psi^*_{\pi \prime}[(k^+ + P_{\pi}^+), ({\bf k}_{\perp} + {\bf P}_{\pi \perp}); ~
 P_{\pi \prime}^+, {\bf P}_{\pi \prime \perp}]$
(see the next Section).
For a finite value of the mass of the pion, let us note that in the limit  $q^2 \rightarrow 0^-$
from Eq. (\ref{kinsl}) one obtains  ${P}^+_{\pi} \rightarrow \infty$
and ${P}^+_{\pi \prime} = \left (M_n + {P}^+_{\pi}\right )\rightarrow \infty$.
Then, since the squared free mass for the final pion is
\be
M^2_{0 \pi \prime}[(k^+ + P_{\pi}^+), ({\bf k}_{\perp} + {\bf P}_{\pi \perp});
 P_{\pi \prime}^+, {\bf P}_{\pi \prime \perp}]
= {P}^+_{\pi \prime} \left ({|{\bf k}_{ \perp}|^2 + m^2 \over
P_{\pi}^+ + k^+} + {|{\bf k}_{ \perp}|^2 + m^2 \over {P}^+_{\pi
\prime} - P_{\pi}^+ - k^+} \right ) \ , \
\label{M0p}
\ee
it becomes
large for ${P}^+_{\pi} \rightarrow \infty$, i.e. $M^2_{0 \pi
\prime} \sim {P}^+_{{\pi \prime}}$.
Then $\psi^*_{\pi \prime}[(k^+ + P_{\pi}^+), ({\bf k}_{\perp} + {\bf P}_{\pi \perp});
 P_{\pi \prime}^+, {\bf P}_{\pi \prime \perp}]$
  becomes a constant for
${P}^+_{\pi \prime} \rightarrow \infty$. Furthermore, for
$P^+_{\pi} \rightarrow \infty$ the traces in Eqs. (\ref{TonIInbb}, \ref{T1IInbb})
are proportional to
$\sim P^+_{\pi}$. Therefore, collecting together the factors $P^+_{\pi}$
in Eq. (\ref{ffpi2}), one concludes that for a finite value of the  pion mass
$\lim_{~ q^2\rightarrow
0^-} ~ f^{II}_n(q^2) \sim \lim_{~ q^2\rightarrow 0^-} ~
1/P^+_{\pi} = 0$. The same result, $\lim_{~q^2\rightarrow 0^-} ~
f^{II}_n(q^2) = 0$, should also hold for pion wave functions which
are eigenfunctions of a Hamiltonian \cite{FPZ02}.

On the contrary,
in the limit of $m_\pi=0$, the longitudinal momenta of the pions  are
${P}^+_\pi=0$ and ${P}^+_{\pi \prime} = M_n$, respectively (see Eq. (\ref{kinsl})). Then,
according to Eq. (\ref{jmuIa}), the  valence contribution to
the space-like pion form factor vanishes, while
 the Z-diagram yields a nonzero contribution.

A comment is appropriate here. In the work of
Ref. \cite{pach02}, where $m_\pi \neq 0$, it was found that the wave function contribution
to the space-like pion form factor  strongly decreases in the frame $q^+=\sqrt{-q^2}$
as the momentum transfer $-q^2$ increases. As a consequence, the Z-diagram contribution,
which is zero at $q^2 = 0$,
becomes the dominant one at high momentum transfer.  As the pion
mass is artificially decreased in that model, we find that the momentum at which
the Z-diagram  starts to dominate the form factor tends toward
zero,
in agreement with the previous
discussion.

Since in this paper we work at the chiral limit of a vanishing pion mass,
in our reference frame,
 the full space-like pion form factor is given
 by $F^{II}(q^2)$ alone.
It has to be noted that,
 as occurs in the time-like region and for the same reasons, for $m_\pi=0$ only
 the instantaneous terms ${\cal T}^{\prime}_{1, (2,n)}$ and
 ${\cal T}^{\prime}_{3, (2,n)}$
 (cf Eqs. (\ref{T1IInbb}) and (\ref{T3IInbb})) give
 contribution to the pion form factor. These terms can be written in the following form :

\begin{eqnarray}
{\cal T}^{\prime}_{1, (2,n)} &=& ~ - ~
 \frac{1} { 2 } ~ \frac{m}{f_\pi} ~
   \left [ \overline\Lambda_{\pi \prime}(k + P_{\pi}, P_{\pi \prime})
 \right ] _{k^- = q^- + (k - q)^-_{on}}
~  ~
 \times
\nonumber \\
 && Tr \left [ \gamma^+  ~ [(\rlap\slash k - \rlap\slash q)_{on} + m] ~
\left [ \widehat{V}_{nz}(k,k - P_n) ~ \right ] _{on}
~(\rlap\slash k_{on} + m)  \right ]
\label{T1IInbL}
\ee

\begin{eqnarray}
{\cal T}^{\prime}_{3, (2,n)} &=& \frac{1} { 2 } ~
\psi^* _{\pi\prime}(k^+ , {\bf k} _{\perp}; M_n, {\bf 0}_{ \perp}) ~ ~
\times \quad \quad \quad
\nonumber \\
&& Tr \left [[- \rlap\slash k _{on} + m]
~ [(\rlap\slash k - \rlap\slash q)_{on} + m]~
\left [ \widehat{V}_{nz}(k,k - P_n) ~ \right ] _{on}
~ \gamma^+ \right ] \quad \ .
\label{T3IInbL}
\ee

\section{A Light-front model}
To evaluate the pion form factor we need :
\begin{itemize}
\item[i)] a model  for the HLFD pion and vector meson
wave functions which appear in  Eqs. (\ref{jmuT}) and (\ref{ffpi2});
\item[ii)] a value for the probability, $P_{q\bar q,n}$, of the VM valence component
(see Appendices D and E);
\item[iii)] an approximation for the pion vertex functions
which represent the pion emission or absorption by a quark;
\item[iv)] to assign a value to the pion and VM vertex functions
with an instantaneous quark leg.
\end{itemize}

The vector-meson resonances are described by an effective light-front model
inspired by QCD \cite{FPZ02}, that can be also applied to the pion.
  The squared mass-operator for the $S$-mesons  contains a harmonic oscillator
  interaction featuring the confinement and a Dirac delta-function that acts
in the $^1S_0$ channel (with a renormalized strength). The wave
functions for the $^3S_1$ channel are
 solutions of the following eigenvalue problem
\be
\left [4(|{\bf \kappa}|^2 + m^2) + {1 \over 64} \omega^2 r^2 + a \right ]
~\Psi^{HO}_{n}({\bf r})= M^2_{n} ~\Psi^{HO}_{n}({\bf r}) \, \, ~~,
\label{model1}
\ee
where $|{\bf \kappa}|^2 = M^2_{0}/4 -m^2$ is the square of the intrinsic quark three-momentum,
$M^2_{n}= n~\omega ~ + ~ M^2_{\rho}$ and the
eigenfunctions $\Psi^{HO}_{n}({\bf r})$ are the three-dimensional
 harmonic oscillator wave functions for zero angular momentum.
The HLFD wave functions, without the Melosh rotations, are
given in the Fourier space by
\be
\psi_{n}(k^+,{\bf k}_{\perp},P^+_n,{\bf P}_{n \perp}) = P^+_n ~ \Psi^{HO}_{n}(|{\bf \kappa}|^2)
\label{model2} \, \, ~~.
\ee
The factor $P^+_n$ comes from the different normalizations used for
$\psi_{n}(k^+,{\bf k}_{\perp},P^+_n,{\bf P}_{n \perp})$ and $\Psi^{HO}_{n}(|{\bf \kappa}|^2)$.
Indeed the function $\Psi^{HO}_{n}(|{\bf \kappa}|^2)$ is normalized through the equation
\be
\int|\Psi^{HO}_{n}(|{\bf \kappa}|^2)|^2  d^3 \kappa = 1 ~~~\ ,
\label{norm3}
\ee
while the function $\psi_{n}(k^+,{\bf k}_{\perp},P^+_n,{\bf P}_{n \perp})$
is normalized through the evaluation of the
charge form factor of a vector meson at $q^2=0$,
i.e. by using the so-called charge normalization (Appendix D),
more appropriate in a relativistic
 context \cite{mandel}. In the actual calculation, we have to consider that,
 after properly integrating the valence component, its probability should be
 recovered. This amounts to construct a schematic model for the probability,
 $P_{n,q \bar q}$, for each excited state ( see Appendix E), and subsequently
to  renormalize $\psi_{n}(k^+,{\bf k}_{\perp},P^+_n,{\bf P}_{n \perp})$ in Eq.
(\ref{model2}) as
follows
\be
\psi_{n}(k^+,{\bf k}_{\perp},P^+_n,{\bf P}_{n \perp})~=
~\sqrt{P_{n,q \bar q}}~P^+_n ~ \Psi^{HO}_{n}(|{\bf \kappa}|^2)
\label{rinmodel2}
\ee
In the model of Ref. \cite{FPZ02} the complete form of the pion wave function
is an eigenstate of the mass operator of Eq. (\ref{model1}) plus
a Dirac-delta interaction (in the configuration space),
which is necessary for producing  a pion with a small mass (i.e. a
collapsing $q\bar{q}$ pair in the $^1S_0$ channel).
The pion wave function is found from the pole of the resolvent, explicitly written in
Ref. \cite{FPZ02}. The result is the following:
\begin{eqnarray}
\psi_{\pi}(k^+,{\bf k}_{\perp},P^+_{\pi} ,{\bf P}_{\pi \perp}) = P^+_{\pi} \sum_{n}
{\Psi^{HO}_{n }(|{\bf \kappa}|^2)\Psi^{HO}_{n }(0) \over
m^2_\pi - M^2_{n}} \ ,
 \label{model5}
\end{eqnarray}
where $\Psi^{HO}_{n}(0)$ is  the S-wave harmonic oscillator
eigenfunction in coordinate space at the origin.

In this model, the pion wave function  approaches the asymptotic
limit, Eq. (\ref{model3}), imposed by the presence of the Dirac
delta-function in the interaction.

The relativistic constituent quark  model of  Ref. \cite{FPZ02}
achieves a satisfactory description of the experimental masses for both singlet and
triplet $S$-wave
mesons, with a natural explanation of the "Iachello-Anisovitch law" \cite{iach,ani},
namely the almost linear relation between the square mass of the excited states
and the radial quantum
number $n$. Since the model does not include the mixing between isoscalar and isovector
mesons, in this paper we include only the contributions of the isovector
$\rho$-like vector mesons.

As already discussed in Sec. VI, we approximate the pion vertex functions
which represent the antipion and pion emission by a quark, as well as
the quark-pion absorption vertex by means of a constant
\be
\overline {\cal{D}}_{\pi} =  {\cal{D}}_{\overline \pi} = \frac{m}{f_\pi} ~ \lambda _{\pi} \ ;
\quad
~~~~{\text{and}}~~~~~~~ {\cal{D}}_\pi
 = {m \over f_{\pi}} ~ \lambda _{\pi}
 \label{model4}
 \ee
 in agreement with the constant form proposed in Ref. \cite{JI01} and successfully tested
in the study of the pseudo-scalar meson decays. The actual value of the
constant $\lambda _{\pi}$
is fixed by the pion charge normalization.

 As anticipated in Sec. VII and Sec. VIII, to simplify our calculations we
are going to use $m_\pi=0$. Within this assumption, for the time-like
form factor only the instantaneous
contributions ${\cal {T}}_{1, (2, n)}$ and ${\cal {T}}_{3, (2, n)}$ survive, while
for the space-like form factor only the instantaneous terms
${\cal T}^{\prime}_{1, (2,n)}$ and ${\cal T}^{\prime}_{3, (2,n)} $ remain.
Then to fully evaluate the pion form factor
in the time-like and in the space-like
region we have still to assign a value to the pion and VM vertex functions
with an instantaneous quark leg, i.e. to the vertex functions
$\left [ \Lambda_{\bar{\pi}}(k - P_{\pi},P_{\bar{\pi}}) \right ] _{k^- = q^- + (k - q)^-_{on}}$
and
$\left [ \overline\Lambda_{\pi \prime}(k + P_{\pi}, P_{\pi \prime})\right ] _{k^- = q^- + (k - q)^-_{on}}$
in Eqs. (\ref{T12VTL}) and (\ref{T1IInbL}), respectively, and
to the vertex function $\left [ \Lambda_{n}(k,P_n) \right ]_{k^- =  q^- + (k - q)^-_{on}}$
of Eqs. (\ref{T32nTL}) and (\ref{T3IInbL}).
The instantaneous contributions to the time-like pion form factor corresponding to the vertex
functions
$\left [ \Lambda_{\bar{\pi}}(k - P_{\pi},P_{\bar{\pi}}) \right ] _{k^- = q^- + (k - q)^-_{on}}$
and $\left [ \Lambda_{n}(k,P_n) \right ]_{k^- =  q^- + (k - q)^-_{on}}$ are represented by
diagrams (a) and (b) of Fig. 8, respectively.

Let us note that the presence of the factors $(k^+ \pm P^+_{\pi})$ and $k^+$ in the
denominators of the two instantaneous terms produces an enhancement of the contributions
around the values $(k^+ \pm P^+_{\pi}) = 0$ and $k^+ = 0$ in the $k^+$ integration.
Within our assumption of a vanishing pion mass, this means that, for both the instantaneous terms,
there is an enhancement of the contribution at the end point $k^+ = 0$,
which corresponds to an infinite value
of the $z$ component of the intrinsic quark three-momentum, $\kappa_z = M_0(2x-1)/2$
($\kappa_z = - \infty$ for $x = 0$, since $M_0 \rightarrow \infty$). Therefore the high momentum part of the meson
vertex functions, i.e. the short-range part in coordinate space, is very relevant.
Then in the vertex functions with an instantaneous quark leg, ${\Lambda}^{ist}_{\pi (n)}$,
we assume that the very short-range part of the one-gluon-exchange interaction,
which includes spin-spin terms \cite{deR}, is the dominant one.
In symbolic notation we  have (see Fig. 8) :
\be
{\Lambda}^{ist}  = {\cal {K}}^{ist} ~ G_0 ~ {\Lambda}^{full}
\label{ista}
\ee
where ${\cal {K}}^{ist}$ is the Bethe-Salpeter kernel for the instantaneous
vertex function ${\Lambda}^{ist}$, $G_0$ the propagator of two free quarks
 and ${\Lambda}^{full}$  the full vertex function.

 The kernel ${\cal {K}}^{ist}$ is assumed to be dominated by the short-range part
of the interaction.
Actually we drastically simplify Eq. (\ref{ista}) as follows :
\be
{\Lambda}^{ist} \sim c ~ {\Lambda}^{full} \, \, ~~.
\label{OGE}
\ee
This amounts to naively assume that ${\Lambda}^{full}$ is an eigenstate of
${\cal {K}}^{ist} ~ G_0$. Furthermore, we  assume that
${\Lambda}^{full}$  is still related to the LF meson wave function
as illustrated in Sec. IV,
i.e. ${\Lambda}^{full}_{\pi (n)} = \psi_{\pi (n)} ~ [M^2_{\pi (n)} - M^2_0] / P^+_{\pi (n)}$.

The constant $c$ is thought to roughly describe the effects of the
short-range interaction. In particular,
if  we  take grossly into account only the spin-spin interaction term,
then the results of Ref. \cite{DFPS} are recovered i) by choosing $c
= - 3/4$ for the pion vertex function (Fig. 8 (a)) and
$c=1/4$ for the VM vertex function (Fig. 8 (b)) and ii) by using
 the probabilities
$P_{q\bar q;n} =
\frac{\delta ~ \omega^\frac12 }{2 ~ \sqrt{ 2 n+ \frac{3}{2}}}$
for the VM valence components with $\delta ~ \omega^\frac12/2 = 1$
(see Appendix E).
 With this choice for the constants $c$'s, the relative weight of the VM instantaneous  terms with
 respect to the  pion instantaneous  terms is equal to $- 1/3$.
 At variance with Ref. \cite{DFPS},
in the present paper we use this relative weight, $w_{VM} = c_{VM} / c_{\pi}$, as a free parameter.

 In conclusion,  we replace the momentum component of
 the pion vertex function  in Eq. (\ref{T12VTL}) as follows
 \be
 {m \over f_{\pi}} ~\left [ \Lambda_{\bar{\pi}}(k - P_{\pi},P_{\bar{\pi}})
 \right ] _{k^- = q^- + (k - q)^-_{on}}
 \rightarrow \nonu
{c_{\pi} \over P^+_{\bar{\pi}}}
\psi_{\bar{\pi}}(k^+ - P^+_{\pi}, {\bf k}_{\perp} - {\bf P}_{\pi \perp};
 P^+_{\bar{\pi}}, {\bf P}_{\bar{\pi} \perp})
 ~ [m^2_\pi - M^2_0(k^+ - P^+_{\pi}, {\bf k}_{\perp} - {\bf P}_{\pi \perp}; P^+_{\bar{\pi}},
{\bf P}_{\bar{\pi} \perp})]
\ee
and  in Eq. (\ref{T1IInbL}) as follows
\be
{m \over f_{\pi}} ~\left [ \overline\Lambda_{\pi \prime}(k + P_{\pi}, P_{\pi \prime})
\right ] _{k^- = q^- + (k - q)^-_{on}}
 \rightarrow \nonu
{ c_{\pi}\over P^+_{\pi \prime}}\psi^*_{\pi \prime}(k^+ + P^+_{\pi}, {\bf k}_{\perp} + {\bf P}_{\pi \perp};
P^+_{\pi \prime}, {\bf P}_{\pi \prime \perp})
 ~ [m^2_\pi - M^2_0(k^+ + P^+_{\pi}, {\bf k}_{\perp} + {\bf P}_{\pi \perp}; P^+_{\pi \prime},
{\bf P}_{\pi \prime \perp})] ~.\ee
The momentum component of the VM vertex function
in Eqs. (\ref{T32nTL}), (\ref{T3IInbL})is approximated by
\be\left [ \Lambda_{n}(k,P_n) \right ]_{k^- =  q^- + (k - q)^-_{on}}
\rightarrow
 {c_{VM} \over P^+_{n}} \psi_{n}(k^+, {\bf k}_{\perp}; P^+_{n}, {\bf P}_{n \perp}) ~
[M^2_n - M^2_0(k^+, {\bf k}_{\perp}; P^+_{n}, {\bf P}_{n \perp})]
. \ee

As explained in the previous sections, in the limit of a vanishing pion mass
both in the time-like and in the space-like case one has $P^+_\pi=0$ and
${P}^+_{\pi \prime} = P^+_{\bar{\pi}} = M_n$. Then,
the quantities $g^+_{Vn}(q^2)$ of Eq. (\ref{jmuT}) and $f^{II}_{n}(q^2)$ of
Eq. (\ref{ffpi2}) acquire the same functional form, despite the sign of $q^2$,
and reduce to the same function $\xi_{n}(q^2)$ :
\begin{eqnarray}
\xi_{n}(q^2) = &&{N_c \over 16 \pi^3}  ~ {m\over f_{\pi}} ~ \lambda _{\pi} ~ c_{\pi}
~{\sqrt{2} \over M^2_{n}} \int_0^{M_{n}}{dk^+\over
(k^+)^2~(M_{n}-k^+)}\int d{\bf k}_\perp
\left[ {\cal T}_{1, n}(k^+,{\bf k}_\perp) + {\cal T}_{3, n} (k^+,{\bf k}_\perp) \right]
~\times \nonumber \\
\nonumber \\
&& \psi^*_{\pi \prime}(k^+, {\bf k}_{\perp}; M_n, {\bf 0}_{ \perp})
~ \left[ M^2_n - M^2_0(k^+, {\bf k}_{\perp}; M_{n}, {\bf 0}_{\perp}) \right] ~
\psi_{n}(k^+, {\bf k}_{\perp}, M_{n}, {\bf 0}_{ \perp }) ~   ,
\label{qsi}
\end{eqnarray}
where ${\cal T}_{1,n}$ and ${\cal T}_{3,n}$ are  given by
 \begin{eqnarray}
 {\cal T}_{1, n} &=&    - ~
 { \left[ m^2_\pi - M^2_0(k^+, {\bf k}_{\perp}; M_n, {\bf 0}_{ \perp}) \right] \over
\left [ q^2 - M^2_0(k^+, {\bf k}_{\perp}; M_n, {\bf 0}_{\perp}) + i\epsilon \right ] } ~
 \times \nonumber \\ &&
Tr \left [ \gamma^+  ~ [(\rlap\slash k - \rlap\slash q)_{on} + m] ~
\left [ \widehat{V}_{nz}(k,k - P_n) ~ \right ] _{on}
~(\rlap\slash k_{on} + m)  \right ] =
\nonumber \\
\nonumber \\
&=& -  ~ 4 ~ {\left[ m^2_\pi - M^2_0(k^+, {\bf k}_{\perp}; M_n, {\bf 0}_{ \perp}) \right] \over
\left [ q^2 - M^2_0(k^+, {\bf k}_{\perp}; M_n, {\bf 0}_{\perp}) + i\epsilon \right ] } ~
 \times
\nonumber \\
&& \left[ k^+(k-q)_{on, z} + (k-q)_{on}\cdot k_{on} + (k^+ - M_{n}) k_{on, z} - m^2\right.
 \nonumber \\
&&\left.  - ~
m ~ (2k^+-M_{n})(k_{on}-(q-k)_{on})_z~H_S(M_0) \right]
\label{T1nbL}
\ee

\begin{eqnarray}
{\cal T}_{3, n} &=& w_{VM} ~ Tr \left [[- \rlap\slash k _{on} + m]
~ [(\rlap\slash k - \rlap\slash q)_{on} + m]~
\left [ \widehat{V}_{nz}(k,k - P_n) ~ \right ] _{on}
~ \gamma^+ \right ] =
\nonumber \\
&=& w_{VM} ~ 4 \left[ - k^+(k-q)_{on, z} + (k-q)_{on}\cdot k_{on} + (k^+ - M_{n}) k_{on, z} - m^2 \right.
\nonumber \\
&&\left. + ~ m ~ M_{n}
\left [ k_{on}-(q-k)_{on} \right ] _z ~ H_S(M_0) \right]
\quad \ .
\label{T3nbL}
\ee
In the last steps in Eqs. (\ref{T1nbL}) and (\ref{T3nbL}) the traces have been explicitly evaluated and
the function $H_S(M_0)$  is given by:
\begin{eqnarray}
H_S(M_0)=\frac1{M_0+2m} \label{hs}  \ .
\end{eqnarray}
Actually the value of $c_{\pi}$ together with the value of $\lambda_{\pi}$ is fixed by the
charge normalization and we have to assign a value only to the relative weight $w_{VM}$.

In Eq. (\ref{qsi}) there is no divergence from the poles at the end points
$k^+ = 0$ and $q^+ - k^+ = 0$, because of the Gaussian decrease of the
VM wave functions at these end points, which correspond to infinite values
of the $z$ component of the intrinsic quark three-momentum, $\kappa_z = M_0(2x-1)/2$
($\kappa_z = - \infty$ or $\kappa_z = + \infty$ for $x = 0$ or $x = 1$, respectively).

Finally, both in the time-like and in the space-like regions,
the pion electromagnetic form factor can be written as
\be
 F_{\pi}(q^2) = \sum_n ~ {f_{Vn}   \over
 \left [ q^2 - M^2_n + \imath M_n \tilde{\Gamma}_n(q^2) \right ]} ~~ \xi_{n}(q^2)
\label{ffactor}
\end{eqnarray}
We stress that the pion form factor is continuous at $q^2 = 0$
in the limit $m_{\pi} \rightarrow 0$ and that only the instantaneous terms contribute in this limit.
We would like to remind the reader that
the vector meson wave functions are normalized to the probability of
the valence component, which can be roughly estimated
in a simple model,
 as shown in Appendix E.
The decreasing probability of the valence
component for the excited vector meson states is essential to make
convergent the sum over the resonances.

\section{Results}

The pion electromagnetic form factor is calculated through
Eqs. (\ref{ffactor}) and (\ref{qsi}), where the pion and vector
meson wave functions are eigenstates of the square mass operator
defined in Eq. (\ref{model1}) ( shown for the vector channel only).

In our calculation we have a small set of  parameters:
i) the constituent quark mass, ii) the oscillator strength $\omega$,
  iii) the widths for the vector mesons, $\Gamma _n$,
and  iv) the relative weight $w_{VM}$ of the two instantaneous contributions.

 The  up-down quark mass is fixed at
0.265 $GeV$ \cite{FPZ02} and the oscillator strength is fixed at
 $\omega$ = 1.556 $GeV^2$ of Ref. \cite{FPZ02}.

For the first four vector mesons, the  masses and  widths, presented in
Table I,  are  used.

The non-trivial $q^2$ dependence of $\xi_{n}(q^2)$
in our microscopical model allows a small shift of the VM masses
with respect to the values obtained in the analyses of the experimental data
by using Breit-Wigner functions
with constant values for $\xi_{n}(q^2)$.

For the radial excitations with $M_n > 2.150$ $GeV$,
the mass values corresponding to the model of Ref. \cite{FPZ02} are used.
For the unknown widths we use a single width as a
fitting parameter.  We choose the value
$\Gamma _n = 0.15$ $GeV$, which presents the best agreement with the compilation of the
experimental data of Ref. \cite{baldini}. We consider 20 resonances in our calculation
to obtain stability of the results up to $q^2 = 10$ $(GeV/c)^2$.

The probabilities $P_{q\bar q,n}$ of the valence component of the VM states are fixed according to the schematic
model of Appendix E (see Eq. (\ref{prob1}) and Table II).

 As we  discussed in the previous Sections, it is also
necessary to know the amplitude for the virtual process where a
constituent quark radiates or absorbs a pion. This unknown
function was first investigated  in a phenomenological study of decay processes
within LF dynamics \cite{JI01},  it was
 approximated  by a constant, obtaining a satisfactory descrption of
 the experimental data.  We followed the approximation proposed in \cite{JI01}
 in the calculation of the decay
amplitude $\xi_n(q^2)$ of Eq. (\ref{qsi}).
The value of the constant $\lambda _{\pi}$,
together with the constant $c_{\pi}$ (see the previous Section),
is fixed by the charge normalization.

  The values of the coupling constants, $f_{Vn}$, are calculated using Eq. (\ref{fV2ap})
  of Appendix A from the model VM wave functions.
   The corresponding
   partial decay width, $\Gamma_{e^+e-}$,
for these mesons are calculated from our values of $f_{Vn}$ using
Eq. (\ref{gee}) \cite{Jaus99} and are reported in Table I. The partial decay widths
for the vector mesons are in good agreement with the data, when available \cite{pdg}.

We perform two sets of calculations, to test the effect of the
pion wave function model. In one set  we use the
asymptotic form of the pion valence wave function,
Eq. (\ref{model3}), and in another one we choose the eigenstate of
the square mass operator of the model of Ref. \cite{FPZ02}, given by the pion wave
function of Eq. (\ref{model5}).

The results for the form factor are shown in Figs. 9, 10, and 11.
In Fig. 9 the results corresponding to the weight $w_{VM} = - 0.7$ are shown, while in
Fig. 11 the results corresponding to $w_{VM} = - 0.7$ and $w_{VM} = - 1.5$ are compared in a
linear scale around the $\rho$ meson peak. In Fig. 9 we also report the results
calculated with
the masses and the widths used in Ref. \cite{DFPS} and reported in Table III. For this
calculations the oscillator strength $\omega = 1.39$ $GeV^2$,  the probabilities
$P_{q\overline q;n} =\frac{1}{ \sqrt{ 2 n+ \frac{3}{2}}}$
and $c_{\pi} = - 3/4$, $c_{VM} = 1/4$ have been used.

Let us note that our results are the same within a few percent, if in Dirac structure of
the $n{\rm th}$ VM vertex (Eq. (\ref{gams1})), the free mass is replaced with $M_n$.

 In Fig. 9, we show our results in a wide region of square
momentum transfers, from -10 up to 10 $(GeV/c)^2$, comparing them with the data
collected by Baldini et al. \cite{baldini} and with the data of
Ref. \cite{JLABp}. A general
qualitative agreement with the data is seen in this wide range of
momentum transfers, independently of the detailed form of the pion
wave function.
It has to be stressed that the heights of the TL bumps directly depend on the calculated
values of $f_{Vn}$ and $\xi_{n}(q^2)$.

The results obtained with the asymptotic pion
wave function and the full model present some difference
only above 3 $(GeV/c)^2$.

The pion form factor is particularly very well described in the space-like region,
both using the weight $w_{VM} = - 0.7$ or the weight $w_{VM} = - 1.5$, as can be clearly seen
 in Fig. 10, where the ratio of the SL form factor to the monopole
factor $M(q^2) = 1/(1~ - ~ q^2/M_{\rho}^2)$ is shown.
The excellent agreement with the experimental form factor
at low momentum transfers is expected, since we have built-in
the generalized $\rho$-meson dominance.

The time-like region between 0 and 3 $(GeV/c)^2$, where $\rho(770)$,
$\rho(1450)$ and $\rho(1700)$ appear, is shown in Fig. 11 in a linear scale. The
$\rho$-meson peak is placed at the right position using a bare
mass of 770 $MeV$. From this figure it is clear that  the
parameter $w_{VM}$ is able to control the region of the  $\rho(770)$ peak, while in
other regions its effect is less relevant. For $w_{VM}=-1.5$, the
$\rho(770)$ peak is very well  described, except for
 the region around 2 $(GeV/c)^2$, where our results underestimate the
experimental data.
This dip is due to a destructive interference between the
contributions of $\rho(770)$,  $\rho(1450)$,  and $\rho(1700)$, and could be
potentially sensitive for a detailed test of the model presently adopted for
the meson wave functions and
other approximations introduced.

 It is clear that the introduction of $\omega$-like and $\phi$-like mesons could improve the
description of the data in the TL region.
However, a consistent dynamical description of the mixing of
isospin states is far beyond the present work, and we leave it for
 future developments of the model.

Finally, we have also calculate the adimensional quantity,
 $F_\pi(q^2)~q^2/.77^2$, up to
$q^2=-1000~(GeV/c)^2$,  observing a smooth decreases from a value of $0.691$
for $q^2=-100~(GeV/c)^2$ to a value of $0.677$ for $q^2=-1000~(GeV/c)^2$.
\section{Summary and Conclusions}

In this work, we are able to give  a unified description
  of the pion electromagnetic
form factor in the space- and time-like regions, thanks to the choice of  a
reference frame where $q^+~>~0$.

The main steps are shortly summarized. Within the covariant approach proposed
by Mandelstam \cite{mandel},  the matrix elements
 of the electromagnetic current are evaluated between pion states,
 in impulse approximation, but with  all the vertexes of
 the triangle diagram properly
dressed. Exploiting a suitable decomposition of the fermionic propagators,
 one singles out
  on-shell and
instantaneous contributions. The  integration over the light-front energy, $k^-$,
in the
momentum loop of the triangle diagram is performed  disregarding
 the effect of possible singularities of the vertex functions
and taking care  of only the singularities in the propagators.

For the photon vertex function, in the processes where a $q\bar q$-pair
in the odd-parity spin-1 channel is produced, we use a  generalization
of the Vector Meson Dominance approach, built up from the VM Bethe-Salpeter
amplitude (phenomenologically determined) and the VM propagator, enlightening
 the relation between the hadronic part
of the photon valence wave function and the pion electromagnetic
form factor.

The obtained expression for the electromagnetic  current  matrix elements
 are carefully discussed and the different contributions are
interpreted  in terms of valence and
nonvalence components of the pion and photon wave functions.

In the valence  components of the pion and VM amplitudes, the momentum part is
described through the corresponding HLFD wave functions, evaluated in a
QCD-inspired model which shows a satisfactory description for the $^1S_0$ and
 $^3S_1$ mesons. A schematic model is used for the probability, $P_{n,q\bar q}$,
  of the valence
 component of the mesons.

The contribution of the
nonvalence component of the photon wave function appears in the
time-like region, while the nonvalence component of the pion appears
 in the space-like region. The nonvalence contributions of the photon and
 pion wave functions, relevant for the process under consideration,
 involve emission/absorption amplitudes, that in
principle can be calculated from the valence components of the
corresponding  particles,  and a suitable  kernel.
 However, since our knowledge of this
kernel is poor, we use a constant vertex approximation for the
emission/absorption amplitudes \cite{JI01}.

 To simplify our calculation, we take advantage of the smallness of the pion
mass, which is put to zero. Then, only the "Z-diagram"
survives in the space-like region.

We point out that, for
$m_\pi=0$,  only the instantaneous terms contribute to the pion
form factor. Therefore, in order to evaluate the pion form factor we need the
instantaneous vertex functions, which we approximate by the full vertex functions
times a constant.

Only a few parameters define our light-front model: the oscillator
strength, the constituent quark mass,  the VM meson
masses and  widths. We use the experimental
width and mass for the vector mesons, when available. For the radial
excitations above $ 2.150$ $GeV$ we use the masses of the theoretical spectrum and
a single width as a fitting parameter. It is worth noting that the results
are not markedly sensitive  upon  different  pion wave functions, like the
asymptotic wave function and the full-model one of Ref. \cite{FPZ02}.
This could be ascribed   to the strong pion binding, that makes
the pion wave function
similar to its  PQCD asymptotic limit \cite{lepag}.

In the space-like region, the
pion electromagnetic form factor  is very well
described  on the whole  experimentally-explored interval, i.e.  up to $q^2 = - 10$ $(GeV/c)^2$.
In the time-like
region, we find a general agreement up to 10 $(GeV/c)^2$,
except near the experimental dip at 2 $(GeV/c)^2$.

Our model can be straightforwardly improved in many respects. For instance: i) more realistic VM wave functions can
be used, as  the ones of Ref. \cite{Isgur}, that take into account, e.g., the D-state
nature of some of the VM resonances, as $\rho(1700)$; ii) the introduction of
 both a dynamical mixing of isospin states and the
contribution of $\phi$ meson.

Other improvements, like  taking care of the non vanishing pion mass, or
considering a more realistic model for the instantaneous vertexes and for the
emission/absorption of a pion by a quark, are highly non trivial.

In summary, our work appears an encouraging step forward in achieving
a detailed investigation
of  important issues, as the light-quark
content of the photon valence light-front wave function,
 through the analysis of the pion electromagnetic form factor in the time-like
region.  The peculiar feature represented by the  smallness of the  pion
mass is the key point to accomplish such an investigation.

\section*{Acknowledgments}
This work was partially supported by the Brazilian agencies CNPq
and FAPESP and by Ministero dell'Istruzione, dell'Universit\`a e della Ricerca.
 J.P.B.C. M. and T.F. acknowledge the hospitality of
the Dipartimento di Fisica, Universit\`a di Roma "Tor Vergata" and
of Istituto Nazionale di Fisica Nucleare, Sezione Tor Vergata and
of Istituto Nazionale di Fisica Nucleare, Sezione Roma I.
\newpage
\appendix

 \section{Vector meson decay constant}

The vector meson decay constant, $f_{Vn}$, of the $n{\rm th}$ state of
the vector meson is defined as \cite{Jaus99}
\be \epsilon
^{\mu}_{\lambda} \sqrt{2} f_{V,n}=\langle 0| \bar{q}(0) \gamma^{\mu}
q(0) |\phi _{n,\lambda}\rangle
\label{fVapp}
\ee
where $\epsilon^{\mu}_{\lambda}$ are the VM polarization vectors and
$|\phi _{n,\lambda}\rangle$ is the VM state.

 Let us begin with the four-dimensional representation of the decay
 amplitude in terms of the Bethe-Salpeter vertex of the vector
 meson, and use the plus component of Eq.
(\ref{fVapp}) and $\lambda=z$ in the rest frame of the vector meson,
 where $P_n^\mu=[M_n,\vec 0]$ and $\epsilon_z^+=1$:
 \begin{eqnarray}
&&f_{Vn} = - ~ \imath ~ \frac{N_c } {4 ~ (2\pi)^4} ~
\int \frac{dk^- dk^+ d{\bf k}_{\perp}}{ k^+ ~ (P_n^+ - k^+)} ~
\frac{\Lambda_{n}(k,P_{n})}{(k^- - k^-_{on} +  \frac{\imath
\epsilon}{k^+}) ~ (P_n^- - k^-  - (P_n - k)^-_{on} +
\frac{\imath\epsilon}{P_n^+ - k^+})}
  ~ \times \nonu
 Tr \left[ (\psla k - \psla P_n + m) ~ \gamma^+ ~ (\psla k  + m)
  \left [ \epsilon_{z} (P_n) \cdot [ \widehat{V}_{n}(k,k-P_n) ]_{on} \right ]
\right ]\ .
\label{fVa}
\ee
  A factor of $\sqrt{2}$ enters in the
denominator of Eq. (\ref{fVa}) from the normalization of the
neutral vector meson, i.e., $\left(u\overline u- d\overline
d\rangle \right)/\sqrt{2}$.
Then integrating over $k^-$, with the assumptions on the VM vertex function
  already presented at the beginning of Sect. III,
  and taking advantage of the identification of Eqs. (\ref{wfn}) and (\ref{wfpn}), one
 arrives at a three-dimensional formula for the
decay constant where the valence component of the
 vector meson wave function appears:
\be
f_{Vn} = - {N_c \over 4 ~ (2\pi)^3}
\int^{M_n}_0 {dk^+ ~ d{\bf k}_{\perp} \over k^+~(M_n-k^+)} ~ \psi
_{n}(k^+, {\bf k}_{\perp}; M_n,{\bf 0}_{\perp})
\times \nonu
Tr \left[ ~ (\psla k - \psla P_n + m) ~ \gamma^+ ~ (\psla k  + m)
 \left (\psla \epsilon_z - {(k_{on} - (P_n - k)_{on})\cdot \epsilon_z \over
M_{0}(k^+, {\bf k}_{\perp}; P^+_{n}, {\bf P}_{n \perp}) + 2m }\right ) \right ] \; .
\label{fV1ap}
\ee

Evaluating the trace in Eq. (\ref{fV1ap}) the final expression of
the decay constant is:
\be
f_{Vn}= - {N_c \over 8 \pi^3}
\int^{M_n}_0 {dk^+ ~ d{\bf k}_{\perp} \over k^+~(M_n-k^+)} ~
\psi_{n}(k^+, {\bf k}_{\perp}; M_n,{\bf 0}_{\perp})
\times \nonu
\left[m^2 -  k_{on} \cdot
(P_n - k)_{on} - (P_n -k)_{on, z} ~ k^+ ~ + ~
k_{on, z} ~ (M_n-k^+) \right.
\nonu
\left. + m ~ (2k^+ - M_n) ~ {(k_{on} - (P_n -
k)_{on})_z \over M_{0} + 2m } \right ] \ ,
\label{fV2ap}
\ee
where $M^2_0=({\bf k}_\perp^2+m^2)/(x(1-x))$
with $x=k^+/M_n$.

From the vector meson decay constant one gets the decay width to
$e^+e^-$ as \cite{Jaus90}:
 \be
\Gamma_{e^+e^-} = \frac{8\pi\alpha^2}{3}~\frac{f_{Vn}^2}{M_n^3}  ~,
\label{gee}
\ee
where $\alpha$ is the fine structure constant.

\section{Current conservation}

Let us define the four quantities
\begin{eqnarray}
 {\cal {V}} _n^{\mu} &=& \overline{V}^{\mu}_{n} ~ {q \cdot P_n \over M_n^2} ~ - ~
{P_n^{\mu} \over  M_n^2 } ~ q \cdot \overline{V}_{n}
\label{1A}
\ee
where, see Eq. (\ref{gams1}) of Sec. II,
\begin{eqnarray}
\overline{V}^{\mu}_{n} = \left [ \widehat{V}^{\mu}_{n}(k,k-q)\right ]_{on} &=& \gamma^{\mu} -
{k^{\mu}_{on}-(q-k)_{on}^{\mu} \over  M_0 + 2 m } \quad \; \; .
\label{gams1A}
\ee
One can immediately  verify that
\begin{eqnarray}
q \cdot  {\cal {V}} _n &=& 0 \quad \ .
\label{2A}
\ee
Since the vector meson propagator \cite{Halzen} is given by
\be
D^{\mu}_{~ \nu} = \left[ - g^{\mu}_{~\nu} ~ +~ {q^\mu q_\nu \over M^2_n} \right ]
~{ 1 \over \left [ q^2 - M^2_n + \imath M_n \tilde{\Gamma}_n(q^2)\right ]} \quad \ ,
\label{3A}
\ee
a possible conserved photon-($q\bar{q}$) dressed vertex can be defined as follows :
\be
{\cal {J}}^{\mu}(k,q) =\sum_n {\cal {J}}_n^{\mu}(k,q)
\ee
with
\be
{\cal {J}}_n^{\mu}(k,q)=\sqrt{2} ~
\left [-g^{\mu}_{~\nu} ~ +~ {q^\mu q_\nu \over M^2_n} \right ] ~ {\cal {V}}^{\nu} _n
~ \Lambda_{n}(k,q) ~ { f_{Vn} \over \left [ q^2 -
M^2_n + \imath M_n \tilde{\Gamma}_n(q^2)\right ]} =
\nonu
= - ~ \sqrt{2} ~
 ~ {\cal {V}}^{\mu} _n
~ \Lambda_{n}(k,q) ~ { f_{Vn} \over \left [ q^2 -
M^2_n + \imath M_n \tilde{\Gamma}_n(q^2)\right ]}
\quad \ .
\label{4A}
\ee
Indeed it is straightforward to show that $q \cdot {\cal {J}}(k,q) = 0$, since
one has $q \cdot  {\cal {V}} _n = 0$.

 For each term ${\cal {J}}_n^{\mu}$,  let us consider the reference frame where
 $q^+ = M_n > 0$ and ${\bf q }_\perp=0$ ( see Sect. VII for the possibility to use different reference frame
 for different terms in the sum). In this reference frame one has $q^- = q^2 / M_n$ and then
\be
q \cdot \overline{V}_{n} =
{1 \over 2}~ \left [ q^- \overline{V}^+_n + q^+ \overline{V}^-_n \right ] =
{1 \over 2}~ \left [ \overline{V}^+_n ~ q^2 / M_n +  M_n \overline{V}^-_n \right ]
\label{7A}
\ee
\be
q \cdot P_n  =
{1 \over 2}~ \left [ q^2  +  M_n^2  \right ] \quad  .
\label{8A}
\ee
Therefore one obtains
\begin{eqnarray}
 {\cal {V}} _n^{+} &=&
 {1 \over 2}~ \left [ \left ( {q^2 \over   M_n^2 } + 1 \right ) \overline{V}^{+}_{n} ~ - ~
{1 \over  M_n } ~
\left ( \overline{V}^+_n ~ q^2 / M_n +  M_n \overline{V}^-_n \right ) \right ] =
\overline{V}_{n,z}
\label{6A}
\ee
and in conclusion we have
\be
{\cal {J}}^{+}_n(k,q) =
- ~ \sqrt{2}
 ~ \overline{V}_{n,z}
~ \Lambda_{n}(k,q) ~ { f_{Vn} \over \left [ q^2 -
M^2_n + \imath M_n \tilde{\Gamma}_n(q^2)\right ]}
\quad .
\label{5A}
\ee
If in Eq. (\ref{1A}) the quantity $\overline{V}^{\mu}_{n}$ is replaced by
\be
\widehat{V}^{\mu}_{n}(k,k-P_n) =  \gamma^{\mu} -
{k^{\mu} + k'^{\mu}  \over  M_n + 2 m }
\label{9A}
\ee
as defined in Eq. (\ref{gams2}), then the current ${\cal {J}}^{\mu}(k,q)$ is a four vector.

Let us note that, if in Eq. (\ref{5A}) the momentum component of the VM vertex function
$\Lambda_{n}(k,q)$ is taken at the VM pole, then the plus component of the current
${\cal {J}}^{+}(k,q)$ coincides with the one used in our calculations (see Eq. (\ref{polz})).

\section{Subtraction of the bare term in the photon LF vertex}

Let us calculate explicitly
the contribution to the current of a $\gamma^{\mu}$ bare term  in the case of massless pions,
 in  collinear kinematics (${\bf q}_{\perp}=0$, ${\bf P}_{\pi \perp}=0$,
 ${\bf P}_{\bar{\pi} \perp}=0$).
 As discussed at length in Section VII and
VIII, only the instantaneous terms can give a contribution in the limit of a vanishing pion mass.
Therefore one has  :
\be
\delta~j^+
\propto \int d^4k~\Lambda_{\bar{\pi}}(k - P_{\pi},P_{\bar{\pi}}) ~
\overline \Lambda_{\pi}(k,P_{\pi}) ~
Tr \left[\gamma^5 ~ \frac{\gamma^+}{2k^+}~ \gamma^5 ~
\left [\psla k - \psla q +m\right ]~ \gamma^+ ~ \left [\psla k +m
\right]\right]
\times \nonu{1\over(k^2-m^2+\imath\epsilon)
~ ((k-q)^2-m^2+\imath\epsilon)}~~~.
\label{deltaj}
\ee
Performing the
trace, the integration on $k^-$ in Eq. (\ref{deltaj}) and by using the identifications
in Eqs. (\ref{wf2}),
(\ref{wfp}) and (\ref{piemi}), o one has:
\be
\delta~j^+ \propto \int^{q^+}_0 d k^+ \int d{\bf k}_\perp ~ \overline  {\cal{D}}_{\pi} ~
 {k^+ ~ (k^+-q^+)\over (k^+)^2 ~ (q^+ - k^+)} ~
{q^+ \over [ q^2 - M^2_0(k^+, {\bf k}_{\perp}; q^+,
{\bf 0}_{ \perp})  + \imath\epsilon ]}
\times \nonu
\psi_{\bar{\pi}}(k^+, {\bf k}_{\perp}; P^+_{\bar{\pi}}, {\bf 0}_{ \perp})
 ~ [m^2_{\pi} - M^2_0(k^+, {\bf k}_{\perp}; P^+_{\bar{\pi}},
{\bf 0}_{ \perp})] / P^+_{\bar{\pi}} ~
~~~,
\label{deltaj1}
\ee
where $P^+_{\bar{\pi}} = q^+$.
Using the fraction $x=k^+/q^+$ in Eq. (\ref{deltaj1}),
one obtains :
\be
\delta~j^+ \propto -  ~ \int^{1}_0 \frac{dx}{x} ~ \int d{\bf k}_\perp  ~
{\overline {\cal{D}}_{\pi} ~
\psi_{\bar{\pi}}(k^+, {\bf k}_{\perp};  P^+_{\bar{\pi}}, {\bf 0}_{ \perp})
~ [m^2_{\pi} - M^2_0(k^+, {\bf k}_{\perp}; P^+_{\bar{\pi}},
{\bf 0}_{ \perp})]
 \over [q^2 - M^2_0(k^+, {\bf k}_{\perp}; q^+, {\bf 0}_{ \perp}) +\imath\epsilon ] }
\label{deltajp}
\ee

 The bare term of the current violates the
 current conservation. Indeed, the matrix element of the minus component of the bare term is
\be
\delta~j^- \propto
\int^{q^+}_0 d k^+ \int d{\bf k}_\perp ~ \overline  {\cal{D}}_{\pi} ~
 {(m^2 + |{\bf k}_{\perp}|^2)\over (k^+)^2 ~ (q^+ - k^+)} ~
{q^+ \over [ q^2 - M^2_0(k^+, {\bf k}_{\perp}; q^+,
{\bf 0}_{ \perp})  + \imath\epsilon ]}
\times \nonu
\psi_{\bar{\pi}}(k^+, {\bf k}_{\perp}; P^+_{\bar{\pi}}, {\bf 0}_{ \perp})
 ~ [m^2_{\pi} - M^2_0(k^+, {\bf k}_{\perp}; P^+_{\bar{\pi}},
{\bf 0}_{ \perp})] / P^+_{\bar{\pi}}
 \nonu
= \int^{1}_0
\frac{dx}{x} ~ \int d{\bf k}_\perp ~ \overline  {\cal{D}}_{\pi} ~
 { M^2_0(k^+, {\bf k}_{\perp}; q^+,
{\bf 0}_{ \perp})
 \over  q^+ ~[ q^2 - M^2_0(k^+, {\bf k}_{\perp}; q^+,
{\bf 0}_{ \perp})  + \imath\epsilon ]}
 \times  \nonu
\psi_{\bar{\pi}}(k^+, {\bf k}_{\perp}; P^+_{\bar{\pi}}, {\bf 0}_{ \perp})
 ~ [m^2_{\pi} - M^2_0(k^+, {\bf k}_{\perp}; P^+_{\bar{\pi}},
{\bf 0}_{ \perp})] / P^+_{\bar{\pi}}
~ ~~.
\label{deltajm}
\ee
Then one has
\be
\delta~j^+ ~ q^- ~ + ~ \delta~j^- ~ q^+ \propto \nonu
\propto ~ - ~ q^2 ~ \int^{1}_0
\frac{dx}{x} ~ \int d{\bf k}_\perp  ~
{\overline  {\cal{D}}_{\pi} ~ \psi_{\bar{\pi}}(k^+, {\bf k}_{\perp};  P^+_{\bar{\pi}}, {\bf 0}_{ \perp})
~ [m^2_{\pi} - M^2_0(k^+, {\bf k}_{\perp}; P^+_{\bar{\pi}},
{\bf 0}_{ \perp})]
 \over [q^2 - M^2_0(k^+, {\bf k}_{\perp}; q^+, {\bf 0}_{ \perp}) +\imath\epsilon ] ~ P^+_{\bar{\pi}}} +
 \nonu
 + ~ \int^{1}_0
\frac{dx}{x} ~ \int d{\bf k}_\perp ~ M^2_0(k^+, {\bf k}_{\perp}; q^+, {\bf 0}_{ \perp}) ~
{ \overline  {\cal{D}}_{\pi} ~
\psi_{\bar{\pi}}(k^+, {\bf k}_{\perp}; P^+_{\bar{\pi}}, {\bf 0}_{ \perp})
 ~ [m^2_{\pi} - M^2_0(k^+, {\bf k}_{\perp}; P^+_{\bar{\pi}},
{\bf 0}_{ \perp})]
 \over  ~[ q^2 - M^2_0(k^+, {\bf k}_{\perp}; q^+,
{\bf 0}_{ \perp})  + \imath\epsilon ] ~ P^+_{\bar{\pi}}}
\nonu =
- ~ \int^{1}_0
\frac{dx}{x} ~ \int d{\bf k}_\perp  ~
{\overline  {\cal{D}}_{\pi} ~ \psi_{\bar{\pi}}(k^+, {\bf k}_{\perp};  P^+_{\bar{\pi}}, {\bf 0}_{ \perp})
~ [m^2_{\pi} - M^2_0(k^+, {\bf k}_{\perp}; P^+_{\bar{\pi}},
{\bf 0}_{ \perp})]
 \over P^+_{\bar{\pi}}} ~ \neq ~ 0 ~ ~~.
\label{deltajc}
\ee

Furthermore, in the present model the pion wave function at large momentum has the asymptotic form given by Eq. (\ref{model3}),
which decays as $1/|{\bf k}_{\perp}|^2$. Therefore the matrix element of the plus component of the bare term
of the electromagnetic current is ultraviolet divergent.
Therefore in the model of the present paper,
which deals with a massless pion, we give out the bare term, which gives an unphysical result.

\section{Normalization of the vector meson wave function}

The light-front wave function of  vector mesons includes the
relativistic spin part, as we have written in Eq. (\ref{wfn}), so
one has to consider the whole structure of the wave function to
obtain the normalization of the valence component of the state. We
normalize the wave function using the good component of the vector
current, imposing that
\begin{eqnarray}
\langle \phi_{n,~\lambda}|j^+(0)|\phi_{n,~\lambda}\rangle \ =2
M_{n} ~P_{q\overline q,n}\ ,
\label{normvec}
\end{eqnarray}
 where it appears the probability of the valence component in the
vector meson, $P_{q\overline q;n}$, which is  estimated in the next
Appendix E.

The matrix element of the good component of the current in impulse
approximation is represented by a Feynman triangle diagram.
After the integration over the light-front energy, performed disregarding
the singularities of the VM vertex in the $k^-$ complex-plane, one
obtains the contribution of the valence wave-function to the
normalization:
\begin{eqnarray}
P_{q\overline q,n}={1\over 2 M_{n}}{N_c \over 16 \pi^3}
\int_0^{M_{n}}{dk^+\over (k^+)^2(M_{n}-k^+)}
\int d{\bf k}_\perp
{\cal N}(k^+,{\bf k}_\perp)  ~\mid\psi_{n}(k^+, {\bf
k}_{\perp},M_{n},{\bf 0}_{ \perp })\mid^2~
  \ .
\label{normvec1}
\end{eqnarray}
In Eq. (\ref{normvec1}) the quantity  ${\cal N}(k^+,{\bf k}_\perp) $ is
the following trace
\begin{eqnarray}
 &&{\cal N}(k^+,{\bf k}_\perp)=
 \nonumber \\
 &&Tr \left [ [(\psla k - \psla P_n)_{on} + m]
~ [\psla \epsilon_\lambda + (k_{on} - (P_n - k)_{on})  \cdot
\epsilon_\lambda ~ H_S(M_0)] ~ (\psla k_{on} +  m)\gamma^+(\psla k_{on}+m) ~ \times
\right. \nonumber \\
&&\left. [ \psla
\epsilon_\lambda +(k_{on} - (P_n - k)_{on})\cdot \epsilon_\lambda
~H_S(M_0)]
\right ]
\end{eqnarray}
where $H_S(M_0^2)$ is defined in Eq. (\ref{hs}).

To evaluate the normalization, we choose the polarization in the
transverse direction, which is free of the pair term contribution in the limit of
zero momentum transfer \cite{pach97,pach98}.

\section{Estimate of the probability of the valence component}

In this Appendix we construct in the Fock space of constituent $q\overline q$
pairs a schematic model for the light-front square
mass operator, $\widehat M^2$, which allows one to roughly estimate
the probability of the valence
component in the $n{\rm th}$ excited vector meson state, $P_{q\overline q;n}$.
We are looking for a square mass operator in the Fock space with
 a spectrum where the mass of the $n {\rm th}th$ vector meson grows with
$ \sqrt{n}$, as occurs experimentally \cite{iach,ani}.

Let us denote by $i \ge 0$ the number of $q\overline q$ pairs
and by $|i\rangle_0$ the noninteracting Fock state with $i$
pairs.

Let us suppose that
the free mass operator, $\widehat M_0$, is additive in the number
of pairs and therefore that
the noninteracting Fock-state
$|i\rangle_0$ is eigenfunction of the free squared mass operator
with eigenvalue $\alpha^2~i^2$:
\be
\widehat M_0^2 |i\rangle_0 =
\alpha^2 \ i^2 ~|i\rangle_0 ~ .
\label{p1}
\ee
where $\alpha$ is the
energy of a free $q\overline q$ state. \

We suppose that the interaction, $M_I^2$, in the squared mass operator
($\widehat M^2 = \widehat M_0^2 + \widehat M_I^2$):

i)  has constant diagonal matrix elements
\be
_0\langle i|\widehat{M}_{I}^2|i\rangle_0= \frac{2}{\delta^2}
- c
\label{iqual1}
~,
\ee

ii) mixes the state
$|i\rangle_0$ with the states $|(i-1)\rangle_0$ and $|(i+1)\rangle_0$, and
is attractive and constant:
\be
_0\langle
(i+1)|\widehat{M}_{I}^2|i\rangle_0= ~_0\langle (i
-1)|\widehat{M}_I^2|i\rangle_0=-\frac{1}{\delta^2}
\label{iqual}
~,
\ee
while the other matrix elements of $\widehat M^2_I$
are supposed to be zero.

 The eigenvalue equation for the squared mass operator is
 \be
 \widehat{M}_0^2 ~ |n\rangle + \widehat{M}_{I}^2 ~ |n\rangle = M_n^2 ~ |n\rangle ~,
  \label{mass}
  \ee
where the $n{\rm th}$ excited state of the meson has mass $M_n$.
The VM wave function for the $n{\rm th}$ excited state  in the Fock space is given by
\be
|n\rangle=\sum_{i~\ge ~ 1}
~a_{n,i}~|i\rangle_0 ~ ~~,
\label{wf}
\ee
and the amplitudes $a_{n,i}$ are normalized as follows :
\be
\sum_{i~\ge ~ 1}
~ |a_{n,i}|^2 = 1 ~ ~~.
\label{norm}
\ee
In the above sums one has $i~\ge ~ 1$, since the vector mesons have quantum numbers
different from the vacuum and then $a_{n,0} = 0$.

Introducing the interaction defined by Eqs. (\ref{iqual1}, \ref{iqual}) into
Eq. (\ref{mass}) and projecting the eigenvalue equation
 in the Fock-space state basis $\{|i\rangle_0 \}$, one has:

\be
\alpha^2 i^2 a_{n,i} -
\frac{a_{n,i+1}-2 a_{n,i} + a_{n,i-1}}{\delta^2} = \left(M_n^2 + c
\right)a_{n,i} ~~.
\label{fineq}
\ee
If we  define $x = \Delta \cdot i$, with $ \Delta > 0$, then
Eq. (\ref{fineq}) can be rewritten as follows
\be
\frac{\delta^2 \alpha^2}{\Delta^4} x^2 a_{n,i} -
\frac{[ a_{n,i+1}-2 a_{n,i} + a_{n,i-1} ]}{\Delta^2} = \frac{\delta^2}{\Delta^2} \left(M_n^2 + c
\right)a_{n,i}~
\label{finequ}
\ee
Then defining $\alpha \delta / \Delta ^2 = \Omega/4$ and going to the
continuous limit
 one gets:
\be
\frac{\Omega^2}{16} x^2 a_n(x)- \frac{d^2 a_n(x)}{dx^2}=
\lambda_n ~ a_n(x) \ ,
\label{difeq}
\ee
with the boundary condition  $a_n(0) = 0$, in order to reflect the constraint  $a_{n,0} = 0$.
Trivially, because of this constraint
 the eigenvalues and the eigenstates of $\widehat M^2$ correspond
to the {\em odd } eigenvalues
and eigenstates of the unidimensional harmonic oscillator, namely
\be
\lambda_n={\Omega \over 2} \left (n+{1\over 2}\right )
\label{aval} \\ &&
a_n(x)= \sqrt{{2 \over 2^n~  n! }}~\left[{\Omega \over 4\pi}\right]^{1/4} ~H_n(\xi)~
e^{-\xi^2/2}
\label{afun}
\ee
where $H_n(\xi)$ are  the Hermite polynomials and $$\xi=\sqrt{\Omega \over 4}x ~~~.$$
The function $a_n(x)$ in our case has to be normalized as follows
\be
\int_0^{\infty} |a_n(x)|^2 dx = 1
\ee
in order to have the proper correspondence with Eq. (\ref{norm}). This
normalization explains the presence of
the factor $\sqrt{2}$ in the definition of $a_n(x)$ (Eq. (\ref{afun})).

Then, defining $\alpha / \delta = \omega/4$, from the eigenvalues $\lambda_n$ of
Eq. (\ref{difeq}) one can obtain the  masses of the vector mesons, viz
\be
M_{n}^2 = \frac{\omega}{2} ~ (n+ \frac12) - c
= {\omega} ~ ( n_{\text {ex}} + \frac 3 4) - c
\; \; ,
\label{avalo}
\ee
 where only the odd values of $n$, namely $n = 2n_{\text {ex}} + 1$, are allowed.
 The number $n_{\text {ex}}$ is the
excitation number 0, 1, 2, 3 ... of the vector mesons, with zero for the meson ground
state. Then the square mass of the vector mesons is given
by:
\be
M_{n_{\text {ex} }}^2 = \omega\cdot n_{\text{ex}} +
M^2_{\text{g.s.}}~,
\label{av}
\ee
where the meson ground state  has mass
$M_{\text{g.s.}}^2 = 3 \omega /4 - c $ (note that, given $\omega$, the constant $c$ is fixed).

  At this point we exactly retrieve the
experimental spectrum law \cite{iach,ani}, which is the motivation of
this  simple model.

The final step is the estimate of the probability of the
valence state, namely $P_{q\bar q,n}$.
 From a comparison of the discrete and the continuum  case
of our model,
one can associate to the probability $|a_{n,i}|^2$ of the wave function
component with $i$ $q\overline q$ pairs, Eq. (\ref{wf}) the quantity
\be
\int_{(i -1)\Delta}^{i\Delta} dx ~|a_n(x)|^2 ={2 \over 2^n~  n! }~\left[{\Omega \over
4\pi}\right]^{1/2} \int_{(i - 1)\Delta}^{i\Delta} dx~|H_n(\xi)|^2~e^{-\xi^2}=
\nonu=
{2 \over 2^n~  n! \sqrt{\pi}}~
\int_{(i -1)\delta \sqrt{\omega}/2}^{i\delta \sqrt{\omega}/2} d\xi~|H_n(\xi)|^2~e^{-\xi^2} ~~.
\label{probi}
\ee
Then the probability
$P_{q\bar q,n_{\text{ex}}}$ of the valence component is given by
the quantity
\be
P_{q\bar q,n_{\text{ex}}} = ~|a_{n,1}|^2 \sim \int_{0}^{\Delta} dx ~|a_n(x)|^2 =
{2 \over 2^n~  n! \sqrt{\pi}}~ \int_0^{\delta \sqrt{\omega}/2} d\xi~|H_n(\xi)|^2~e^{-\xi^2} ~~.
\label{prob1}
\ee

Imposing that  $P_{q\overline q;0}$,
i.e. the valence component probability in the
ground state of the vector mesons, is about the same as the one
found in constituent quark models of the pion \cite{pach02}, i.e. it is equal to
$0.77$,  we obtain  $\omega^\frac12 \delta =2.94 $.

Alternatively, let us evaluate the average number of $q\overline q$ pairs in the vector meson:
\be
\sqrt{ <i^2>} =
\frac{\sqrt{<x^2>}}{\Delta} =
\frac{2}{\Omega^\frac12 \Delta} \sqrt{n + \frac{1}{2}} ~ = ~
\frac{2}{\omega^\frac12 \delta} ~ \sqrt{2 n_{ex}+ \frac{3}{2}}
\ee
Then we can estimate the probability for the lowest Fock component
to be roughly given by
\be P_{q\overline q;n_{\text{ex}}} =
\frac{\delta ~ \omega^\frac12 }{2 ~ \sqrt{ 2 n_{ex}+ \frac{3}{2}}}
\label{pest} ~ \ .
\ee
If the value $\omega^\frac12 \delta = 2$ is used in this last estimate, the probabilities
considered in Ref. \cite{DFPS} are obtained. With this choice, the valence component probability
in the
ground state of the vector meson is equal to
$\sim 0.8$


\newpage

\begin{table}
\caption{ Known vector meson
 masses, $M_n$, and widths, $\Gamma_n$, used in the model. The corresponding
decay widths  into $e^-e^+$ pairs, calculated with the VM valence probabilities
$P_{q\overline q;n}$ obtained in Appendix E (see also Table II) and the
oscillator strength $\omega = 1.556$ $GeV^2$, are compared with
 the experimental values from
\cite{pdg}.
( See text for details)
}
\vspace{0.5cm}
\begin{tabular}{|c|c|c|c|c|c|c|}\hline
Meson&$M_n$ (MeV) &$M_n^{\text{exp}}$ (MeV) \cite{pdg} &$\Gamma_n$ (MeV) &
$\Gamma^{\text{exp}}_n$ (MeV) \cite{pdg} &
$\Gamma_{e^+ e^-}$ (KeV) & $\Gamma_{e^+ e^-}^{\text{exp}}$ (KeV) \cite{pdg} \\
\hline\hline $\rho$(770) & 770  & 775.8 $\pm$ 0.5     &  146.4  & 146.4
$\pm$ 1.5
& 6.98 & 7.02 $\pm$ 0.11  \\ 
${\rho}$(1450)& 1497 \cite{Akhm} & 1465.0 $\pm$ 25.0  &  226 \cite{Akhm} & 400 $\pm$ 60
& 1.04  & 1.47 $\pm$ 0.4\\ 
${\rho}$(1700)& 1720  & 1720.0 $\pm$ 20.0  & 220  & 250 $\pm$ 100
& 0.98  & $>$ 0.23 $\pm$ 0.1  \\ 
${\rho}$(2150) & 2149  & 2149.0 $\pm$ 17  &230 \cite{anis} & 363 $\pm$ 50 &
0.65 & -
\\  \hline
\end{tabular}
\end{table}

\clearpage
\newpage

\begin{table}
\caption{The vector-meson  valence probabilities
$P_{q\overline q;n}$ for the first 10 resonances. } \vspace{0.5cm}
\begin{tabular}{|c|c|c|c|c|c|c|c|c|c|c|}\hline
n&0&1 &2 &3 &4 &5 &6 &7 &8 &9\\
\hline
$P_{q\overline q;n}$&~0.77 ~&~0.31 ~&~0.29 ~&~0.27 ~&~0.22 ~&~0.18 ~&~0.18~ &~0.18~ &~0.17~&~0.16~
\\  \hline
\end{tabular}
\end{table}
\clearpage

\newpage

\begin{table}
\caption{ Vector meson masses, $M_n$, and widths,  $\Gamma_n$, used in Ref. \cite{DFPS}.
The corresponding  decay width into $e^-e^+$ pairs calculated with
  the VM  valence probabilities
$P_{q\overline q;n} =\frac{1 }{ ~ \sqrt{ 2 n+ \frac{3}{2}}}$
and the oscillator strength $\omega = 1.39$ $GeV^2$, are reported in the sixth column.
($^\dagger$The value of 180 MeV for the
width of $\rho(2150)$ is the lower bound of the value obtained by Anisovitch
et al quoted in \cite{anis}.)} \vspace{0.5cm}
\begin{tabular}{|c|c|c|c|c|c|c|}\hline
Meson&$M_n$ (MeV) &$M_n^{\text{exp}}$ (MeV) \cite{pdg}&$\Gamma_n$ (MeV) &
$\Gamma^{\text{exp}}_n$ (MeV) \cite{pdg}&
$\Gamma_{e^+ e^-}$ (KeV) & $\Gamma_{e^+ e^-}^{\text{exp}}$ (KeV)  \cite{pdg}\\
\hline\hline $\rho$(770) & 750  & 775.8 $\pm$ 0.5     &  149  & 146.4
$\pm$ 1.5
& 6.37 & 7.02 $\pm$ 0.11  \\ 
${\rho}$(1450)& 1465 & 1465.0 $\pm$ 25.0  &  310  & 400 $\pm$ 60
& 1.61  & 1.47 $\pm$ 0.40  \\ 
${\rho}$(1700)& 1723  & 1720.0 $\pm$ 20.0  & 240  & 250 $\pm$ 100
& 1.23  & $>$ 0.23 $\pm$ 0.1  \\ 
${\rho}$(2150) & 2150 & 2149.0 $\pm$ 17  &180$^\dagger$ & 363 $\pm$ 50 &
0.78 & -
\\  \hline
\end{tabular}
\end{table}

\clearpage

\begin{figure}[thb]
\centering
\includegraphics[width=5. in,angle=0]{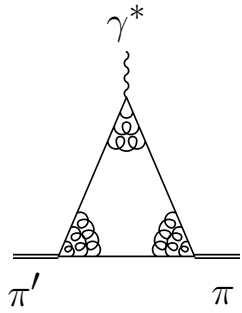}
\vskip -11 cm
\caption{Covariant amplitude for $\pi~\gamma^*
\rightarrow \pi'$, or $\gamma^* \rightarrow \pi \pi'$. The final
meson $\pi'$ is a pion in the elastic case or an antipion in
the production process.} \label{fig1}
\end{figure}

\vskip -4.3  cm

 \newpage

\begin{figure}[thb]
\centering
\includegraphics[width=5. in,angle=0]{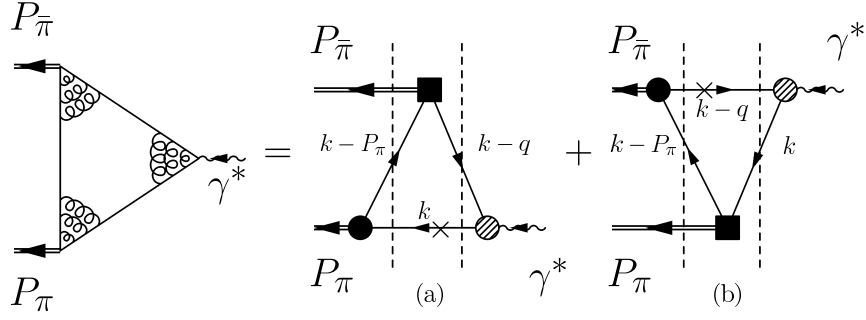}
\vskip -12 cm
\caption{Dressed photon-decay amplitude ($\gamma^*
\rightarrow \pi \bar{\pi}$) with two possible $x^+$ time (light-front time)
orderings, represented by diagrams (a) and (b). The two dashed vertical lines
represent different light-front times (the light-front time flows from the
right to the left).  Diagrams $(a)$ and
diagram $(b)$ contain the processes $ q \rightarrow q \bar{\pi}$ and
$ q \rightarrow q {\pi}$, respectively, represented by a full square. The crosses indicate
the quark lines which are on shell, after the $k^-$ integration. The
dashed circle represents the dressed photon vertex(see Fig. 4 for details). }
\label{fig2}
\end{figure}

 \newpage

\begin{figure}[thb]
\centering
\includegraphics[width=5. in,angle=0]{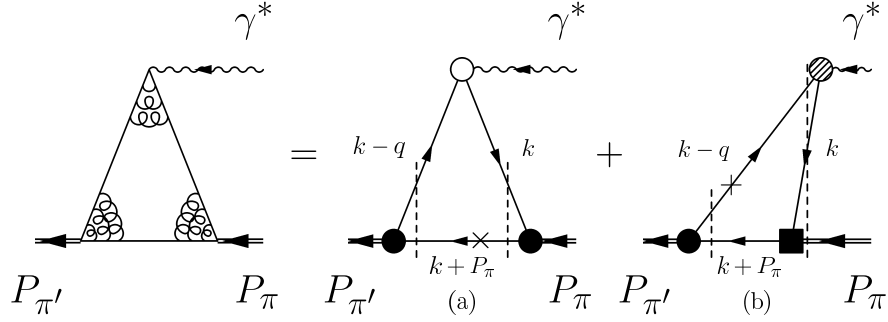}
\vskip -12 cm
\caption{Diagrammatic representation of the space-like
elastic form factor of the pion for $q^+ >0$.
The light-front time ordering allows one to single out two-quark
and four-quark configurations at different light-front times,
 as indicated by the dashed vertical lines.
Diagram $(a)$,
where $0\le -k^+ \le P^+_{\pi}$, represents the contribution of the
valence component in the  wave function of the initial pion.
Diagram $(b)$, where $0 \le k^+ \le q^+$, represents
the non-valence contribution to the pion form factor (pair production process). Both
processes contain the contribution from the dressed photon
vertex. The full square is the vertex function which describes a pion absorption by a quark.
The
dashed circle represents the dressed photon vertex(see Fig. 4 for details).}
\label{fig3}
\end{figure}

 \newpage

\begin{figure}[thb]
\centering
\includegraphics[width=5. in,angle=0]{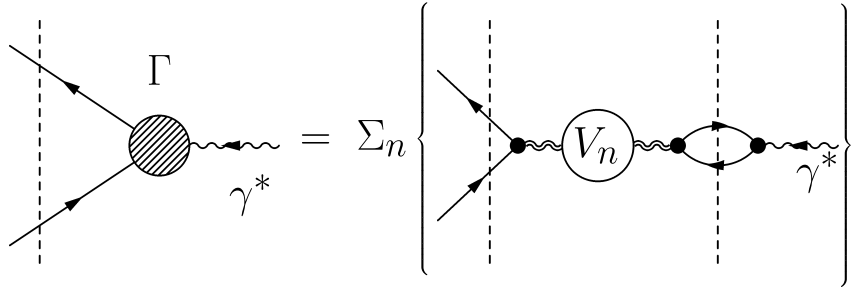}
\vskip -12 cm
\caption{Dressed photon vertex. The double-wiggly
lines represent the Green function describing the
 propagation of the vector meson $V_n$. The loop on the right represents
 the VM decay constant, $f_{Vn}$ (see Appendix A).}
\label{fig7}
\end{figure}

\newpage

\begin{figure}[thb]
\centering
\includegraphics[width=5. in,angle=0]{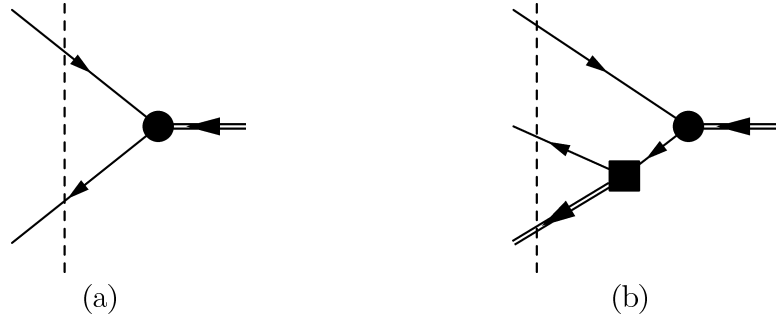}
\vskip -12  cm
\caption{ (a) Valence component, $|q\bar{q} \rangle$, of the wave function of an incoming
system (photon or meson). (b) Non valence component,
 $|q \bar{q} q \bar{q}\rangle$, of the wave function of the same incoming
 system. The extra $q \bar{q}$ pair is radiatively emitted by a quark in the valence
 component }
 \label{fig4}
\end{figure}

\newpage

\begin{figure}[thb]
\centering
\includegraphics[width=5. in,angle=0]{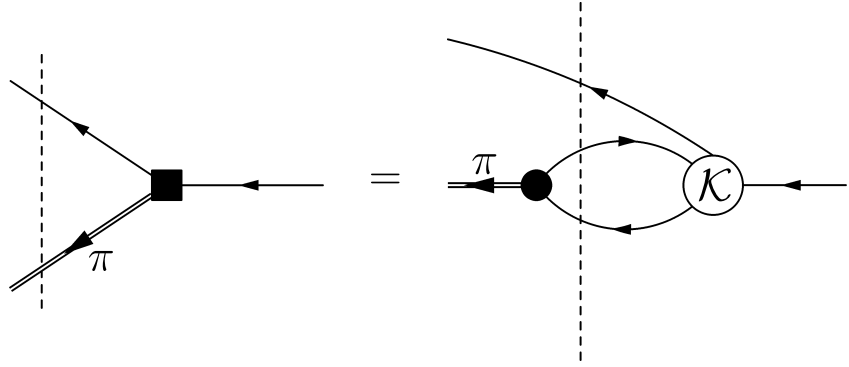}
\vskip -12 cm
\caption{Virtual decay amplitude for a pion emission from a quark ($q
\rightarrow q\pi$) produced by the operator ${\cal K}$, see Eq. (\ref{piemi}).
 The analogous diagram for pion absorption by a quark
can be easily obtained by replacing the final pion leg with an initial pion leg.}
\label{fig8}
\end{figure}

 \newpage

\begin{figure}[thb]
\centering
\includegraphics[width=5. in,angle=0]{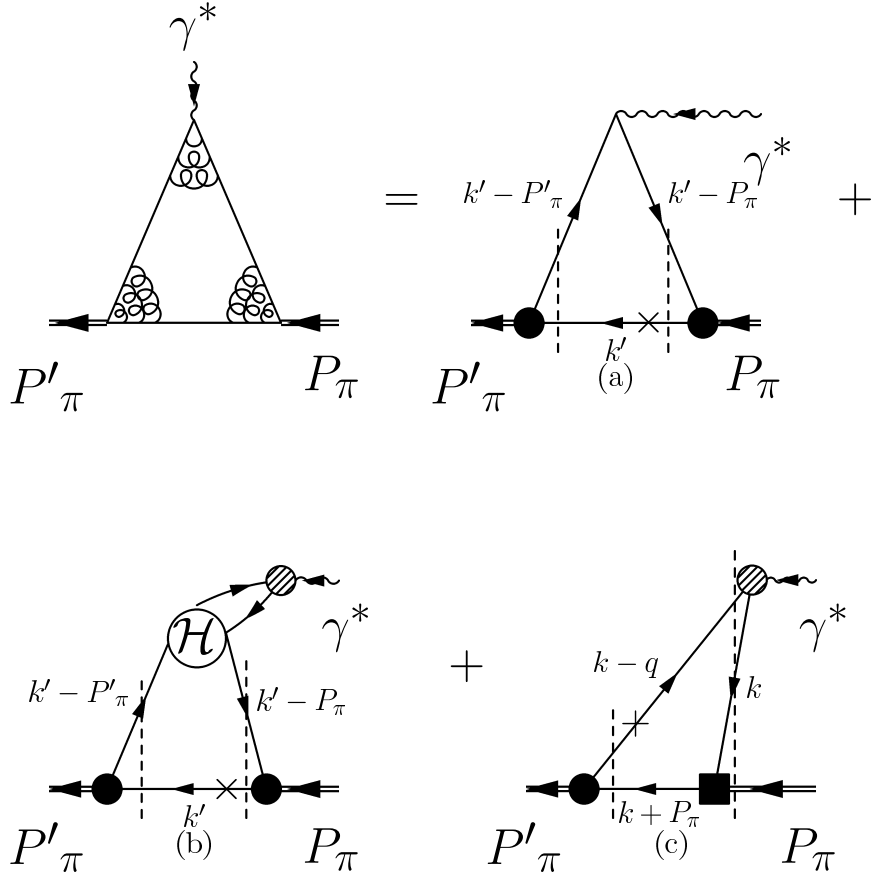}
\vskip -6 cm
 \caption{ Space-like em form factor of the pion, $\pi~\gamma^*
\rightarrow \pi'$, for $q^+ >0$. Two possible light-front time orderings are shown.
 The first one allows to single out the following processes: (a),  where
 a point-like quark-photon interaction occurs, and
 (b), where the absorption of a $q\bar{q}$ pair by a quark proceeds through the kernel ${\cal H}$.
Diagram (c), where the process $\gamma^* ~ \rightarrow ~ q\bar{q}$ appears
(pair production process) with the
subsequent absorption of the initial pion by a quark, corresponds to the second time ordering. }
\label{fig5}
\end{figure}

 \newpage

\begin{figure}[thb]
\centering
\includegraphics[width=5. in,angle=0]{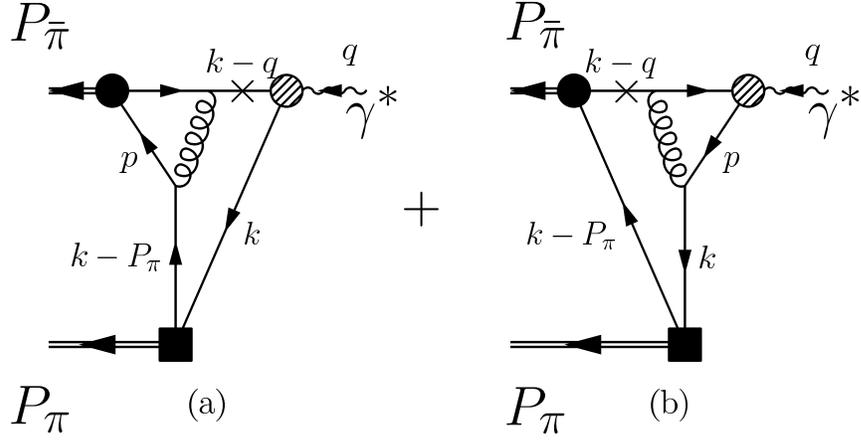}
\vskip -10 cm
\caption{ Instantaneous contributions to the time-like em form factor of a massless
pion. The instantaneous quark
leg is attached to the pion vertex in (a) and to VM vertex in (b).  The
dashed circle represents the dressed photon vertex (see Fig. 4 for details). }
\label{fig6}
\end{figure}

 \newpage

\begin{figure}[thb]
\centering
\includegraphics[width=6.3 in,angle=0]{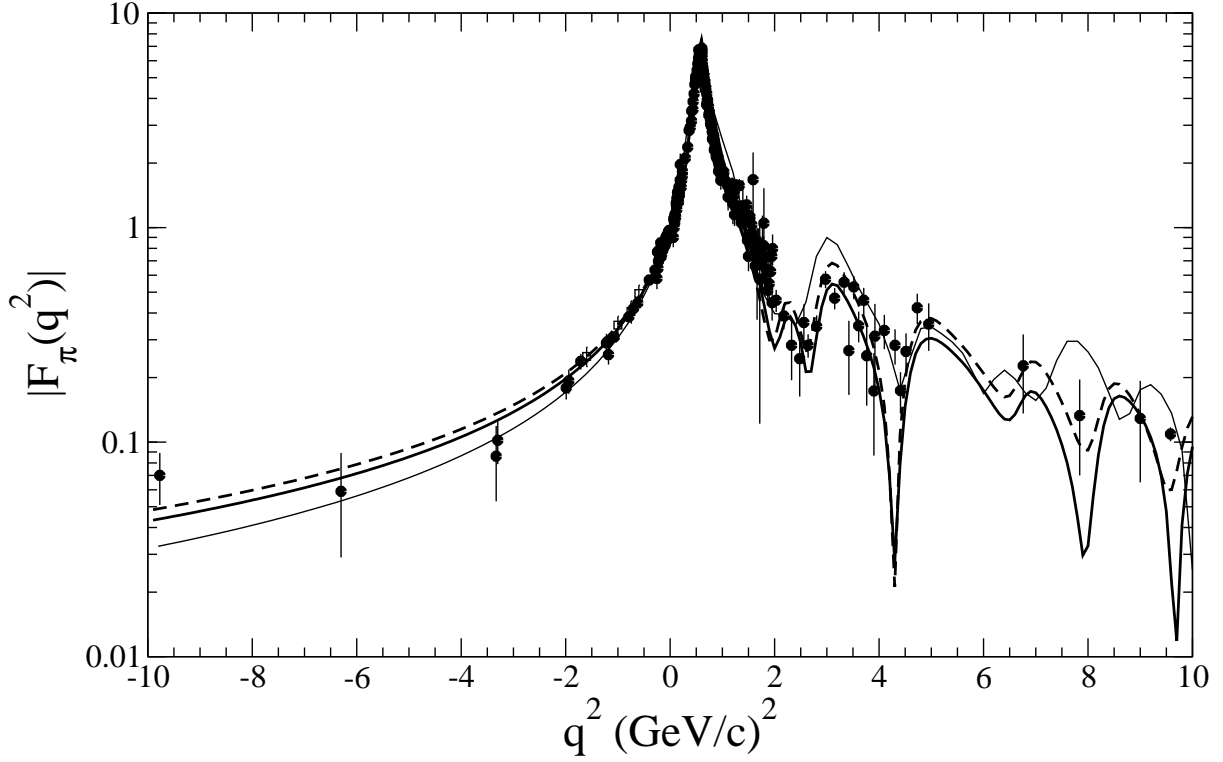}
\caption{Pion electromagnetic form factor as a function of the
momentum squared $q^2$. Results  for the asymptotic and the full pion
wave functions, obtained with $w_{VM}=-0.7$ (see Sect. IX) and the quantities
shown in Table I,
 are indicated by dashed and solid curves, respectively.
The thin solid line represents the result with $w_{VM}=-1/3$ and the parameters of Table III.
Experimental data are from Ref. \cite{baldini} (full dots)
and Ref. \cite{JLABp} (open squares). }
\label{fig9}
\end{figure}

\newpage

\begin{figure}[thb]
\centering
\includegraphics[width=6.3 in,angle=0]{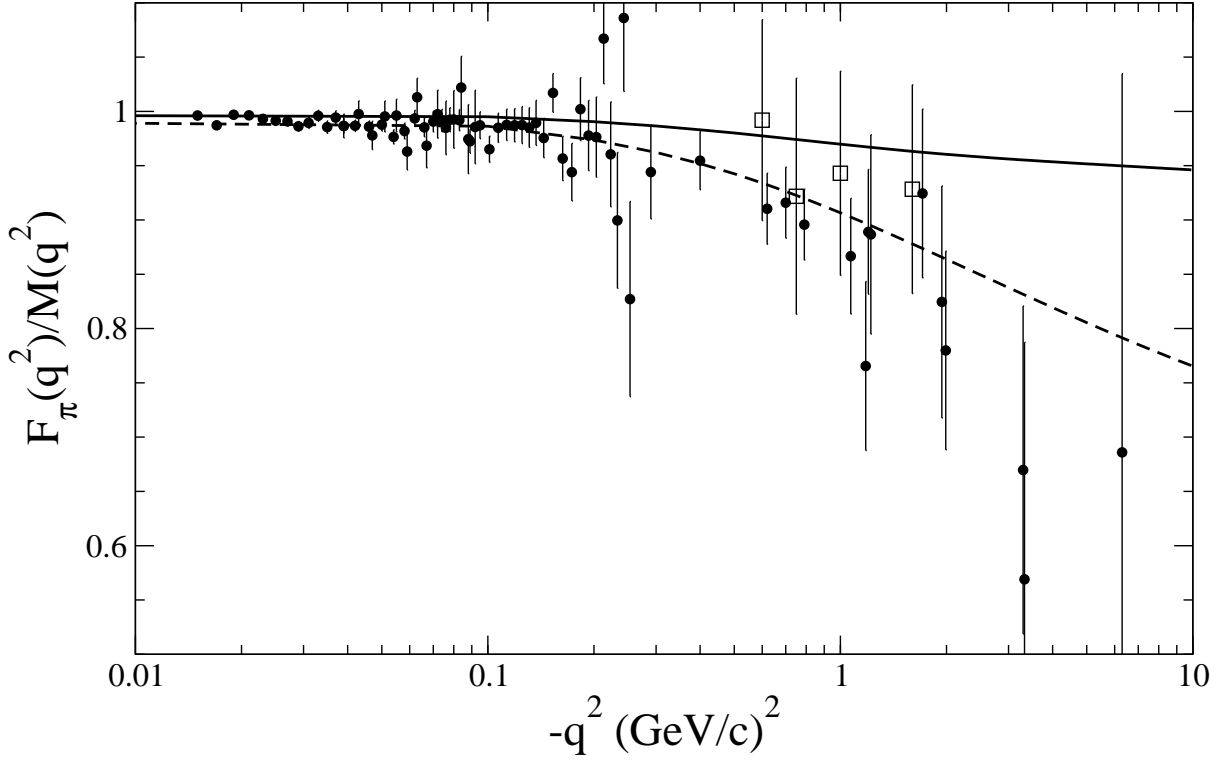}
\caption{Space-like pion electromagnetic form factor divided by the monopole
$M(q^2) = 1/(1 - q^2/M_{\rho}^2)$ vs the momentum squared $q^2$. The solid curve corresponds to
$w_{VM} = - 1.5$ and the dashed line to $w_{VM} = - 0.7$, all the other
quantities are according to Table I.
The experimental data are as in Fig. \ref{fig9}.}
\label{fig10}
\end{figure}

\begin{figure}[thb]
\centering
\includegraphics[width=6.3 in,angle=0]{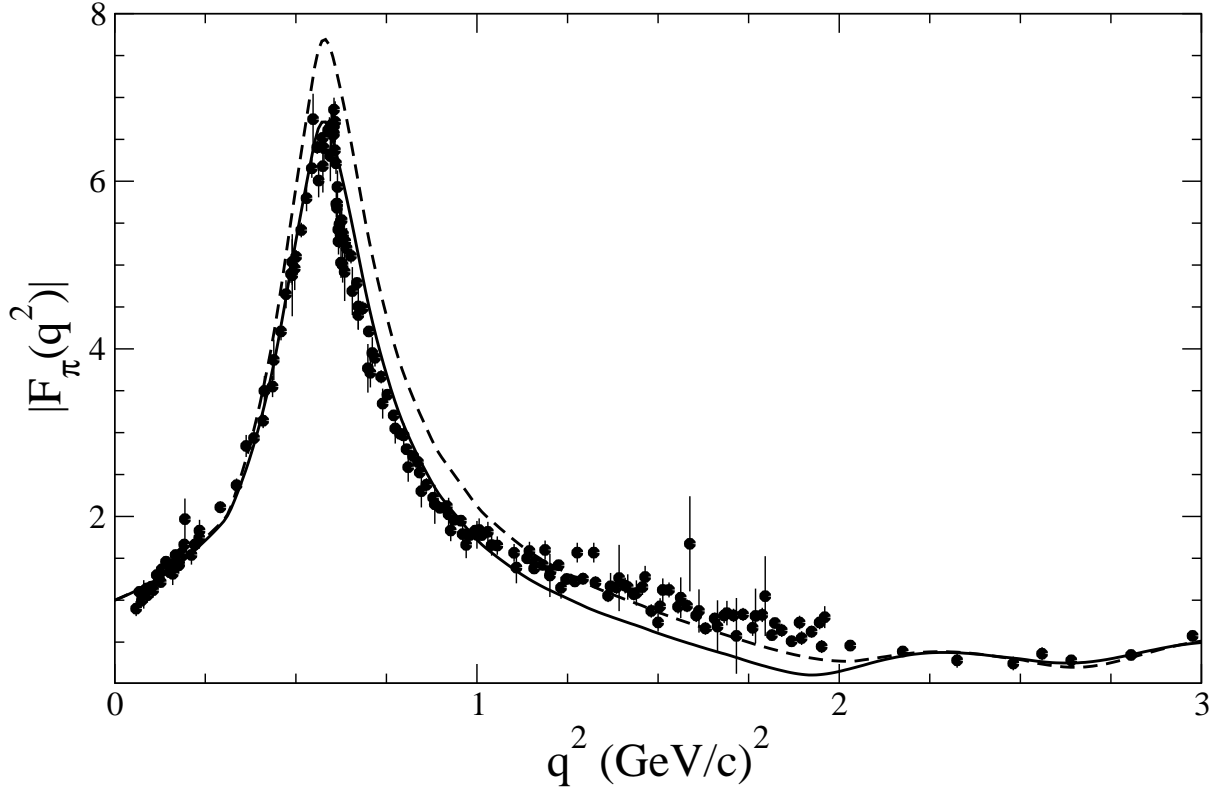}
\caption{Pion electromagnetic form factor as a function of the
momentum squared $q^2$. The solid curve corresponds to
$w_{VM} = - 1.5$ and the dashed line to $w_{VM} = - 0.7$, all the other
quantities are according to Table I.
The experimental data are as in Fig. \ref{fig9}. }
\label{fig11}
\end{figure}

\end{document}